\documentclass[aps,pre,onecolumn,preprint,showpacs,superscriptaddress,groupedaddress]{revtex4}
\usepackage{graphicx}
\usepackage{amsmath,amssymb}
\usepackage{xcolor}

\newcommand{\Del}{\Delta}
\newcommand{\del}{\delta}
\newcommand{\al}{\alpha}
\newcommand{\Lam}{\Lambda}
\newcommand{\lam}{\lambda}

\newcommand{\ga}{\gamma}
\newcommand{\vth}{\vartheta}
\newcommand{\vsa}{\varsigma}
\newcommand{\vphi}{\varphi}

\newcommand{\sech}{\mbox{ sech}}
\begin{document}

\title{Nondegenerate Solitons and their Collisions in Manakov System}%
\author{R. Ramakrishnan}%
\author{  S. Stalin}
\author{M. Lakshmanan \footnote{Corresponding author E-mail: lakshman@cnld.bdu.ac.in}}
\affiliation{Department of Nonlinear Dynamics, School of Physics, Bharathidasan University, Tiruchirapalli--620 024, India}

\date{December 2018}%
\begin{abstract}
  Recently, we have shown that the Manakov equation can admit a more general class of nondegenerate vector solitons, which can undergo collision without any intensity redistribution in general among the modes, associated with distinct wave numbers, besides the already known energy exchanging solitons corresponding to identical wave numbers. In the present comprehensive paper, we discuss in detail the various special features of the reported nondegenerate vector solitons.  To bring out these details, we derive the exact forms of such vector one-, two- and three-soliton solutions through Hirota bilinear method and they are rewritten in more compact forms using Gram determinants.  The presence of distinct wave numbers allows the nondegenerate fundamental soliton to admit various profiles such as double-hump, flat-top and  single-hump structures. We explain the formation of double-hump structure in the fundamental soliton when the relative velocity of the two modes tends to zero.  More critical analysis shows that the nondegenerate fundamental solitons can undergo shape preserving as well as shape altering collisions under appropriate conditions.  The shape changing collision occurs between the modes of nondegenerate solitons when the parameters are fixed suitably. Then we observe the coexistence of degenerate and nondegenerate solitons when the wave numbers are restricted appropriately in the obtained two-soliton solution.  In such a situation we find the degenerate soliton induces shape changing behavior of nondegenerate soliton during the collision process. By performing suitable asymptotic analysis we analyze the consequences that occur in each of the collision scenario. Finally we point out that the previously known class of energy exchanging vector bright solitons, with identical wave numbers, turns out to be a special case of the newly derived nondegenerate solitons.
\end{abstract}	
\maketitle
\section{Introduction}
The propagation of light pulses in optical Kerr media is still one of the active areas of research in nonlinear optics \cite{d1}. In particular the fascinating dynamics of light in multi-mode fibers and fiber arrays has stimulated the investigation on temporal multi-component/vector solitons over different aspects, especially from the applications point of view \cite{gp}. In the nonlinear optics context, temporal vector solitons are formed due to the balance between dispersion and Kerr nonlinearity. Mathematically these vector solitons are nothing but the solutions of certain integrable coupled nonlinear Schr\"{o}dinger family of equations.  There exist many types of vector solitons which have been reported so far in  the literaure and their dynamics have also been investigated in various physical situations. For instance,  bright-bright solitons \cite{i,j,m1}, bright-dark solitons \cite{k00,bd1,bd2,dd1} and dark-dark solitons \cite{k00,dd} are some of the solitons which have been investigated in these systems. These vector solitons have also received considerable attention  in other areas of science including Bose-Einstein condensates (BECs) \cite{dj1,dj2}, bio-physics \cite{sc}, plasma physics \cite{cr} and so on.  Apart from the above, partially coherent solitons/soliton complexes have been reported in self-induced multi-mode waveguide system \cite{v,h}, while polarization locked solitons and phase locked solitons in fiber lasers \cite{new1} and dissipative vector solitons in certain dissipative systems \cite{new2,new2a,new2b} have also been analyzed in the literature.  

From the above studies on vector solitons we have noted that the intensity profiles of multi-component solitons reported, especially in the integrable coupled nonlinear Schr\"{o}dinger systems, are defined by identical wave numbers in all the components. We call these vector solitons as degenerate class of solitons. As a consequence of degeneracy in the wave numbers, single-hump strcutured intensity profiles only emerge in these systems in general \cite{ss1}. In the coherently coupled system even degenerate fundamental soliton can also admit double-hump profile when the four wave mixing process is taken into account  \cite{k1,k11}. However, in this case one can not expect more than a double-hump profile. Very interestingly our theoretical \cite{i,j} and other experimental \cite{k,l,l11} studies confirm that the degenerate vector solitons undergo in general energy redistribution among the modes during the collision, except for the special case of polarization parameters satisfying specific restrictions, for example in the case of two component Manakov systems as $\frac{\alpha_1^{(1)}}{\alpha_2^{(1)}}=\frac{\alpha_1^{(2)}}{\alpha_2^{(2)}}$ where $\alpha _i^{(j)}$'s, $i, j = 1,2$, are complex numbers related to the polarization vectors. By exploiting the fascinating shape changing collision scenario of degenerate Manakov solitons, it has been theoretically suggested that the construction of optical logic gates is indeed possible, leading to all optical computing \cite{ma}. We also note that logic gates have been implemented using two stationary dissipative solitons of complex Ginzburg-Landau equation \cite{lgd}.  

Recently in Refs. \cite{new3,new3a,new3b} it has been reported that multi-hump structured dispersion managed solitons/double-hump intensity profile of soliton molecule may be useful for application in optical communications because they may provide alternative coding schemes for transmitting information with enhanced data-carrying capacity. Multi-hump solitons have also been identified in the literature in various physical situations \cite{n,o,p,q,t,u,yang}. They have been observed experimentally in a dispersive nonlinear medium \cite{q}. Theoretically frozen double-hump states have been predicted in birefringent dispersive nonlinear media \cite{n,o}. These  solitons have been found in various nonlinear coupled field models also \cite{t}. In the case of saturable nonlinear medium, stability of double and triple-hump optical solitons has also been investigated \cite{u}. Multi-humped partially coherent solitons have also been investigated in photorefractive medium \cite{v}.  In addition to the above, the dynamics of double-hump solitons have also been studied in mode-locked fiber lasers \cite{new1,new2,new2a,new2b}. A double hump soliton has been observed  during the buildup process of soliton molecules in deployed fiber systems and fiber laser cavities \cite{new3,new4}. 

From the above studies, we observe that the various properties associated with the degenerate vector bright solitons of many integrable coupled field models have been well understood.  However, to our knowledge, studies on fundamental solitons with nonidentical wave numbers in all the modes have  not been considered so far and multi-hump structure solitons have also not been explored in the integrable coupled nonlinear Schr\"{o}dinger type systems except in our recent work \cite{ss} and that of Qin et al \cite{ss2} on the following Manakov system \cite{d,mja},    
\begin{equation}
iq_{jz}+q_{jtt}+2\sum_{p=1}^{2}|q_{p}|^{2}q_j=0,~~~~j=1,2,
\label{e1}
\end{equation}
where $q_{j}$, $j=1,2$,  describe orthogonally polarized complex waves in a birefringent medium. Here the subscripts $z$ and $t$ represent normalized distance and retarded time, respectively. Based on the above studies we are motivated to look for a new class of fundamental solitons, which possess nonidentical wave numbers as well as multi-hump profiles, which are useful for optical soliton based applications. We have successfully identified such a new class of solitons in \cite{ss}. We call the fundamental solitons with nonidentical wave numbers as nondegenerate vector solitons \cite{ss,ss1}. Surprisingly this new class of vector bright solitons    exhibit multi-hump structure (double-hump soliton arises in the present Manakov system and one can also observe $N$-hump soliton in the case of $N$-coupled Manakov type system) which may be useful for transmitting information in a highly packed manner.  Therefore it is very important to investigate the role of additional wave number(s) on the new class of fundamental soliton structures and collision scenario as well, which were briefly discussed in \cite{ss}. 
In the present comprehensive version we discuss the various properties associated with the nondegenerate solitons in a detailed manner by finding their exact analytical forms through Hirota bilinearization method. Then 
we discuss how the presence of additional distinct wave numbers and the cross phase modulation $(|q_1|^2+|q_2|^2)q_j$, $j=1,2$, among the modes bring out double-hump profile in the structure of nondegenerate fundamental soliton. We find that the nondegenerate solitons undergo shape preserving collision generally, as reported by us in \cite{ss}, and shape altering and shape changing collisions for specific parametric values. Further,  we figured out the coexistence of degenerate and nondegenerate solitons in the Manakov system. Such coexisting solitons undergo novel shape changing collision scenario leading to useful soliton based signal amplification application. Finally, we show that the degenerate class of vector solitons reported in \cite{i,j} can be deduced from the obtained nondegenerate two-soliton solution.

The structure of the paper is organized as follows: In section II, we discuss the Hirota bilinear procedure in order to derive nondegenerate soliton solutions for Eq. (\ref{e1}). Using this procedure we obtained nondegenerate one- and two-soliton solutions in Gram determinant forms and also identified the coexistence of degenerate and nondegenerate solitons in Section III. In Section IV  we discuss the various collision properties of nondegenerate solitons. Section V deals with the collision between degenerate and nondegenerate solitons. In Section VI we recovered the degenerate one- and two-soliton solutions from the nondegenerate one- and two-soliton solutions by suitably restricting the wave numbers and in Section VII we point out the possible experimental observations of nondegenerate solitons.  In Section VIII we summarize the results and discuss possible extension of this work. Finally in the Appendix A we present the three soliton solution in Gram determinant forms for completion while in Appendix B we discuss about certain asymptotic forms of solitons. In Appendix C, we introduce explicit forms of certain parameters appearing in the text.  Finally in Appendix D we discuss the numerical stability analysis of nondegenerate solitons under different strength of white noise as perturbation. 


\section{ Bilinearization}
To derive the nondegenerate soliton solutions for the Manakov system we adopt the same Hirota bilinear procedure that has been already used to get degenerate vector bright soliton solutions but with appropriate form of initial seed solutions. We point out later how such a simple form  of new seed solutions will produce remarkably new physically important class of soliton solutions. In general, the exact soliton solutions of Eq. (\ref{e1}) can be obtained by introducing the bilinearizing transformation, which can be identified from the singularity structure analysis of Eq. (\ref{e1}) \cite{saha} as
\begin{equation}
q_j(z,t)=\frac{g^{(j)}(z,t)}{f(z,t)}, ~~j=1,2,
\label{2.1}
\end{equation} 
to Eq. (\ref{e1}). This results in the following set of bilinear forms of Eq. (\ref{e1}),
\begin{subequations}\begin{eqnarray}
	(iD_z+D^2_t)g^{(j)} \cdot f&=&0, j=1,2, \label{2.2a}\\
	D^2_t f \cdot f&=&2 \sum_{n=1}^{2}g^{(n)}g^{(n)*}.
	\label{2.2b}
	\end{eqnarray}\end{subequations}
Here $g^{(j)}$'s are complex functions whereas $f$ is a real function and $*$ denotes complex conjugation. The Hirota's bilinear operators $D_z$ and $D_t$ are defined \cite{wa} by the expressions
$D_z^mD_t^n (a\cdot b)=\bigg(\frac{\partial}{\partial z}-\frac{\partial}{\partial z'}\bigg)^m\bigg(\frac{\partial}{\partial t}-\frac{\partial}{\partial t'}\bigg)^n a(z,t)b(z',t')_{\big|z=z',~t=t'}$. Substituting the standard expansions for the unknown functions $g^{(j)}$ and $f$,
\begin{eqnarray}
g^{(j)}&=&\epsilon g_1^{(j)}+\epsilon^3 g_3^{(j)}+...,~ j=1,2,\nonumber \\
f&=&1+\epsilon^2 f_2+\epsilon^4 f_4+...,
\label{2.4}
\end{eqnarray}
in the bilinear Eqs. (\ref{2.2a})-(\ref{2.2b}) one can get a system of linear partial differential equations (PDEs). Here $\epsilon$ is a formal series expansion parameter. The set of linear PDEs arises after collecting the coefficients of same powers of  $\epsilon$. By solving these linear PDEs recursively (at an appropriate order of $\epsilon$), the resultant associated explicit forms of $g^{(j)}$'s and $f$ constitute the soliton solutions to the underlying system (\ref{e1}). We note that the truncation of series expansions (\ref{2.4}) for the nondegenerate soliton solutions is different from degenerate soliton solutions. This is essentially due to the general form of seed solutions assigned to the lowest order linear PDEs. 

\section{A new class of nondegenerate soliton solutions}
To study the role of additional wave numbers on the structural, propagational and collisional properties of nondegenerate soliton it is very much important to find the exact analytical form of it systematically. In this section by exploiting the procedure described above we intend to construct nondegenerate   one- and two-soliton solutions which can be generalized to arbitrary $N$-soliton case (For $N=3$, see Appendix A below). In principle this is possible  because of the existence of nondegenerate $N$-soliton solution ensured by the complete integrability property of Manakov Eq. (\ref{e1}). Then we point out the possibility of coexistence of degenerate and nondegenerate solitons by imposing certain restriction on the wave numbers in the obtained nondegenerate two-soliton solution. Further we also point out the possibility of deriving this partially nondegenerate two-soliton solution through Hirota bilinear method. We note that to avoid too many mathematical details we provide the final form of solutions only since the NDS solution construction process is a lengthy one. 
\subsection{Nondegenerate fundamental soliton solution}
In order to deduce the exact form of nondegenerate one-soliton solution we consider two different seed solutions for the two modes as 
\begin{eqnarray}
g_1^{(1)}=\alpha_{1}^{(1)}e^{\eta_1}, ~~g_1^{(2)}=\alpha_{1}^{(2)}e^{\xi_1},
\label{3.1}
\end{eqnarray}
where $\eta_{1}=k_{1}t+ik_{1}^{2}z$ and $\xi_{1}=l_{1}t+il_{1}^{2}z$, to the following linear PDEs 
\begin{equation}
ig_{1z}^{(j)}+g_{1tt}^{(j)}=0,~j=1,2. \label{3.2}
\end{equation}
In (\ref{3.1}) the complex parameters $\alpha^{(j)}_1$, $j = 1, 2$, are arbitrary. The above equations arise in the lowest order of $\epsilon$.  The presence of two distinct complex wave numbers $k_1$ and $l_1$ ($k_1\neq l_1$, in general) in the seed solutions (\ref{3.1})  makes the final solution as nondegenerate one. This construction procedure is different from the standard one that has been followed in earlier works on degenerate vector bright soliton solutions \cite{i,j} where identical seed solutions of Eq. (\ref{e1}) (solutions (\ref{3.1}) with $k_1=l_1$ and distinct $\alpha_1^{(j)}$'s, $j=1,2$) have been used as starting seed solutions for Eq. (\ref{3.2}). We note that such degenerate seed solutions only yield degenerate class of vector bright soliton solutions \cite{i,j,ss}.    

With the starting solutions (\ref{3.1}) we allow the series expansions (\ref{2.4}) to terminate by themselves while solving the system of linear PDEs. From this recursive process, we find that the expansions (\ref{2.4}) get terminated for the nondegenerate fundamental sliton solution as, $g^{(j)}=\epsilon g_1^{(j)}+\epsilon^3 g_3^{(j)}$ and  $f=1+\epsilon^2 f_2+\epsilon^4 f_4$. The explicit expressions of $g_1^{(j)}$, $g_3^{(j)}$, $f_2$ and $f_4$ constitute a general form of new fundamental one-soliton solution to Eq. (\ref{e1}) as
\begin{eqnarray}
q_{1}=\frac{g_1^{(1)}+g_3^{(1)}}{1+f_2+f_4}=(\alpha_{1}^{(1)} e^{\eta_{1}}+e^{\eta_{1}+\xi_{1}+\xi_{1}^*+\Delta_{1}^{(1)}})/D_1 \nonumber\\ 
q_{2}=\frac{g_1^{(2)}+g_3^{(2)}}{1+f_2+f_4}=(\alpha_{1}^{(2)} e^{\xi_{1}}+e^{\eta_{1}+\eta_{1}^*+\xi_{1}+\Delta_{1}^{(2)}})/D_1.
\label{3.3}
\end{eqnarray}
Here $D_1=1+e^{\eta_{1}+\eta_{1}^{*}+\delta_{1}}+e^{\xi_{1}+\xi_{1}^{*}+\delta_{2}}+e^{\eta_{1}+\eta_{1}^{*}+\xi_{1}+\xi_{1}^{*}+\delta_{11}}$, $e^{\Delta_{1}^{(1)}}=\frac{(k_{1}-l_{1})\alpha_{1}^{(1)}|\alpha_{1}^{(2)}|^2}{(k_{1}+l_{1}^*)(l_{1}+l_{1}^*)^{2}}$, $e^{\Delta_{1}^{(2)}}=-\frac{(k_{1}-l_{1})|\alpha_{1}^{(1)}|^2\alpha_{1}^{(2)}}{(k_{1}+k_{1}^*)^{2}(k_{1}^*+l_{1})}$,  $e^{\delta_{1}}=\frac{|\alpha_{1}^{(1)}|^2}{(k_{1}+k_{1}^{*})^{2}}$, $e^{\delta_{2}}=\frac{ |\alpha_{1}^{(2)}|^2}{(l_{1}+l_{1}^*)^{2}}$ and 
$e^{\delta_{11}}=\frac{|k_{1}-l_{1}|^{2} |\alpha_{1}^{(1)}|^{2}|\alpha_{1}^{(2)}|^{2}}{(k_{1}+k_{1}^*)^{2}(k_{1}^*+l_{1})(k_{1}+l_{1}^*)(l_{1}+l_{1}^*)^{2}}$. 
\begin{figure}
	\centering
	\includegraphics[width=0.9\linewidth]{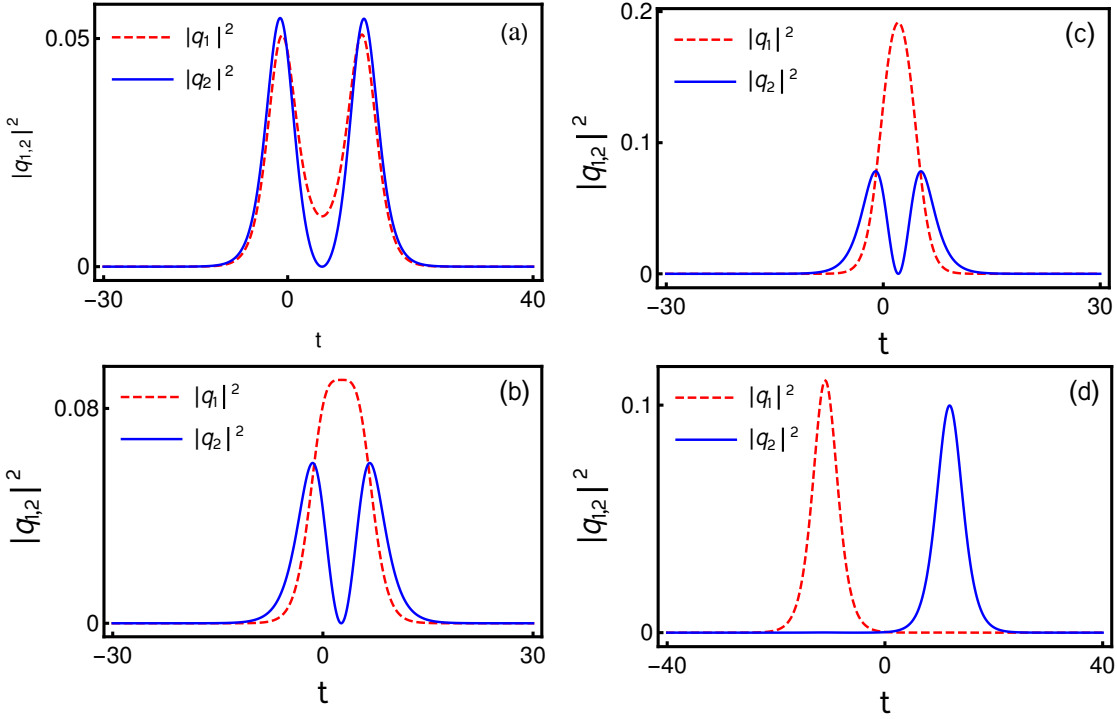}
	\caption{Various symmetric intensity profiles of nondegenerate fundamental soliton: While (a) denotes double-hump solitons in both the modes (b) and (c) represent flat-top-double-hump solitons and single-hump-double-hump solitons, respectively. Single-hump solitons in both the modes are illustrated in (d). The parameter values of each figures are: (a): $k_1=0.333+0.5i$, $l_1=0.315+0.5i$,  $\alpha_{1}^{(1)}=0.45+0.45i$, $\alpha_{1}^{(2)}=0.49+0.45i$. (b): $k_1=0.425+0.5i$, $l_1=0.3+0.5i$,  $\alpha_{1}^{(1)}=0.44+0.51i$, $\alpha_{1}^{(2)}=0.43+0.5i$. (c): $k_1=0.55+0.5i$, $l_1=0.333+0.5i$,  $\alpha_{1}^{(1)}=0.5+0.5i$, $\alpha_{1}^{(2)}=0.5+0.45i$. (d): $k_1=0.333+0.5i$, $l_1=-0.316+0.5i$,  $\alpha_{1}^{(1)}=0.45+0.5i$, $\alpha_{1}^{(2)}=0.5+0.5i$.  }
	\label{f1}
\end{figure}
In the above one-soliton solution two distinct complex wave numbers, $k_1$ and $l_1$, occur in both the expressions of $q_1$ and $q_2$ simultanously. This confirms that the obtained solution is nondegenerate. We also note that the solution (\ref{3.3}) can be rewritten in a more compact form using Gram determinants as
\begin{subequations}
	\begin{eqnarray}
	g^{(1)} =
	\begin{vmatrix}
	\frac{e^{\eta_1+\eta_1^*}}{(k_1+k_1^*)} & \frac{e^{\eta_1+\xi_1^*}}{(k_1+l_1^*)} & 1 & 0 & e^{\eta_1} \\ 
	\frac{e^{\xi_1+\eta_1^*}}{(l_1+k_1^*)} & \frac{e^{\xi_1+\xi_1^*}}{(l_1+l_1^*)}  & 0 & 1 &   e^{\xi_1}\\
	-1 & 0 & \frac{|\al_1^{(1)}|^2}{(k_1+k_1^*)} & 0 & 0\\
	0 & -1 & 0 &  \frac{|\al_1^{(2)}|^2}{(l_1+l_1^*)} & 0\\
	0 & 0& -\al_1^{(1)} & 0 &0
	\end{vmatrix},~~
	g^{(2)} =
	\begin{vmatrix}
	\frac{e^{\eta_1+\eta_1^*}}{(k_1+k_1^*)} & \frac{e^{\eta_1+\xi_1^*}}{(k_1+l_1^*)} & 1 & 0 & e^{\eta_1} \\ 
	\frac{e^{\xi_1+\eta_1^*}}{(l_1+k_1^*)} & \frac{e^{\xi_1+\xi_1^*}}{(l_1+l_1^*)}  & 0 & 1 &   e^{\xi_1}\\
	-1 & 0 & \frac{|\al_1^{(1)}|^2}{(k_1+k_1^*)} & 0 & 0\\
	0 & -1 & 0 &  \frac{|\al_1^{(2)}|^2}{(l_1+l_1^*)} & 0\\
	0 & 0& 0 & -\al_1^{(2)} &0
	\end{vmatrix},~~
	\end{eqnarray}
	
	\begin{eqnarray}
	f=
	\begin{vmatrix}
	\frac{e^{\eta_1+\eta_1^*}}{(k_1+k_1^*)} & \frac{e^{\eta_1+\xi_1^*}}{(k_1+l_1^*)} & 1 & 0  \\ 
	\frac{e^{\xi_1+\eta_1^*}}{(l_1+k_1^*)} & \frac{e^{\xi_1+\xi_1^*}}{(l_1+l_1^*)}  & 0 & 1 \\
	-1 & 0 & \frac{|\al_1^{(1)}|^2}{(k_1+k_1^*)} & 0 \\
	0 & -1 & 0 &  \frac{|\al_1^{(2)}|^2}{(l_1+l_1^*)}
	\end{vmatrix}.
	\end{eqnarray}
	The above Gram determinant forms satisfy the bilinear Eqs. (\ref{2.2a}) and (\ref{2.2b}) as well as Manakov Eq. (\ref{e1}). 
\end{subequations}
To investigate the various properties associated with the above fundamental soliton solution, we rewrite Eq. (\ref{3.3}) as
\begin{subequations}
\begin{eqnarray}
&&q_1=e^{i\eta_{1I}} e^{\frac{\Delta_1^{(1)}+\rho_1}{2}}\{\cosh(\xi_{1R}+\frac{\phi_{1R}}{2})\cos(\frac{\phi_{1I}}{2})+i\sinh(\xi_{1R}+\frac{\phi_{1R}}{2})\sin(\frac{\phi_{1I}}{2})\}/D_2,\label{3.4a}\\
&&q_2=e^{i\xi_{1I}} e^{\frac{\Delta_1^{(2)}+\rho_2}{2}}\{\cosh(\eta_{1R}+\frac{\phi_{2R}}{2})\cos(\frac{\phi_{2I}}{2})+i\sinh(\eta_{1R}+\frac{\phi_{2R}}{2})\sin(\frac{\phi_{2I}}{2})\}/D_2,
\label{3.4b}
\end{eqnarray} 
\end{subequations}
where $D_2=e^{\frac{\delta_{11}}{2}}\cosh(\eta_{1R}+\xi_{1R}+\frac{\delta_{11}}{2})+e^{\frac{\delta_{1}+\delta_{2}}{2}}\cosh(\eta_{1R}-\xi_{1R}+\frac{\delta_{1}-\delta_{2}}{2})$,  $\eta_{1R}=k_{1R}(t-2k_{1I}z)$, $\eta_{1I}=k_{1I}t+(k_{1R}^2-k_{1I}^2)z$, $\xi_{1R}=l_{1R}(t-2l_{1I}z)$, $\xi_{1I}=l_{1I}t+(l_{1R}^2-l_{1I}^2)z$, $\rho_j=\log \alpha_1^{(j)}$, $j=1,2$. Here, $\phi_{1R}$, $\phi_{1I}$, $\phi_{2R}$ and $\phi_{2I}$ are real and imaginary parts of $\phi_1=\Delta_{1}^{(1)}-\rho_1$ and $\phi_2=\Delta_{1}^{(2)}-\rho_2$, respectively, and also $k_{1R}$, $l_{1R}$, $k_{1I}$ and  $l_{1I}$ are the real and imaginary parts of $k_1$ and $l_1$, respectively. From the above, we can write $\phi_{1R}=\frac{1}{2}\log\frac{|k_1-l_1|^2|\alpha_1^{(2)}|^4}{|k_1+l_1^*|^2(l_1+l_1^*)^4}$, $\phi_{1I}=\frac{1}{2}\log\frac{(k_1-l_1)(k_1^*+l_1)}{(k_1^*-l_1^*)(k_1+l_1^*)}$, $\phi_{2R}=\frac{1}{2}\log\frac{|l_1-k_1|^2|\alpha_1^{(1)}|^4}{|k_1+l_1^*|^2(k_1+k_1^*)^4}$ and $\phi_{2I}=\frac{1}{2}\log\frac{(l_1-k_1)(k_1+l_1^*)}{(l_1^*-k_1^*)(k_1^*+l_1)}$. The profile structures of solution (\ref{3.4a})-(\ref{3.4b})  are described by the four complex parameters $k_1$ , $l_1$ and $\alpha_1^{(j)}$, $j=1,2$. 
For the nondegenerate fundamental soliton in the first mode,  the amplitude, velocity and central position are found from Eq. (\ref{3.4a}) as $2k_{1R}$, $2l_{1I}$ and $\frac{\phi_{1R}}{2l_{1R}}$, respectively. Similarly for the soliton in the second mode they are found from Eq. (\ref{3.4b}) as $2l_{1R}$, $2k_{1I}$ and $\frac{\phi_{2R}}{2k_{1R}}$, respectively. Note that $\alpha_1^{(j)}$, $j=1,2$, are related to  the unit polarization vectors of the nondegenerate fundamental solitons in the two modes. They constitute  different phases for the nondegenerate soliton in the two modes as   $A_1=(\alpha_1^{(1)}/\alpha_1^{(1)*})^{1/2}$ and $A_2=(\alpha_1^{(2)}/\alpha_1^{(2)*})^{1/2}$.

To explain the various properties associated with solution (\ref{3.4a})-(\ref{3.4b}) further we consider two physically important special cases where the imaginary parts of the wave numbers $k_1$ and $l_1$ are either identical with each other ($k_{1I}=l_{1I}$) or nonidentical with each other ($k_{1I}\neq l_{1I}$). Physically this implies that the former case corresponds to solitons in the two modes travelling with identical velocities $v_{1}=v_{2}=2k_{1I}$ but with $k_{1}\neq l_{1}$ whereas the latter case corresponds to solitons which propagate in the two modes with non-identical velocities $v_{1}\neq v_{2}$. In the identical velocity case, the quantity $\phi_{jI}$, $j=1,2$ becomes zero in  (\ref{3.4a})-(\ref{3.4b}) when $k_{1I}=l_{1I}$. This results in the following expression for the fundamental soliton propagating with single velocity, $v_{1,2}=2k_{1I}$, in the two modes,  
\begin{eqnarray}
&&q_1=e^{i\eta_{1I}} e^{\frac{\Delta_1^{(1)}+\rho_1}{2}}\cosh(\xi_{1R}+\frac{\phi_{1R}}{2})/D_2,\nonumber\\
&&q_2=e^{i\xi_{1I}} e^{\frac{\Delta_1^{(2)}+\rho_2}{2}}\cosh(\eta_{1R}+\frac{\phi_{2R}}{2})/D_2,
\label{3.6}
\end{eqnarray}
where $D_2=e^{\frac{\delta_{11}}{2}}\cosh(\eta_{1R}+\xi_{1R}+\frac{\delta_{11}}{2})+e^{\frac{\delta_{1}+\delta_{2}}{2}}\cosh(\eta_{1R}-\xi_{1R}+\frac{\delta_{1}-\delta_{2}}{2})$ with $\eta_{1R}=k_{1R}(t-2k_{1I}z)$, $\eta_{1I}=k_{1I}t+(k_{1R}^2-k_{1I}^2)z$, $\xi_{1R}=l_{1R}(t-2k_{1I}z)$, $\xi_{1I}=k_{1I}t+(l_{1R}^2-k_{1I}^2)z$.
Note that the constants that appear in the above solution becomes equivalent to the one that appear in the solution (\ref{3.4a})-(\ref{3.4b}) after imposing the condition $k_{1I}=l_{1I}$ in it. The solution (\ref{3.6}) admits four types of symmetric profiles (satisfying appropriate conditions on parameters, see below) and also their corresponding asymmetric profiles. The symmetric profiles are: (i)  double-humps in both the modes (or a double-hump in $q_1$ mode and a M-type double-hump in $q_2$ mode), (ii) a flat-top in one mode and a double-hump in the other mode, (iii) a single-hump in the first mode and a double-hump in the second mode (or vice versa), (iv) single-humps in both the modes.  The corresponding four types of asymmetric wave profiles can be obtained by tuning the real parts of wave numbers $k_1$ and $l_1$ and the arbitrary complex parameters $\alpha_1^{(j)}$'s, $j=1,2$.

\begin{figure}
	\centering
	\includegraphics[width=0.9\linewidth]{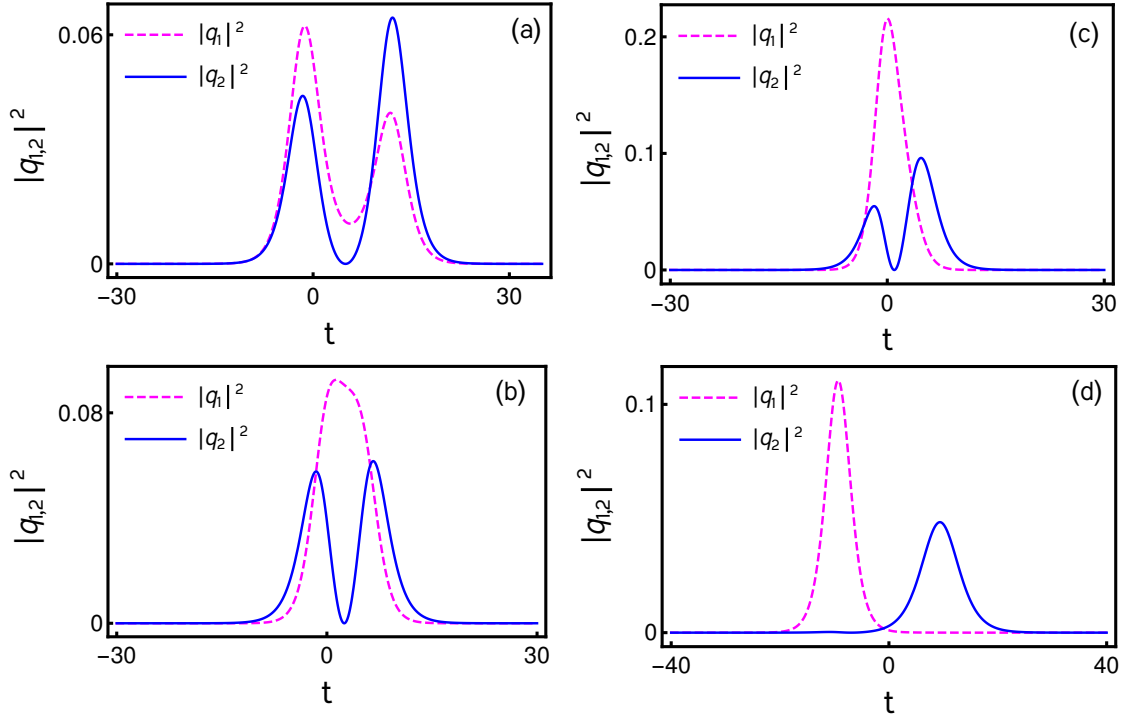}
	\caption{Various asymmetric intensity profiles of nondegenerate fundamental soliton: Figures (a), (b), (c) and (d) represent each of figures asymmetric  intensity profiles as against the symmetric profiles of Figs.1(a)-(d). The corresponding parameter values of each figures are: (a): $k_1=0.333+0.5i$,$l_1=0.315+0.5i$,  $\alpha_{1}^{(1)}=0.65+0.45i$, $\alpha_{1}^{(2)}=0.49+0.45i$. (b): $k_1=0.425+0.5i$,$l_1=0.3+0.5i$,  $\alpha_{1}^{(1)}=0.5+0.51i$, $\alpha_{1}^{(2)}=0.43+0.5i$. (c): $k_1=0.55+0.5i$,$l_1=0.333+0.5i$,  $\alpha_{1}^{(1)}=1.2+0.5i$, $\alpha_{1}^{(2)}=0.5+0.45i$. (d): $k_1=0.333+0.5i$,$l_1=-0.22+0.5i$,  $\alpha_{1}^{(1)}=0.45+3i$, $\alpha_{1}^{(2)}=0.5+0.5i$.}
	\label{f2}
\end{figure}

To illustrate the symmetric and asymmetric nature of the nondegenerate soliton in the identical velocity case we fix $k_{1I}=l_{1I}=0.5$ in Figs. \ref{f1} and \ref{f2}. The symmetric profiles are displayed in Fig. 1. 
The asymmetric profiles are depicted in Fig. \ref{f2} for the values of parameters indicated in Fig. 2. From Figs. 1 and 2 we observe that the transition which occurs from double-hump to single-hump is through a special flat-top profile. The flat-top profile has been considered as an intermediate soliton state.  It is noted that flattop soliton is also observed in a complex Ginzburg-Landau equation \cite{ak}. In Ref. \cite{ss} we have discussed symmetric and asymmetric nature of solution (\ref{3.6}) by incorporating the condition $k_{1R}<l_{1R}$ \cite{w2}. However to exhibit the generality of these structures, in the present paper, we discuss these properties for $k_{1R}>l_{1R}$. It should be pointed out here that in Ref. \cite{ss2} the authors have derived this solution in the context of multi-component BEC using Darboux transformation and they have classified density profiles as we have reported in Ref. \cite{ss} for $k_{1R}<l_{1R}$ in the context of nonlinear optics. They have also studied the stability of double-hump soliton using Bogoliubov-de Gennes excitation spectrum.                       
\begin{figure}
	\centering
	\includegraphics[width=0.9\linewidth]{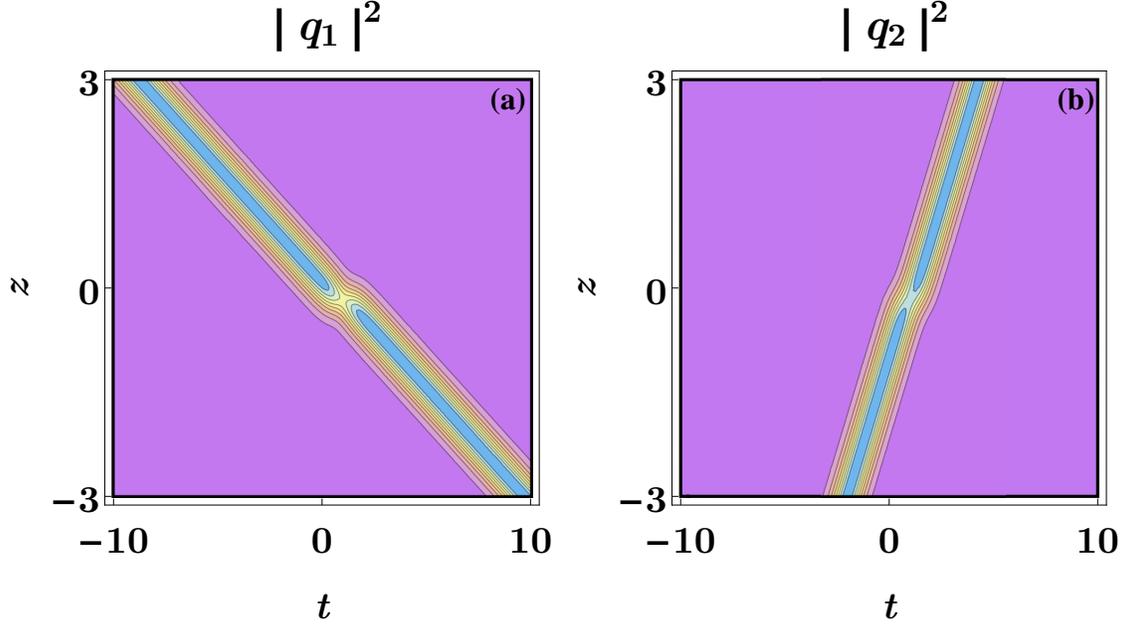}
	\caption{Node formation in the nonidentical velocity case. The parameter values are $k_1=1+1.5i$,$l_1=1.5+0.5i$,  $\alpha_{1}^{(1)}=1.5+0.5i$, $\alpha_{1}^{(2)}=0.45+0.5i$. }
	\label{f3}
\end{figure}

The symmetric nature of all the four cases can be confirmed  by finding the  extremum points of the nondegenerate one-soliton solution (\ref{3.6}). For instance, to show that the double-hump soliton profile displayed in Fig. 1(a) is symmetric, we find the corresponding local maximum and minium points by applying the first derivative test ($\{|q_j|^2\}_t=0$) and the second derivative test ($\{|q_j|^2\}_{tt}<0$ or $>0$) to the expression of $|q_j|^2$, $j=1,2$, at $z=0$. For the first mode, the three three extremal points are identified, namely $t_1=-0.9$, $t_2=5.5$ and $t_3=11.9$. We find another set of three extremal points for the second mode, namely $t_4=-1.2$, $t_5=5.5$ and $t_6=12.2$ by setting $\{|q_2|^2\}_t=0$.  The points $t_1$ and $t_3$ correspond to the maxima (at which $\{|q_1|^2\}_{tt}<0$) of the double hump soliton whereas $t_2$  corresponds to the minimum of the double hump soliton. Similarly the extremal points $t_4$ and $t_6$ represent the maxima and $t_5$ corresponds to the minimum of the double hump soliton in the $q_2$ mode. In the first component the two maxima $t_1$ and $t_3$ are symmetrically located about the minimum point $t_2$. This can be easily confirmed by finding the difference between $t_2$ and $t_1$ and  $t_3$ and $t_2$, that is $t_2-t_1=6.4=t_3-t_2$. This is true for the second component also, that is $t_5-t_4=6.7=t_6-t_5$. This implies that the two maxima $t_4$ and $t_6$ are located symmetrically from the minimum point $t_5$. Then the magnitude ($|q_1|^2$) of each hump (of the double hump soliton) corresponding to the maxima $t_1$ is equal to $0.051$ and $t_3$ is equal to $0.051$. In the second mode, the magnitude ($|q_2|^2$) corresponding to $t_4$ is equal to $0.054$  and $t_6$ is equal to $0.054$. This confirms that the magnitude of each hump of double hump soliton in both the modes are equal. Therefore  it is evident that the double hump soliton drawn in Fig. 1(a) is symmetric. One can easily verify from the Figs. 1(c) and 1(d) that the single-hump soliton is symmetric about the local maximum point (and checking the half widths as well). As far as the flat-top soliton case is concerned, we have confirmed that the first derivative $\{|q_j|^2\}_t$ very slowly tends to zero near the corresponding maximum for certain number of $t$ values. This also confirms that the presence of almost flatness and symmetric nature of the one-soliton. 

We also derive the conditions analytically to corroborate the symmetric and asymmetric nature of soliton solution (\ref{3.6}) in another way. For this purpose, we intend to calculate the relative separation distance $\Delta t_{12}$ between the minima of the two components (modes)
\begin{eqnarray}
\Delta t_{12}&=&\bar{t}_1-\bar{t}_2=(t-t_1)-(t-t_2), \nonumber\\ &=&\frac{\phi_{1R}}{2l_{1R}}-\frac{\phi_{2R}}{2k_{1R}}. \label{3.5}
\end{eqnarray}
If the above quantity $\Delta t_{12}=0$ then the solution (\ref{3.6}) exhibits symmetric profiles otherwise it admits asymmetric profiles.

The explicit form of relative separation distance turns out to be 
\begin{eqnarray}
\Delta t_{12}&=&\frac{1}{2l_{1R}}\log\frac{(k_{1R}-l_{1R})|\alpha_1^{(2)}|^2}{4l_{1R}^2(k_{1R}+l_{1R})} -\frac{1}{2k_{1R}}\log \frac{(l_{1R}-k_{1R}) |\alpha_1^{(1)}|^2}{4k_{1R}^2(k_{1R}+l_{1R})}.
\label{3.8}
\end{eqnarray} 

 We have explicitly calculated the relative separation distance values and confirmed the displayed profiles in Fig. 1 and 2 are symmetric and asymmetric, respectively. For instance, the $\Delta t_{12}$ value corresponding to the symmetric double-hump soliton in both the modes (Fig. 1(a)) is $0.002$ (to get the perfect zero value one has to fine tune the parameters suitably) and for asymmetric double-hump solitons the value is equal to $0.6493$. The above calculated values reaffirm that the obtained figures are symmetric in Fig. 1(a) and asymmetric in Fig. 2(a).  Similarly one can easily confirm the symmetric and asymmetric nature of other profiles in Figs. 1 and 2 also. 
\begin{figure}
	\centering
	\includegraphics[width=0.4\linewidth]{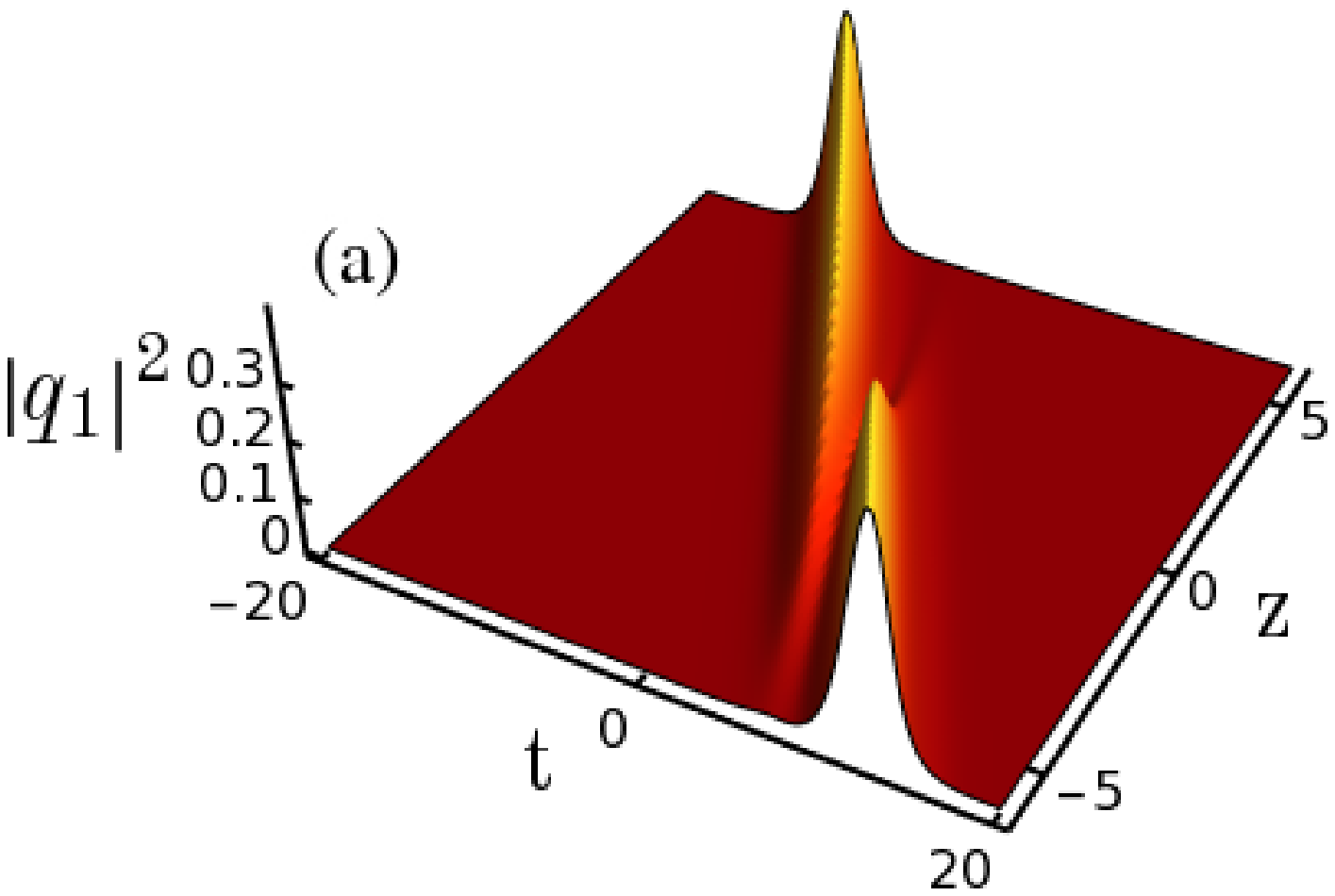}~~
	\includegraphics[width=0.4\linewidth]{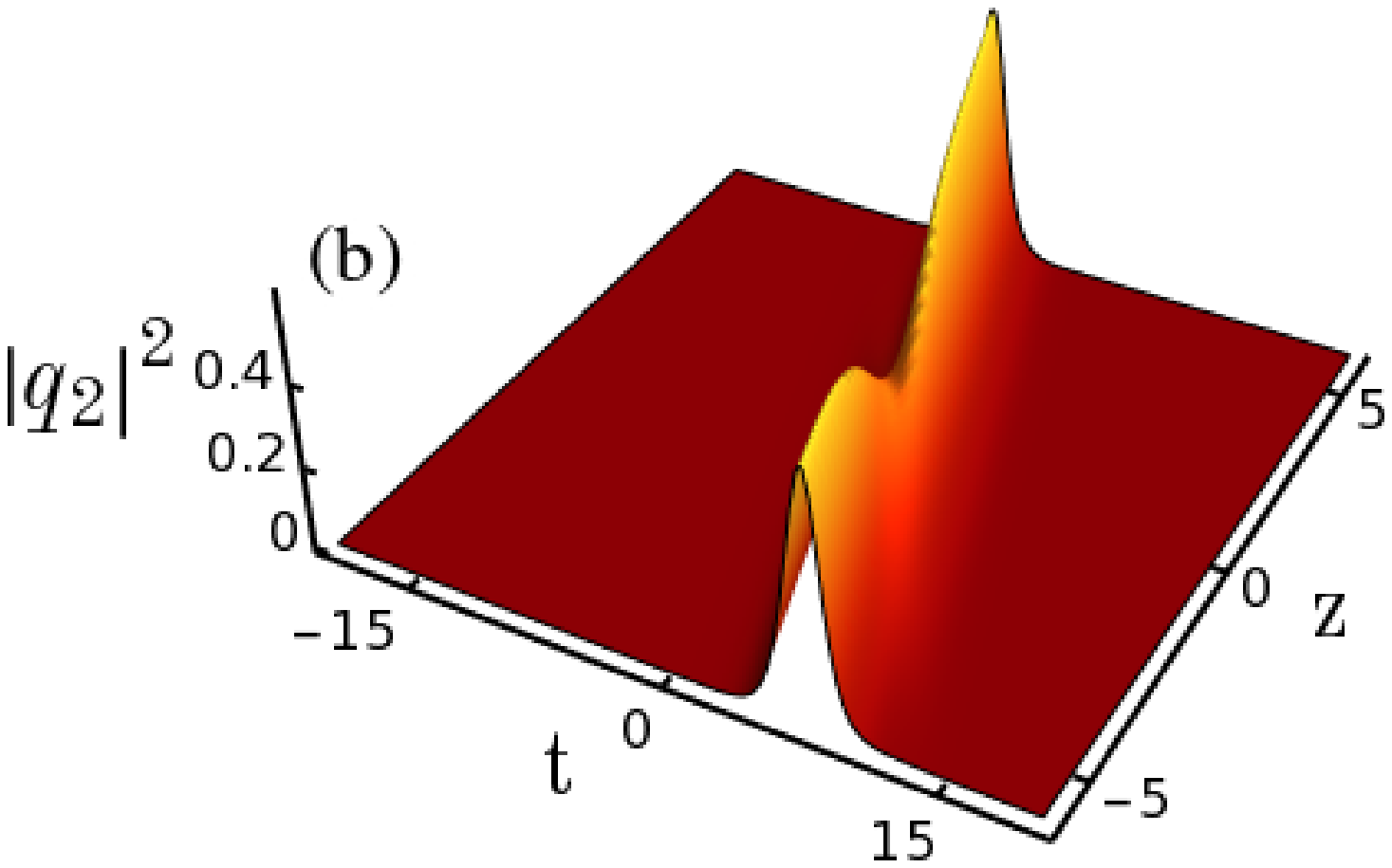}\\
	\includegraphics[width=0.4\linewidth]{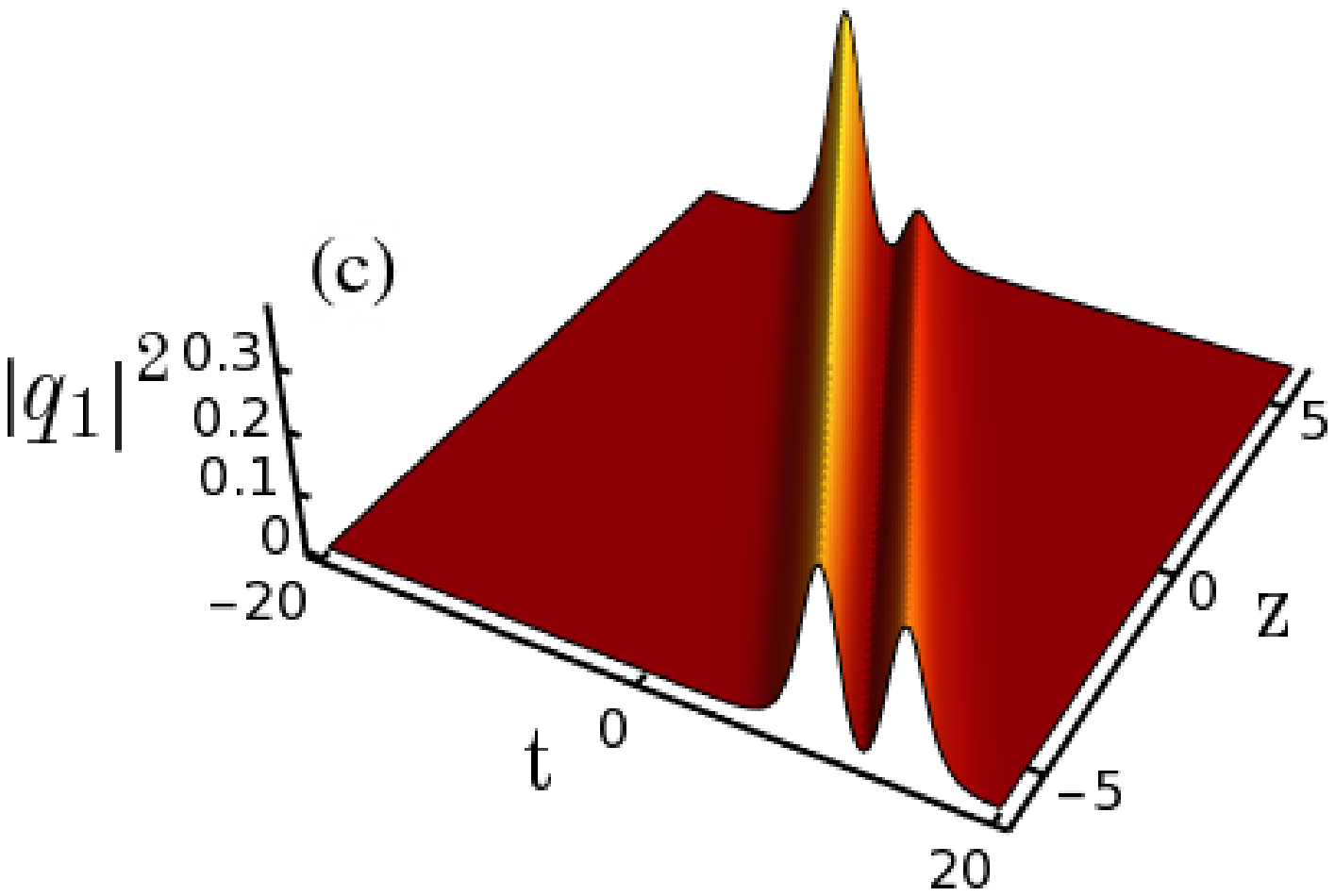}~~
	\includegraphics[width=0.4\linewidth]{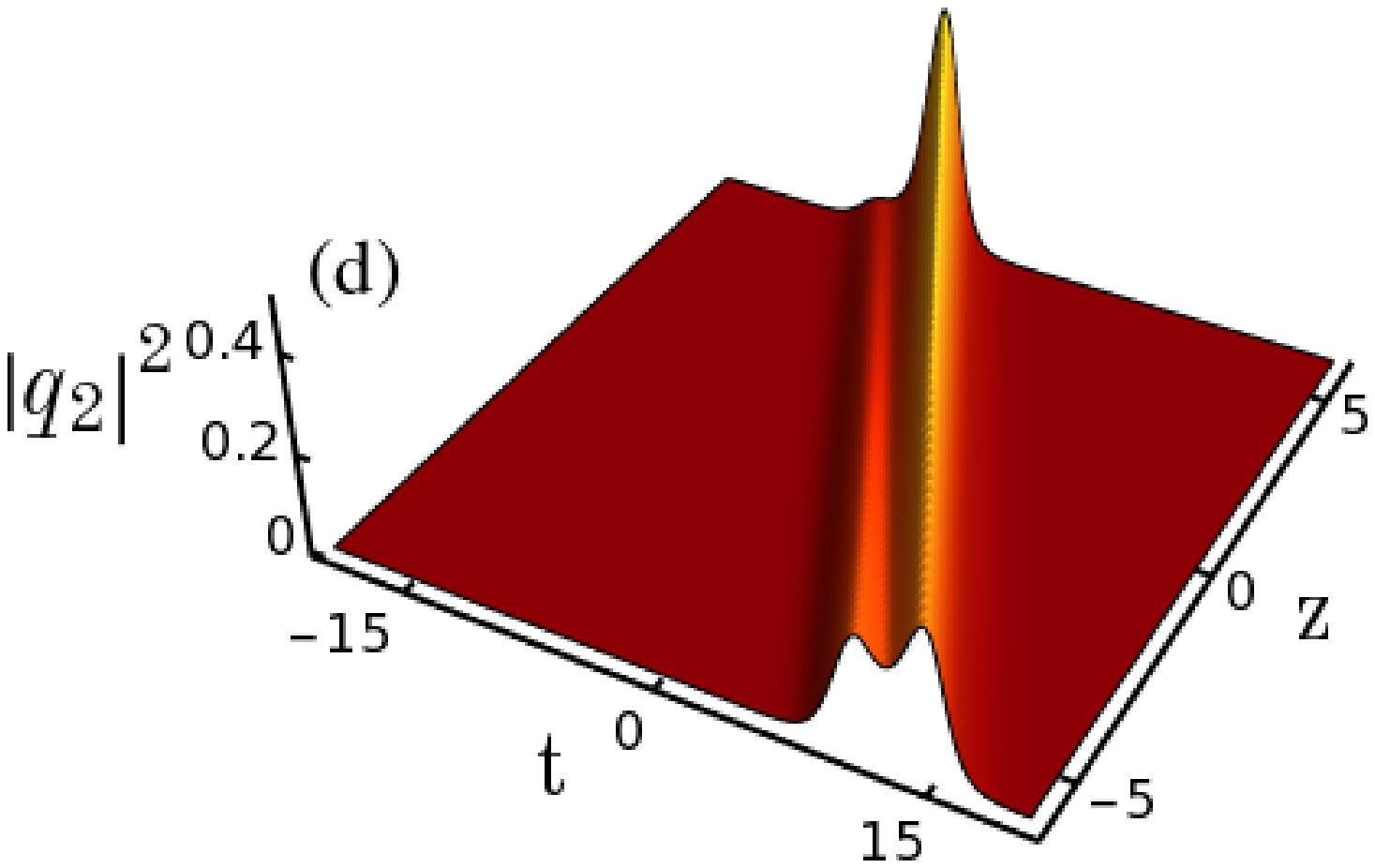}
	\caption{Double-hump formation in the profile structure of nondegenerate fundamental soliton: (a) and (b) represent the node formation in soliton profiles. (c) and (d) denote the emergence of double-hump in both the modes.  The corresponding parameter values for (a) and (b) are:  $k_{1}=0.65-0.85i$, $l_{1}=0.78-0.5i$, $\alpha_{1}^{(1)}=1$ and $\alpha_{1}^{(2)}=0.5$; For figures (c) and (d) the values are chosen as $k_{1}=0.65-0.8i$, $l_{1}=0.78-0.8i$, $\alpha_{1}^{(1)}=1$ and $\alpha_{1}^{(2)}=0.5$.}
	\label{f4}
\end{figure}

In addition to the above, for the general nonidentical velocity case ($k_{1I}\neq l_{1I}$), $v_1\neq v_2$, the distinct wave numbers $k_1$ and $l_1$ influence drastically the propagation of nondegenerate solitons in the two modes. If the relative velocity ($\Delta v_{12}=v_1-v_2$) of the solitons between the two modes is large, then there is a node created in the structure of the fundamental solitons of both the modes \cite{ss2}. This is due to the cross phase modulation between the modes. In this situation the intensity of the fast moving soliton ($v_1=2l_{1I}>0$) in the first mode starts to decrease and it gets completely suppressed after $z=0$. At the same value of $z$ the fast moving soliton reappears in the second mode after a finite time. Similarly this fact is true in the case of slow moving soliton ($v_2=2k_{1I}<0$) as well. Consequently the intensity of solitons is unequally distributed among the two modes. This is clearly  demonstrated in Fig. \ref{f3} and Figs. 4(a)-4(b). On the otherhand, if the relative velocity tends to zero ($\Delta v_{12}\rightarrow 0$), then the total intensity, $I_{\text{total}}=|q_1|^2+|q_2|^2$, of nondegenerate solitons starts to get distributed equally among the two components. As a consequence of this, a double-hump profile starts to emerge in each of the modes as displayed in Fig. 4(c)-4(d). At perfect zero relative velocity ($\Delta v_{12}=0$), the double-hump fundamental soliton emerges completely in both the modes. 
 As we have already pointed out in \cite{ss} the nondegenerate soliton solution exhibits symmetric and asymmetric profiles in the nonidentical velocity case also but the relative velocity of the solitons should be minimum. We have not displayed their plots here for brevity.

Recently we found that the occurence of multi-humps depends on the number of distinct wave numbers and modes \cite{ss3} apart from the nonlinearities. In the present two component case, the resultant nondegenerate fundamental soliton solution (\ref{3.4a})-(\ref{3.4b}) yields only a double-hump soliton. However a triple-hump soliton and a quadruple hump soliton are also observed in the cases of 3 and 4 component Manakov system cases, respectively. For the $N$-component case one may expect a more complicated profile, as mentioned in the case of theory of incoherent solitons \cite{sny,has},  involving $N$-number of humps which are characterized by $2N$-complex parameters. These results will be published elsewhere. Very recently we have also reported the existence of nondegenerate fundamental solitons and their various novel profile structures in other integrable coupled NLS type systems \cite{ss1} as well. It should be pointed out that the multi-hump nature of nondegenerate fundamental soliton is somewhat analogous to partially coherent solitons/soliton complexes \cite{h,v} where such partially coherent solitons can be obtained when the number of modes is equal to the number of degenerate vector soliton solution \cite{j,m}. We also note here that the 2-partially coherent soliton can be deduced from the double-humped nondegenerate fundamental soliton (\ref{3.4a})-(\ref{3.4b}) in the Manakov system by imposing the restrictions $\alpha_1^{(1)}=e^{\eta_{10}}$, $\alpha_1^{(2)}=-e^{\eta_{20}}$, $k_1=k_{1R} $, $l_1=k_{2R}$,  $k_{1I}=l_{1I}=0$, where $\eta_{10}$ and $\eta_{20}$ are real constants, in solution (\ref{3.3}) \cite{m}. The soliton complex reported in \cite{57} is a special case of nondegenerate fundamental soliton solution (\ref{3.3}) when the parameters $k_1$ and $l_1$ are  chosen as real constants and  $\alpha_{1}^{(1)}=\alpha_{1}^{(2)}=1$.  
\subsection{Nondegenerate two-soliton solution}
In order to investigate the collision dynamics of nondegenerate soliton of the form (\ref{3.3}), it is essential to derive the expression for the corresponding two soliton solution. To construct it, we consider the seed solutions as $g_1^{(1)}=\al_1^{(1)}e^{\eta_1}+\al_2^{(1)}e^{\eta_2}$ and $g_1^{(2)}=\al_1^{(2)}e^{\xi_1}+\al_2^{(2)}e^{\xi_2}$, $\eta_{j}=k_{j}t+ik_{j}^{2}z$ and $\xi_{j}=l_{j}t+il_{j}^{2}z$, $j=1,2$, for Eqs. (\ref{3.2}).  By proceeding with the procedure given in the previous subsection along with these seed solutions we find that the series expansions for $g^{(j)}$, $j=1,2$ and $f$ get terminated as $g^{(j)}=\epsilon g_1^{(j)}+\epsilon^3 g_3^{(j)}+\epsilon^5 g_5^{(j)}+\epsilon^7 g_7^{(j)}$ and $f=1+\epsilon^2 f_2+\epsilon^4 f_4+\epsilon^6 f_6+\epsilon^8 f_8$. The other unknown functions, $g_9^{(j)}$, $g_{11}^{(j)}$, $f_{10}$,  $f_{12}$ and etc., are found to be identically zero. We further note here that the termination of these perturbation series occurs at the order of $\epsilon^3$ in $g^{(j)}$'s and at the level of $\epsilon^4$ in $f$ for deriving the degenerate two-soliton solution.  The resulting explicit forms of the unknown functions in the truncated series expansions constitute the following nondegenerate two-soliton solution, in Gram determinant form, to Eq. (\ref{e1}), 
\begin{subequations}
\begin{eqnarray}
g^{(1)} =
\begin{vmatrix}
\frac{e^{\eta_1+\eta_1^*}}{(k_1+k_1^*)} & \frac{e^{\eta_1+\eta_2^*}}{(k_1+k_2^*)} & \frac{e^{\eta_1+\xi_1^*}}{(k_1+l_1^*)} & \frac{e^{\eta_1+\xi_2^*}}{(k_1+l_2^*)} & 1 & 0 & 0& 0 & e^{\eta_1} \\ 
\frac{e^{\eta_2+\eta_1^*}}{(k_2+k_1^*)} & \frac{e^{\eta_2+\eta_2^*}}{(k_2+k_2^*)} & \frac{e^{\eta_2+\xi_1^*}}{(k_2+l_1^*)} & \frac{e^{\eta_2+\xi_2^*}}{(k_2+l_2^*)} & 0 & 1 & 0& 0 & e^{\eta_2} \\ 
\frac{e^{\xi_1+\eta_1^*}}{(l_1+k_1^*)} & \frac{e^{\xi_1+\eta_2^*}}{(l_1+k_2^*)} & \frac{e^{\xi_1+\xi_1^*}}{(l_1+l_1^*)} & \frac{e^{\xi_1+\xi_2^*}}{(l_1+l_2^*)} & 0 & 0 & 1& 0 & e^{\xi_1} \\ 
\frac{e^{\xi_2+\eta_1^*}}{(l_2+k_1^*)} & \frac{e^{\xi_2+\eta_2^*}}{(l_2+k_2^*)} & \frac{e^{\xi_2+\xi_1^*}}{(l_2+l_1^*)} & \frac{e^{\xi_2+\xi_2^*}}{(l_2+l_2^*)} & 0 & 0 & 0& 1 & e^{\xi_2} \\ 
-1 & 0 & 0& 0& \frac{|\al_1^{(1)}|^2}{(k_1^*+k_1)} & \frac{\al_1^{(1)*}\al_2^{(1)}}{(k_1^*+k_2)} & 0 &0 & 0\\
0 & -1 & 0 &  0& \frac{\al_1^{(1)}\al_2^{(1)*}}{(k_2^*+k_1)}& \frac{|\al_2^{(1)}|^2}{(k_2+k_2^*)} & 0 & 0 & 0\\
0 & 0 & -1 &  0& 0& 0 & \frac{|\al_1^{(2)}|^2}{(l_1^*+l_1)} & \frac{\al_1^{(2)*}\al_2^{(2)}}{(l_1^*+l_2)} & 0\\
0 & 0 & 0 & -1& 0& 0& \frac{\al_1^{(2)}\al_2^{(2)*}}{(l_2^*+l_1)} & \frac{|\al_2^{(2)}|^2}{(l_2^*+l_2)} & 0\\
0 & 0& 0& 0& -\al_1^{(1)} & -\al_2^{(1)} &  0 &0  &0
\end{vmatrix},\label{3.7a}
\end{eqnarray}

\begin{eqnarray}
g^{(2)} =
\begin{vmatrix}
\frac{e^{\eta_1+\eta_1^*}}{(k_1+k_1^*)} & \frac{e^{\eta_1+\eta_2^*}}{(k_1+k_2^*)} & \frac{e^{\eta_1+\xi_1^*}}{(k_1+l_1^*)} & \frac{e^{\eta_1+\xi_2^*}}{(k_1+l_2^*)} & 1 & 0 & 0& 0 & e^{\eta_1} \\ 
\frac{e^{\eta_2+\eta_1^*}}{(k_2+k_1^*)} & \frac{e^{\eta_2+\eta_2^*}}{(k_2+k_2^*)} & \frac{e^{\eta_2+\xi_1^*}}{(k_2+l_1^*)} & \frac{e^{\eta_2+\xi_2^*}}{(k_2+l_2^*)} & 0 & 1 & 0& 0 & e^{\eta_2} \\ 
\frac{e^{\xi_1+\eta_1^*}}{(l_1+k_1^*)} & \frac{e^{\xi_1+\eta_2^*}}{(l_1+k_2^*)} & \frac{e^{\xi_1+\xi_1^*}}{(l_1+l_1^*)} & \frac{e^{\xi_1+\xi_2^*}}{(l_1+l_2^*)} & 0 & 0 & 1& 0 & e^{\xi_1} \\ 
\frac{e^{\xi_2+\eta_1^*}}{(l_2+k_1^*)} & \frac{e^{\xi_2+\eta_2^*}}{(l_2+k_2^*)} & \frac{e^{\xi_2+\xi_1^*}}{(l_2+l_1^*)} & \frac{e^{\xi_2+\xi_2^*}}{(l_2+l_2^*)} & 0 & 0 & 0& 1 & e^{\xi_2} \\ 
-1 & 0 & 0& 0& \frac{|\al_1^{(1)}|^2}{(k_1^*+k_1)} & \frac{\al_1^{(1)*}\al_2^{(1)}}{(k_1^*+k_2)} & 0 &0 & 0\\
0 & -1 & 0 &  0& \frac{\al_1^{(1)}\al_2^{(1)*}}{(k_2^*+k_1)}& \frac{|\al_2^{(1)}|^2}{(k_2+k_2^*)} & 0 & 0 & 0\\
0 & 0 & -1 &  0& 0& 0 & \frac{|\al_1^{(2)}|^2}{(l_1^*+l_1)} & \frac{\al_1^{(2)*}\al_2^{(2)}}{(l_1^*+l_2)} & 0\\
0 & 0 & 0 & -1& 0& 0& \frac{\al_1^{(2)}\al_2^{(2)*}}{(l_2^*+l_1)} & \frac{|\al_2^{(2)}|^2}{(l_2^*+l_2)} & 0\\
0 & 0& 0& 0& 0& 0& -\al_1^{(2)} & -\al_2^{(2)}  &0
\end{vmatrix},\label{3.7b}
\end{eqnarray}
\begin{eqnarray}
f=
\begin{vmatrix}
\frac{e^{\eta_1+\eta_1^*}}{(k_1+k_1^*)} & \frac{e^{\eta_1+\eta_2^*}}{(k_1+k_2^*)} & \frac{e^{\eta_1+\xi_1^*}}{(k_1+l_1^*)} & \frac{e^{\eta_1+\xi_2^*}}{(k_1+l_2^*)} & 1 & 0 & 0& 0 \\ 
\frac{e^{\eta_2+\eta_1^*}}{(k_2+k_1^*)} & \frac{e^{\eta_2+\eta_2^*}}{(k_2+k_2^*)} & \frac{e^{\eta_2+\xi_1^*}}{(k_2+l_1^*)} & \frac{e^{\eta_2+\xi_2^*}}{(k_2+l_2^*)} & 0 & 1 & 0& 0 \\ 
\frac{e^{\xi_1+\eta_1^*}}{(l_1+k_1^*)} & \frac{e^{\xi_1+\eta_2^*}}{(l_1+k_2^*)} & \frac{e^{\xi_1+\xi_1^*}}{(l_1+l_1^*)} & \frac{e^{\xi_1+\xi_2^*}}{(l_1+l_2^*)} & 0 & 0 & 1& 0\\ 
\frac{e^{\xi_2+\eta_1^*}}{(l_2+k_1^*)} & \frac{e^{\xi_2+\eta_2^*}}{(l_2+k_2^*)} & \frac{e^{\xi_2+\xi_1^*}}{(l_2+l_1^*)} & \frac{e^{\xi_2+\xi_2^*}}{(l_2+l_2^*)} & 0 & 0 & 0& 1 \\ 
-1 & 0 & 0& 0& \frac{|\al_1^{(1)}|^2}{(k_1^*+k_1)} & \frac{\al_1^{(1)*}\al_2^{(1)}}{(k_1^*+k_2)} & 0 &0\\
0 & -1 & 0 &  0& \frac{\al_1^{(1)}\al_2^{(1)*}}{(k_2^*+k_1)}& \frac{|\al_2^{(1)}|^2}{(k_2+k_2^*)} & 0 & 0\\
0 & 0 & -1 &  0& 0& 0 & \frac{|\al_1^{(2)}|^2}{(l_1^*+l_1)} & \frac{\al_1^{(2)*}\al_2^{(2)}}{(l_1^*+l_2)}\\
0 & 0 & 0 & -1& 0& 0& \frac{\al_1^{(2)}\al_2^{(2)*}}{(l_2^*+l_1)} & \frac{|\al_2^{(2)}|^2}{(l_2^*+l_2)}
\end{vmatrix}.\label{3.7c}
\end{eqnarray}	
\end{subequations}

 In the above, the eight arbitrary complex parameters $k_{j}$, $l_{j}$, $\alpha_{1}^{(j)}$ and $\alpha_{2}^{(j)}$, $j=1,2$, define the profile shapes of the nondegenerate solitons and their various interesting collision scenarios. By generalizing the above given  procedure, the nondegenerate $N$-soliton solution of the Manakov system can be obtained. To derive the $N$-nondegenerate soliton solution, the power series expansion should be as in the following form $g^{(j)}=\sum_{n=1}^{2N-1}\epsilon^{2n-1}g_{2n-1}^{(j)}$ and $f=1+\sum_{n=1}^{2N}\epsilon^{2n}f_{2n}$.  The $4N$ complex parameters, which are present in the $N$-soliton solution, determine the shape of the $N$-solitons. In Appendix A, we have given the three-soliton solution form explicitly using the Gram determinants. 
\subsection{Partially nondegenerate two-soliton solution}
To show the possibility of occurrence of degenerate and nondegenerate solitons simultanously in the Manakov system (\ref{e1}), we restrict the wave numbers $k_1$ and $l_1$ (or $k_2$ and $l_2$ ) as $k_1=l_1$ (or $k_2=l_2$) but $k_2\neq l_2$ (or $k_1\neq l_1$) in the obtained completely nondegenerate two-soliton solution (\ref{3.7a})-(\ref{3.7c}). As a consequence of this restriction, the wave variables $\eta_1$ and $\xi_1$ automatically get restricted  as $\xi_1=\eta_1$. By imposing such a restriction in the fully nondegenerate two-soliton solution (\ref{3.7a})-(\ref{3.7c}) we deduce the following form of partially nondegenerate two-soliton solution as
\begin{subequations}
\begin{eqnarray}
g^{(1)} =
\begin{vmatrix}
\frac{e^{\eta_1+\eta_1^*}}{(k_1+k_1^*)} & \frac{e^{\eta_1+\eta_2^*}}{(k_1+k_2^*)} & \frac{e^{\eta_1+\eta_1^*}}{(k_1+k_1^*)} & \frac{e^{\eta_1+\xi_2^*}}{(k_1+l_2^*)} & 1 & 0 & 0& 0 & e^{\eta_1} \\ 
\frac{e^{\eta_2+\eta_1^*}}{(k_2+k_1^*)} & \frac{e^{\eta_2+\eta_2^*}}{(k_2+k_2^*)} & \frac{e^{\eta_2+\eta_1^*}}{(k_2+k_1^*)} & \frac{e^{\eta_2+\xi_2^*}}{(k_2+l_2^*)} & 0 & 1 & 0& 0 & e^{\eta_2} \\ 
\frac{e^{\eta_1+\eta_1^*}}{(k_1+k_1^*)} & \frac{e^{\eta_1+\eta_2^*}}{(k_1+k_2^*)} & \frac{e^{\eta_1+\eta_1^*}}{(k_1+k_1^*)} & \frac{e^{\eta_1+\xi_2^*}}{(k_1+l_2^*)} & 0 & 0 & 1& 0 & e^{\eta_1} \\ 
\frac{e^{\xi_2+\eta_1^*}}{(l_2+k_1^*)} & \frac{e^{\xi_2+\eta_2^*}}{(l_2+k_2^*)} & \frac{e^{\xi_2+\eta_1^*}}{(l_2+k_1^*)} & \frac{e^{\xi_2+\xi_2^*}}{(l_2+l_2^*)} & 0 & 0 & 0& 1 & e^{\xi_2} \\ 
-1 & 0 & 0& 0& \frac{|\al_1^{(1)}|^2}{(k_1^*+k_1)} & \frac{\al_1^{(1)*}\al_2^{(1)}}{(k_1^*+k_2)} & 0 &0 & 0\\
0 & -1 & 0 &  0& \frac{\al_1^{(1)}\al_2^{(1)*}}{(k_2^*+k_1)}& \frac{|\al_2^{(1)}|^2}{(k_2+k_2^*)} & 0 & 0 & 0\\
0 & 0 & -1 &  0& 0& 0 & \frac{|\al_1^{(2)}|^2}{(k_1^*+k_1)} & \frac{\al_1^{(2)*}\al_2^{(2)}}{(k_1^*+l_2)} & 0\\
0 & 0 & 0 & -1& 0& 0& \frac{\al_1^{(2)}\al_2^{(2)*}}{(l_2^*+k_1)} & \frac{|\al_2^{(2)}|^2}{(l_2^*+l_2)} & 0\\
0 & 0& 0& 0& -\al_1^{(1)} & -\al_2^{(1)} &  0 &0  &0
\end{vmatrix},\label{3.8a}
\end{eqnarray}
\begin{eqnarray}
g^{(2)} =
\begin{vmatrix}
\frac{e^{\eta_1+\eta_1^*}}{(k_1+k_1^*)} & \frac{e^{\eta_1+\eta_2^*}}{(k_1+k_2^*)} & \frac{e^{\eta_1+\eta_1^*}}{(k_1+k_1^*)} & \frac{e^{\eta_1+\xi_2^*}}{(k_1+l_2^*)} & 1 & 0 & 0& 0 & e^{\eta_1} \\ 
\frac{e^{\eta_2+\eta_1^*}}{(k_2+k_1^*)} & \frac{e^{\eta_2+\eta_2^*}}{(k_2+k_2^*)} & \frac{e^{\eta_2+\eta_1^*}}{(k_2+k_1^*)} & \frac{e^{\eta_2+\xi_2^*}}{(k_2+l_2^*)} & 0 & 1 & 0& 0 & e^{\eta_2} \\ 
\frac{e^{\eta_1+\eta_1^*}}{(k_1+k_1^*)} & \frac{e^{\eta_1+\eta_2^*}}{(k_1+k_2^*)} & \frac{e^{\eta_1+\eta_1^*}}{(k_1+k_1^*)} & \frac{e^{\eta_1+\xi_2^*}}{(k_1+l_2^*)} & 0 & 0 & 1& 0 & e^{\eta_1} \\ 
\frac{e^{\xi_2+\eta_1^*}}{(l_2+k_1^*)} & \frac{e^{\xi_2+\eta_2^*}}{(l_2+k_2^*)} & \frac{e^{\xi_2+\eta_1^*}}{(l_2+k_1^*)} & \frac{e^{\xi_2+\xi_2^*}}{(l_2+l_2^*)} & 0 & 0 & 0& 1 & e^{\xi_2} \\ 
-1 & 0 & 0& 0& \frac{|\al_1^{(1)}|^2}{(k_1^*+k_1)} & \frac{\al_1^{(1)*}\al_2^{(1)}}{(k_1^*+k_2)} & 0 &0 & 0\\
0 & -1 & 0 &  0& \frac{\al_1^{(1)}\al_2^{(1)*}}{(k_2^*+k_1)}& \frac{|\al_2^{(1)}|^2}{(k_2+k_2^*)} & 0 & 0 & 0\\
0 & 0 & -1 &  0& 0& 0 & \frac{|\al_1^{(2)}|^2}{(k_1^*+k_1)} & \frac{\al_1^{(2)*}\al_2^{(2)}}{(k_1^*+l_2)} & 0\\
0 & 0 & 0 & -1& 0& 0& \frac{\al_1^{(2)}\al_2^{(2)*}}{(l_2^*+k_1)} & \frac{|\al_2^{(2)}|^2}{(l_2^*+l_2)} & 0\\
0 & 0& 0& 0& 0& 0& -\al_1^{(2)} & -\al_2^{(2)}  &0	
\end{vmatrix},\label{3.8b}
\end{eqnarray}
\begin{eqnarray}
f =
\begin{vmatrix}
\frac{e^{\eta_1+\eta_1^*}}{(k_1+k_1^*)} & \frac{e^{\eta_1+\eta_2^*}}{(k_1+k_2^*)} & \frac{e^{\eta_1+\eta_1^*}}{(k_1+k_1^*)} & \frac{e^{\eta_1+\xi_2^*}}{(k_1+l_2^*)} & 1 & 0 & 0& 0 \\ 
\frac{e^{\eta_2+\eta_1^*}}{(k_2+k_1^*)} & \frac{e^{\eta_2+\eta_2^*}}{(k_2+k_2^*)} & \frac{e^{\eta_2+\eta_1^*}}{(k_2+k_1^*)} & \frac{e^{\eta_2+\xi_2^*}}{(k_2+l_2^*)} & 0 & 1 & 0& 0 \\ 
\frac{e^{\eta_1+\eta_1^*}}{(k_1+k_1^*)} & \frac{e^{\eta_1+\eta_2^*}}{(k_1+k_2^*)} & \frac{e^{\eta_1+\eta_1^*}}{(k_1+k_1^*)} & \frac{e^{\eta_1+\xi_2^*}}{(k_1+l_2^*)} & 0 & 0 & 1& 0 \\ 
\frac{e^{\xi_2+\eta_1^*}}{(l_2+k_1^*)} & \frac{e^{\xi_2+\eta_2^*}}{(l_2+k_2^*)} & \frac{e^{\xi_2+\eta_1^*}}{(l_2+k_1^*)} & \frac{e^{\xi_2+\xi_2^*}}{(l_2+l_2^*)} & 0 & 0 & 0& 1 \\ 
-1 & 0 & 0& 0& \frac{|\al_1^{(1)}|^2}{(k_1^*+k_1)} & \frac{\al_1^{(1)*}\al_2^{(1)}}{(k_1^*+k_2)} & 0 &0 \\
0 & -1 & 0 &  0& \frac{\al_1^{(1)}\al_2^{(1)*}}{(k_2^*+k_1)}& \frac{|\al_2^{(1)}|^2}{(k_2+k_2^*)} & 0 & 0\\
0 & 0 & -1 &  0& 0& 0 & \frac{|\al_1^{(2)}|^2}{(k_1^*+k_1)} & \frac{\al_1^{(2)*}\al_2^{(2)}}{(k_1^*+l_2)}\\
0 & 0 & 0 & -1& 0& 0& \frac{\al_1^{(2)}\al_2^{(2)*}}{(l_2^*+k_1)} & \frac{|\al_2^{(2)}|^2}{(l_2^*+l_2)}
\end{vmatrix},\label{3.8c}
\end{eqnarray}	
\end{subequations}
The above new class of solution (\ref{3.8a})-(\ref{3.8c}) can be derived through Hirota bilinear method with the following seed solutions, $g_1^{(1)}=\al_1^{(1)}e^{\eta_1}+\al_2^{(1)}e^{\eta_2}$ and $g_1^{(2)}=\al_1^{(2)}e^{\eta_1}+\al_2^{(2)}e^{\xi_2}$, $\eta_{j}=k_{j}t+ik_{j}^{2}z$ and $\xi_{2}=l_{2}t+il_{2}^{2}z$, $j=1,2$, for Eqs. (\ref{3.2}).
Such coexistence of degenerate and nondegenerate solitons and their dynamics are characterized by seven complex parameters $k_{j}$, $l_{2}$, $\alpha_{1}^{(j)}$ and $\alpha_{2}^{(j)}$, $j=1,2$.  The interesting collision behaviour of the coexisting degenerate and nondegenerate solitons is discussed in section V. 
\section{Various shape preserving and shape changing collisions of  nondegenerate solitons}
The several interesting collision properties associated with the nondegenerate solitons can be explored by analyzing the asymptotic forms of the two-soliton solution (\ref{3.7a})-(\ref{3.7c}) of Eq. (\ref{e1}).  By doing so, we observe that the nondegenerate solitons undergo three types of collision scenarios. For either of the two cases (i) Equal velocities: $k_{1I}=l_{1I}$, $k_{2I}=l_{2I}$ and  (ii) Unequal velocities: $k_{1I}\neq l_{1I}$, $k_{2I}\neq l_{2I}$, the nondegenerate two solitons undergo shape preserving, shape altering and shape changing collision behaviours. Here we present the asymptotic analysis for the case of shape preserving collision only and it can be carried out for other cases also in a similar manner..  
\subsection{Asymptotic analysis}
In order to study the interaction dynamics of  nondegenerate solitons completely, we perform a careful asymptotic analysis for the nondegenerate two soliton solution (\ref{3.7a})-(\ref{3.7c}) and we deduce the explicit forms of individual solitons at the limits $z\rightarrow \pm\infty$. To explore this, we consider
$k_{jR},l_{jR}>0$, $j=1,2$, $k_{1I}>k_{2I}$, $l_{1I}>l_{2I}$,  $k_{1I}=l_{1I}$ and $k_{2I}=l_{2I}$, which corresponds to the case of a head-on collision between the two symmetric nondegenerate solitons. In this situation the two symmetric fundamental solitons $S_1$ and $S_2$ are well separated and subsequently the asymptotic forms of the individual solitons can be deduced from the solution (\ref{3.7a})-(\ref{3.7c}) by incorporating the asymptotic nature of the wave variables $\eta_{jR}=k_{jR}(t-2k_{jI}z)$ and $\xi_{jR}=l_{jR}(t-2l_{jI}z)$, $j=1,2$, in it. The wave variables $\eta_{jR}$ and $\xi_{jR}$ behave asymptotically as (i) Soliton 1 ($S_1$): $\eta_{1R}$, $\xi_{1R}\simeq 0$, $\eta_{2R}$, $\xi_{2R}\rightarrow\mp \infty$ as $z\mp\infty$ and (ii) Soliton 2 ($S_2$): $\eta_{2R}$, $\xi_{2R}\simeq 0$, $\eta_{1R}$, $\xi_{1R}\rightarrow\mp \infty$ as $z\pm\infty$. Correspondingly these results lead to the following asymptotic forms of nondegenerate individual solitons.  \begin{widetext}\vspace{0.21cm}
	\underline{(a) Before collision}: $z\rightarrow -\infty$\\
	\underline{Soliton 1}: In this limit, the asymptotic forms of $q_1$ and $q_2$ are deduced from the two soliton solution (\ref{3.7a})-(\ref{3.7c}) for soliton 1 as below:
	\begin{subequations}
		\begin{eqnarray}
		&&q_{1}\simeq \frac{2A_1^{1-}k_{1R}e^{i\eta_{1I}}\cosh(\xi_{1R}+\phi_1^-)}{\big[{\frac{(k_{1}^{*}-l_{1}^{*})^{\frac{1}{2}}}{(k_{1}^{*}+l_{1})^{\frac{1}{2}}}}\cosh(\eta_{1R}+\xi_{1R}+\phi_3^-)+\frac{(k_{1}+l_{1}^{*})^{\frac{1}{2}}}{(k_{1}-l_{1})^{\frac{1}{2}}}\cosh(\eta_{1R}-\xi_{1R}+\phi_4^-)\big]},\\
		&&q_{2}\simeq\frac{2A_2^{1-}l_{1R}e^{i\xi_{1I}}\cosh(\eta_{1R}+\phi_2^-)}{\big[\frac{(k_{1}^{*}-l_{1}^{*})^{\frac{1}{2}}}{(k_{1}+l_{1}^{*})^{\frac{1}{2}}}\cosh(\eta_{1R}+\xi_{1R}+\phi_3^-)+\frac{(k_{1}^{*}+l_{1})^{1/2}}{(k_{1}-l_{1})^{1/2}}\cosh(\eta_{1R}-\xi_{1R}+\phi_4^-)\big]}.
		\end{eqnarray}\end{subequations}\\
	Here, $\phi_1^-=\frac{1}{2}\log\frac{(k_1-l_1)|\alpha_1^{(2)}|^2}{(k_1+l_1^*)(l_1+l_1^*)^2}$,  $\phi_2^-=\frac{1}{2}\log\frac{(l_1-k_1)|\alpha_1^{(1)}|^2}{(k_1^*+l_1)(k_1+k_1^*)^2}$, $\phi_3^-=\frac{1}{2}\log\frac{|k_1-l_1|^2|\alpha_1^{(1)}|^2|\alpha_1^{(2)}|^2}{|k_1+l_1^*|^2(k_1+k_1^*)^2(l_1+l_1^*)^2}$, $\phi_4^-=\frac{1}{2}\log\frac{|\alpha_1^{(1)}|^2(l_1+l_1^*)^2}{|\alpha_1^{(2)}|^2(k_1+k_1^*)^2}$, $A_{1}^{1-}=[\alpha_{1}^{(1)}/\alpha_{1}^{(1)^*}]^{1/2}$ and $A_{2}^{1-}=i[\alpha_{1}^{(2)}/\alpha_{1}^{(2)^*}]^{1/2}$. In the latter, superscript ($1-$) represents soliton  $S_1$ before collision and subscript $(1,2)$ denotes the two modes $q_1$ and $q_2$ respectively.     \\
	\underline{Soliton 2}: The asymptotic expressions for  soliton 2 in the two modes before collision turn out to be
	\begin{subequations}
		\begin{eqnarray}
		&&q_{1}\simeq \frac{2k_{2R}A_1^{2-}e^{i(\eta_{2I}+\theta_1^-)}\cosh(\xi_{2R}+\varphi_1^-)}{\big[\frac{(k_{2}^{*}-l_{2}^{*})^{\frac{1}{2}}}{(k_{2}^{*}+l_{2})^{\frac{1}{2}}}\cosh(\eta_{2R}+\xi_{2R}+\varphi_3^-)+\frac{(k_{2}+l_{2}^{*})^{\frac{1}{2}}}{(k_{2}-l_{2})^{\frac{1}{2}}}\cosh(\eta_{2R}-\xi_{2R}+\varphi_4^-)\big]},\\
		&&q_2\simeq \frac{2l_{2R}A_2^{2-}e^{i(\xi_{2I}+\theta_2^-)}\cosh(\eta_{2R}+\varphi_2^-)}{\big[\frac{(k_{2}^{*}-l_{2}^{*})^{\frac{1}{2}}}{(k_{2}+l_{2}^*)^{\frac{1}{2}}}\cosh(\eta_{2R}+\xi_{2R}+\varphi_3^-)+\frac{(k_{2}^*+l_{2})^{\frac{1}{2}}}{(k_{2}-l_{2})^{\frac{1}{2}}}\cosh(\eta_{2R}-\xi_{2R}+\varphi_4^-)\big]}.
		\end{eqnarray} \end{subequations}
	In the above,
	\begin{eqnarray}
	&&\vphi_1^-=\frac{1}{2}\log\frac{(k_2-l_2)|\alpha_{2}^{(2)}|^2}{(k_2+l_2^*)(l_2+l_2^*)^2}+\frac{1}{2}\log\frac{|k_1-l_2|^2|l_1-l_2|^4}{|k_1+l_2^*|^2|l_1+l_2^*|^4},\nonumber\\
	&&\vphi_2^-=\frac{1}{2}\log\frac{(l_2-k_2)|\alpha_{2}^{(1)}|^2}{(k_2^*+l_2)(k_2+k_2^*)^2}+\frac{1}{2}\log\frac{|k_2-l_1|^2|k_1-k_2|^4}{|k_2+l_1^*|^2|k_1+k_2^*|^4},\nonumber\\
	&&\vphi_3^-=\frac{1}{2}\log\frac{|k_2-l_2|^2|\alpha_{2}^{(1)}|^2|\alpha_{2}^{(2)}|^2}{|k_2+l_2^*|^2(k_2+k_2^*)^2(l_2+l_2^*)^2}+\frac{1}{2}\log\frac{|k_1-k_2|^4|l_1-l_2|^4|k_2-l_1|^2|k_1-l_2|^2}{|k_1+k_2^*|^4|k_2+l_1^*|^2|k_1+l_2^*|^2|l_1+l_2^*|^4},\nonumber\\
		&&\vphi_4^-=\frac{1}{2}\log\frac{|\alpha_{2}^{(1)}|^2(l_2+l_2^*)^2}{|\alpha_{2}^{(2)}|^2(k_2+k_2^*)^2}+\frac{1}{2}\log\frac{|k_1-k_2|^4|l_1+l_2^*|^4|k_2-l_1|^2|k_1+l_2^*|^2}{|k_1+k_2^*|^4|k_2+l_1^*|^2|k_1-l_2|^2|l_1-l_2|^4},\nonumber\\
&&e^{i\theta_1^-}=\frac{(k_{1}-k_{2})(l_{1}-l_{2})(l_{1}^*+l_{2})(k_{2}-l_{1})^{\frac{1}{2}}(k_{1}+k_{2}^{*})(k_{2}^{*}+l_{1})^{\frac{1}{2}}}{(k_{1}^{*}-k_{2}^{*})(l_{1}+l_{2}^*)(l_{1}^*-l_{2}^*)(k_{2}^{*}-l_{1}^{*})^{\frac{1}{2}}(k_{1}^{*}+k_{2})(k_{2}+l_{1}^{*})^{\frac{1}{2}}},~A_{1}^{2-}=[\alpha_{2}^{(1)}/\alpha_{2}^{(1)^*}]^{1/2},\nonumber \\
	&&e^{i\theta_2^-}=\frac{(l_{1}-l_{2})(k_{1}-l_{2})^{\frac{1}{2}}(k_{1}+l_{2}^{*})^{\frac{1}{2}}(l_{1}+l_{2}^{*})}{(k_{1}^{*}-l_{2}^{*})^{\frac{1}{2}}(l_{1}^{*}-l_{2}^{*})(k_{1}^{*}+l_{2})^{\frac{1}{2}}(l_{1}^{*}+l_{2})},~A_{2}^{2-}=[\alpha_{2}^{(2)}/\alpha_{2}^{(2)^*}]^{1/2}.\nonumber
	\end{eqnarray}
	Here, superscript ($2-$) refers to soliton $S_2$ before collision. 
	\\
	\underline{(b) After collision}: $z\rightarrow +\infty$\\
	\underline{Soliton 1}: The asymptotic forms for soliton 1 after collision deduced as,
	\begin{subequations}
		\begin{eqnarray}
		&&q_{1}\simeq \frac{2k_{1R}A_1^{1+}e^{i(\eta_{1I}+\theta_1^+)}\cosh(\xi_{1R}+\phi_1^+)}{\big[\frac{(k_{1}^{*}-l_{1}^{*})^{\frac{1}{2}}}{(k_{1}^{*}+l_{1})^{\frac{1}{2}}}\cosh(\eta_{1R}+\xi_{1R}+\frac{\del_{18}-\vsa_{22}}{2})+\frac{(k_{1}+l_{1}^{*})^{\frac{1}{2}}}{(k_{1}-l_{1})^{\frac{1}{2}}}\cosh(\eta_{1R}-\xi_{1R}+\frac{\phi_{22}-\del_{16}}{2})\big]},\\
		&&q_2\simeq \frac{2l_{1R}A_1^{2+}e^{i(\xi_{1I}+\theta_2^+)}\cosh(\eta_{1R}+\phi_{2}^+)}{\big[\frac{(k_{1}^{*}-l_{1}^{*})^{\frac{1}{2}}}{(k_{1}+l_{1}^*)^{\frac{1}{2}}}\cosh(\eta_{1R}+\xi_{1R}+\frac{\del_{18}-\vsa_{22}}{2})+\frac{(k_{1}^*+l_{1})^{\frac{1}{2}}}{(k_{1}-l_{1})^{\frac{1}{2}}}\cosh(\eta_{1R}-\xi_{1R}+\frac{\phi_{22}-\del_{16}}{2})\big]}.
		\end{eqnarray} \end{subequations}
	Here,
	\begin{eqnarray}
	&&\phi_1^+=\phi_1^-+\frac{1}{2} \log\frac{|k_2-l_1|^2|l_1-l_2|^4}{|k_2+l_1^*|^2|l_1+l_2^*|^4},~\phi_3^+=\phi_3^-+\frac{1}{2}\log\frac{|k_1-k_2|^4|k_2-l_1|^2|k_1-l_2|^2|l_1-l_2|^4}{|k_1+k_2^*|^4|k_2+l_1^*|^2|k_1+l_2^*|^2|l_1+l_2^*|^4},\nonumber\\
	&&\phi_2^+=\phi_2^-+\frac{1}{2}\log\frac{|k_1-l_2|^2|k_1-k_2|^4}{|k_1+l_2^*|^2|k_1+k_2^*|^4},~\phi_4^+=\phi_4^-+\frac{1}{2}\log\frac{|k_1-k_2|^4|k_2+l_1^*|^2|k_1-l_2|^2|l_1+l_2^*|^4}{|k_1+k_2^*|^4|k_2-l_1|^2|k_1+l_2^*|^2|l_1-l_2|^4},\nonumber\\
	&&e^{i\theta_1^+}=\frac{(k_{1}-k_{2})(k_{1}-l_{2})^{\frac{1}{2}}(k_{1}^*+k_{2})(k_{1}^{*}+l_{2})^{\frac{1}{2}}}{(k_{1}^{*}-k_{2}^{*})(k_{1}^{*}-l_{2}^{*})^{\frac{1}{2}}(k_{1}+k_{2}^*)(k_{1}+l_{2}^{*})^{\frac{1}{2}}},~e^{i\theta_2^+}=\frac{(l_{1}-l_{2})(k_{2}-l_{1})^{\frac{1}{2}}(k_{2}+l_{1}^{*})^{\frac{1}{2}}(l_{1}^*+l_{2})}{(k_{2}^{*}-l_{1}^{*})^{\frac{1}{2}}(l_{1}^{*}-l_{2}^{*})(k_{2}^{*}+l_{1})^{\frac{1}{2}}(l_{1}+l_{2}^*)},\nonumber
	\end{eqnarray} 
    $A_{1}^{1+}=[\alpha_{1}^{(1)}/\alpha_{1}^{(1)^*}]^{1/2}$ and $A_{2}^{1+}=[\alpha_{1}^{(2)}/\alpha_{1}^{(2)^*}]^{1/2}$,  in which superscript ($1+$) denotes soliton $S_1$ after collision. 
	\\
	\underline{Soliton 2}: The expression for soliton 2 after collision deduced from the two soliton solution is
	\begin{subequations}
		\begin{eqnarray}
		&&q_{1}\simeq\frac{2A_2^{1+}k_{2R}e^{i\eta_{2I}}\cosh(\xi_{2R}+\varphi_1^+)}{\big[{\frac{(k_{2}^{*}-l_{2}^{*})^{\frac{1}{2}}}{(k_{2}^{*}+l_{2})^{\frac{1}{2}}}}\cosh(\eta_{2R}+\xi_{2R}+\varphi_3^+)+\frac{(k_{2}+l_{2}^{*})^{\frac{1}{2}}}{(k_{2}-l_{2})^{\frac{1}{2}}}\cosh(\eta_{2R}-\xi_{2R}+\varphi_4^+)\big]},\\
		&&q_{2}\simeq\frac{2A_2^{2+}l_{2R}e^{i\xi_{2I}}\cosh(\eta_{2R}+\varphi_2^+)}{\big[\frac{i(k_{2}^{*}-l_{2}^{*})^{\frac{1}{2}}}{(k_{2}+l_{2}^{*})^{\frac{1}{2}}}\cosh(\eta_{2R}+\xi_{2R}+\varphi_3^+)+\frac{(k_{2}^{*}+l_{2})^{\frac{1}{2}}}{(l_{2}-k_{2})^{\frac{1}{2}}}\cosh(\eta_{2R}-\xi_{2R}+\varphi_4^+)\big]},
		\end{eqnarray} \end{subequations} \end{widetext}
where $\varphi_1^+=\frac{1}{2}\log\frac{(k_2-l_2)|\alpha_{2}^{(2)}|^2}{(k_2+l_2^*)(l_2+l_2^*)^2}$, $\varphi_2^+=\frac{1}{2}\log\frac{(l_2-k_2)|\alpha_{2}^{(1)}|^2}{(k_2^*+l_2)(k_2+k_2^*)^2}$, $\varphi_3^+=\frac{1}{2}\log\frac{|k_2-l_2|^2|\alpha_{2}^{(1)}|^2|\alpha_{2}^{(2)}|^2}{|k_2+l_2^*|^2(k_2+k_2^*)^2(l_2+l_2^*)^2}$, $\varphi_4^+=\frac{1}{2}\log\frac{|\alpha_{2}^{(1)}|^2(l_2+l_2^*)^2}{|\alpha_{2}^{(2)}|^2(k_2+k_2^*)^2}$,   $A_{1}^{2+}=[\alpha_{2}^{(1)}/\alpha_{2}^{(1)^*}]^{1/2}$ and $A_{2}^{2+}=i[\alpha_{2}^{(2)}/\alpha_{2}^{(2)^*}]^{1/2}$. In the latter, superscript ($2+$) represents soliton $S_2$ after collision.  

In the above,  $\eta_{jR}=k_{jR}(t-2k_{jI}z)$,   $\eta_{jI}=k_{jI}t+(k_{jR}^{2}-k_{jI}^{2})z$, $\xi_{jR}=l_{jR}(t-2l_{jI}z)$, $\xi_{jI}=l_{jI}t+(l_{jR}^{2}-l_{jI}^{2})z$, $j=1,2,$ and that the phase terms $\varphi_j^-$, $j=1,2,3,4$ can also be rewritten as
$
\vphi_1^-=\vphi_1^++\frac{1}{2}\log\frac{|k_1-l_2|^2|l_1-l_2|^4}{|k_1+l_2^*|^2|l_1+l_2^*|^4}$, $\vphi_4^-=\vphi_4^++\frac{1}{2}\log\frac{|k_1-k_2|^4|l_1+l_2^*|^4|k_2-l_1|^2|k_1+l_2^*|^2}{|k_1+k_2^*|^4|k_2+l_1^*|^2|k_1-l_2|^2|l_1-l_2|^4}$, $\vphi_2^-=\vphi_2^++\frac{1}{2}\log\frac{|k_2-l_1|^2|k_1-k_2|^4}{|k_2+l_1^*|^2|k_1+k_2^*|^4}$, $\vphi_3^-=\vphi_3^++\frac{1}{2}\log\frac{|k_1-k_2|^4|l_1-l_2|^4|k_2-l_1|^2|k_1-l_2|^2}{|k_1+k_2^*|^4|k_2+l_1^*|^2|k_1+l_2^*|^2|l_1+l_2^*|^4}$. The above asymptotic analysis clearly shows that the shape preserving collision always occur among the nondegenerate solitons whenever the phase terms obey the conditions,\begin{equation}\phi_j^-=\phi_j^+, ~ \varphi_j^-=\varphi_j^+,~ j=1,2,3,4. \label{19}\end{equation}
\subsection{Shape preserving and altering collisions: Elastic collision}
From the above analysis, we observe that the intensities of nondegenerate solitons $S_{1}$ and $S_{2}$ in the two modes are the same before and after collision whenever the phase conditions (\ref{19}) are satisfied. This implies that  the initial amplitudes do not get altered after collision $j=1,2$. It is also evident from the transition amplitude calculations,  $T_{j}^{l}=\frac{A_{j}^{l+}}{A_{j}^{l-}}$, $j,l=1,2$, where the subscript $j$ represents the modes and the superscript $l\pm$ denotes the nondegenerate soliton numbers 1 and 2 in the asymptotic regimes $z \rightarrow \pm \infty$.  
\begin{figure}
	\centering
	\includegraphics[width=0.95\linewidth]{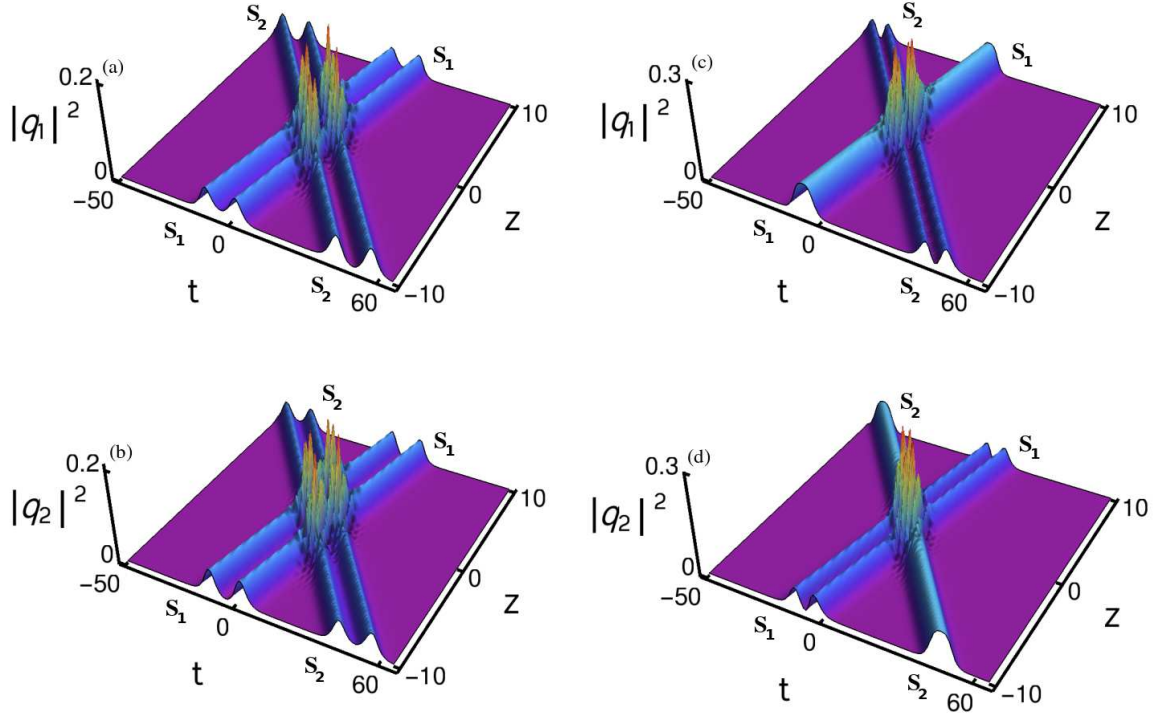}
	\caption{Shape preserving collision of symmetric nondegenerate solitons - The energy does not get exchanged among the nondegenerate solitons during the shape preserving collision process: (a) and (b) represent collision between two symmetric double-hump solitons. (c) and (d) denote interaction among flattop and symmetric double-hump soliton. The parameter values: (a) and (b): $k_{1}=0.333+0.5i$, $l_{1}=0.315+0.5i$, $k_{2}=0.315-2.2i$, $l_{2}=0.333-2.2i$, $\alpha_{1}^{(1)}=0.45+0.45i$, $\alpha_{2}^{(1)}=0.49+0.45i$, $\alpha_{1}^{(2)}=0.49+0.45i$ and $\alpha_{2}^{(2)}=0.45+0.45i$. (c) and (d): $k_{1}=0.43+0.5i$, $l_{1}=0.3+0.5i$, $k_{2}=0.3-2.2i$, $l_{2}=0.43-2.2i$, $\alpha_{1}^{(1)}=0.45+0.5i$, $\alpha_{2}^{(1)}=0.43+0.5i$, $\alpha_{1}^{(2)}=0.43+0.5i$ and $\alpha_{2}^{(2)}=0.45+0.5i$.}
	\label{f5}
\end{figure}
Again to confirm that the intensities of the nondegenerate solitons are preserved during the collision process, we calculate the transition intensities as well, $|T_{j}^{l}|^2$, $l,j=1,2$, which can be obtained by taking the absolute squares of the transition amplitudes $T_{j}^{l}$'s. The transition intensities turn out to be unimodular, that is $|T_{j}^{l}|^2=1$, $l,j=1,2$. Physically this implies that the nondegenerate solitons, for $k_{1I}=l_{1I}$, $k_{2I}=l_{2I}$, $k_1 \neq l_1$, corresponding to two distinct wave numbers undergo elastic collision without any intensity redistribution between the modes $q_1$ and $q_2$ except for a finite phase shift. The latter confirms that the polarization vectors associated with the nondegenerate fundamental solitons do not contribute to the energy redistribution among the modes. Consequently the nondegenerate solitons in each mode exhibit elastic collision.  The total intensity of each soliton is conserved which can be verfied from $|A_j^{l-}|^2=|A_j^{l+}|^2$, $j,l=1,2$. In addition to this, the total intensity in each of the modes is also conserved $|A_j^{1-}|^2+|A_j^{2-}|^2=|A_j^{1+}|^2+|A_j^{2+}|^2=\text{constant}$.   

During the collision process, the initial phase of each of the soliton is also changed. The phase shift of soliton $S_1$ in the two modes gets modified after collision as
\begin{eqnarray}
\Phi_1^1&=&\phi_1^+-\phi_1^-=\log \frac{|k_2-l_1||l_1-l_2|^2}{|k_2+l_1^*||l_1+l_2^*|^2},\nonumber\\
\Phi_2^1&=&\phi_2^+-\phi_2^-=\log \frac{|k_1-l_2||k_1-k_2|^2}{|k_1+l_2^*||k_1+k_2^*|^2}. \label{4.6}
\end{eqnarray} 
Similarly the phase shift suffered by soliton $S_2$ in the two modes are given by 
\begin{eqnarray}
\Phi_1^2&=&\varphi_1^+-\varphi_1^-=\log \frac{|k_1+l_2^*||l_1+l_2^*|^2}{|k_1-l_2||l_1-l_2|^2},\nonumber\\
\Phi_2^2&=&\varphi_2^+-\varphi_2^-=\log \frac{|k_2+l_1^*||k_1+k_2^*|^2}{|k_2-l_1||k_1-k_2|^2}.\label{4.7}
\end{eqnarray} 
From the above expressions we conclude that the phases of all the solitons are mainly influenced by the wave numbers $k_j$ and $l_j$, $j=1,2$, and not by the complex parameters $\al_1^{(j)}$'s and $\al_2^{(j)}$'s, $j=1,2$. This peculiar property of nondegenerate solitons is different in the case of degenerate vector bright solitons (see Sec. V below)  where the complex parameters $\al_1^{(j)}$'s and $\al_2^{(j)}$'s, associated with polarization constants, play a crucial role in shifting the position of solitons after collision.

Further, to confirm that the profile shapes of the nondegenerate solitons $S_1$ and $S_2$ are invariant under the above elastic collision, we explicitly deduce the relative separation distance between the modes of the solitons. This is similar to the analysis which we have  already discussed for the one-soliton solution to confirm the symmetric and asymmetric profile natures of the fundamental soliton. As a consequence of this analysis, one would expect that the relative separation distance values corresponding to solitons $S_1$ and $S_2$ before collision should be equal to the values after collision in order to ensure the shape preserving nature of the collision. For this purpose first we deduce the following expressions for relative separation distance for the solitons $S_1$ and $S_2$ before and after collisions from the asymptotic forms as
\begin{subequations}
	\begin{eqnarray}
	\Delta t_{12}^{1-}&=&\frac{1}{l_{1R}}\log \frac{|\al_1^{(2)}|(k_1-l_1)^{1/2}}{2l_{1R}(k_1+l_{1}^*)^{1/2}}-\frac{1}{k_{1R}}\log \frac{(l_1-k_1)^{1/2}|\al_1^{(1)}|}{2k_{1R}(k_1^*+l_{1})^{1/2}},\\
	\Delta t_{12}^{2-}&=&\frac{1}{l_{2R}}\log \frac{|\al_2^{(2)}||k_1-l_2|(k_2-l_2)^{1/2}|l_1-l_2|^2}{2l_{2R}|k_1+l_2^*|(k_2+l_2^*)^{1/2}|l_1+l_2^*|^2}\nonumber\\
	&&
	-\frac{1}{k_{2R}}\log\frac{|\al_2^{(1)}||k_1-k_2|^2|k_2-l_1|(l_2-k_2)^{1/2}}{2k_{2R}|k_1+k_2^*|^2|k_2+l_1^*|(k_2^*+l_2)^{1/2}},\end{eqnarray} \end{subequations}
\begin{subequations}
	\begin{eqnarray}
	\Delta t_{12}^{1+}&=&\frac{1}{l_{1R}}\log \frac{|\al_1^{(2)}||k_2-l_1|(k_1-l_1)^{1/2}|l_1-l_2|^2}{2l_{1R}|k_2+l_1^*|(k_1+l_1^*)^{1/2}|l_1+l_2^*|^2}~~\nonumber\\
	&&-\frac{1}{k_{1R}}\log\frac{|\al_1^{(1)}||k_1-k_2|^2|k_1-l_2|(l_1-k_1)^{1/2}}{2k_{1R}|k_1+k_2^*|^2|k_1+l_2^*|(k_1^*+l_1)^{1/2}},~~~~~~~~~\\
	\Delta t_{12}^{2+}&=&\frac{1}{l_{2R}}\log \frac{|\al_2^{(2)}|(k_2-l_2)^{1/2}}{2l_{2R}(k_2+l_{2}^*)^{1/2}}-\frac{1}{k_{2R}}\log \frac{(l_2-k_2)^{1/2}|\al_2^{(1)}|}{2k_{2R}(k_2^*+l_{2})^{1/2}}.\label{4.8}
	\end{eqnarray}  
\end{subequations}
To identify the profile change of a given soliton $S_1$ (or $S_2$) during the collision, we analytically find the total change in relative separation distance by subtracting the quantity $\Delta t_{12}^{n-}$ from $\Delta t_{12}^{n+}$, $n=1,2$. This results in the following expressions for soliton $S_1$,   
\begin{eqnarray}
\Delta t_1=\Delta t_{12}^{1+}-\Delta t_{12}^{1-}=\frac{1}{l_{1R}}\log \frac{|k_2-l_1||l_1-l_2|^{2}}{|k_2+l_1^*||l_1+l_2^*|^{2}}-\frac{1}{k_{1R}}\log \frac{|k_1-l_2||k_1-k_2|^{2}}{|k_1+l_2^*||k_1+k_2^*|^{2}},\label{4.9}
\end{eqnarray}
and for soliton $S_2$,
\begin{eqnarray}
\Delta t_2=\Delta t_{12}^{2+}-\Delta t_{12}^{2-}=\frac{1}{l_{2R}}\log \frac{|k_1-l_2||l_1-l_2|^{2}}{|k_1+l_2^*||l_1+l_2^*|^{2}}-\frac{1}{k_{2R}}\log \frac{|k_2-l_1||k_1-k_2|^{2}}{|k_2+l_1^*||k_1+k_2^*|^{2}}.\label{4.10}
\end{eqnarray}
\begin{figure}
	\centering
	\includegraphics[width=0.95\linewidth]{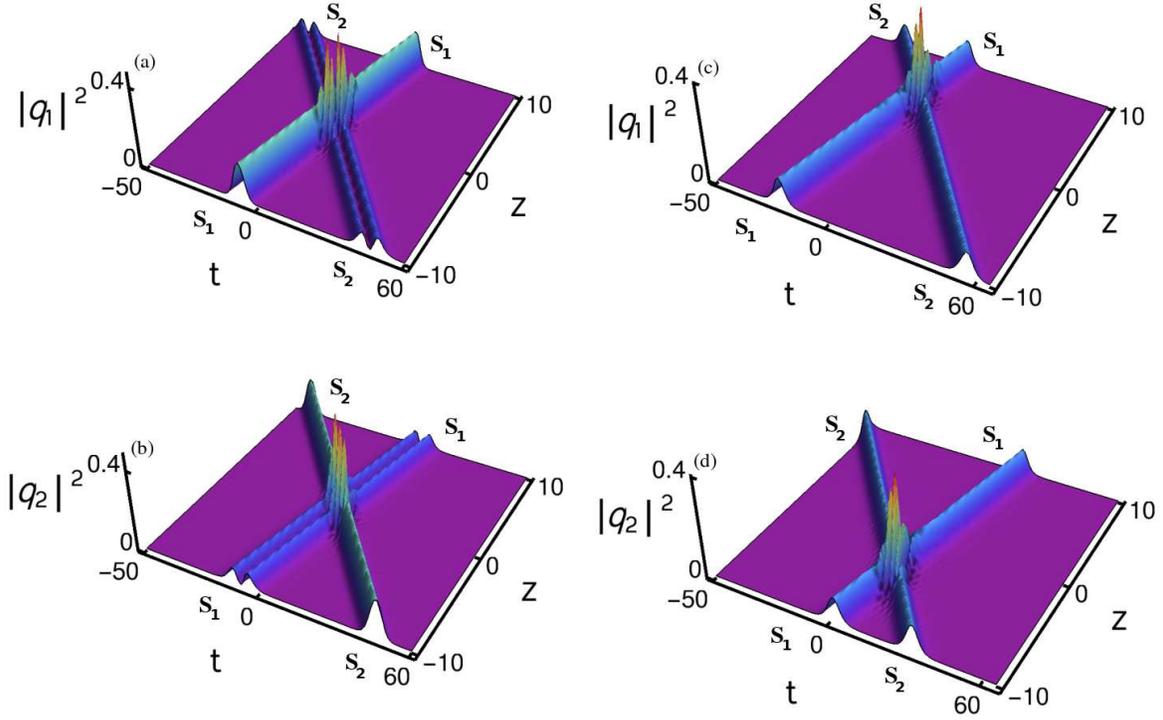}
	\caption{Shape preserving collision of symmetric nondegenerate solitons: (a) and (b) deonte collision between single-hump and double-hump solitons: The values corresponding to this collision scenario are $k_{1}=0.55+0.5i$, $l_{1}=0.333+0.5i$, $k_{2}=0.333-2.2i$, $l_{2}=0.55-2.2i$, $\alpha_{1}^{(1)}=0.45+0.5i$, $\alpha_{2}^{(1)}=0.43+0.5i$, $\alpha_{1}^{(2)}=0.43+0.5i$ and $\alpha_{2}^{(2)}=0.45+0.5i$. (c) and (d) denote two single-hump solitons interaction: The corresponding parameter values are chosen as  $k_{1}=0.333+0.5i$, $l_{1}=-0.316+0.5i$, $k_{2}=-0.316-2.2i$, $l_{2}=0.333-2.2i$, $\alpha_{1}^{(1)}=0.45+0.51i$, $\alpha_{2}^{(1)}=0.5+0.5i$, $\alpha_{1}^{(2)}=0.5+0.5i$ and $\alpha_{2}^{(2)}=0.45+0.51i$.}
	\label{f6}
\end{figure}
To demonstrate the shape preserving collision property of nondegenerate solitons, for the case $k_{1I}=l_{1I}$, $k_{2I}=l_{2I}$, we start with various symmetric profiles as initial conditions. In Figs. 5(a) and 5(b) we set two well separated symmetric double-hump soliton profiles as initial profiles in both the modes. From these figures, we observe that the symmetric nature of double-hump soliton $S_1$ is preserved in both the modes after collision while interacting with another symmetric double-hump soliton $S_2$ except for a finite phase shift, which is already deduced in Eqs. (\ref{4.6}) and (\ref{4.7}). This can be easily verified from the asymptotic analysis itself. Further, in order to ensure the shape preserving collision scenario of symmetric double-hump solitons we explicitly compute the numerical value of relative separation distance between the modes of each double-hump solitons by substituting all the parameter values in Eqs. (\ref{4.9}) and (\ref{4.10}). This action yields  the final values as $\Delta t_{1}=-0.0051$ and $\Delta t_{2}=-0.0051$ (here we provide the values with two decimal accuracy, to get perfect zero, one has to fine tune the parameters suitably). The values reaffirm that symmetric profile struture of double-hump solitons are indeed preserved during the collision. This ensures further that the relative separation distance values are consistent with the shape preserving collision condition $\phi_j^-=\phi_j^+$ and $\varphi_j^-=\varphi_j^+$, $j=1,2,3,4$, given by Eq. (19). 
  We also show the shape preserving collision between flattop soliton and double-hump soliton occurs in Figs. 5(c) and 5(d). The same type of collision behaviour is also observed while the symmetric single-hump soliton collides with the symmetric double-hump soliton, which is illustrated in Figs. 6(a) and 6(b). In Figs. 6(c) and 6(d) we depict the elastic collision between two symmetric single-hump solitons. From Figs. 6, we find that  each soliton retains its structure during the collision scenario.             
\begin{figure}
	\centering
	\includegraphics[width=0.95\linewidth]{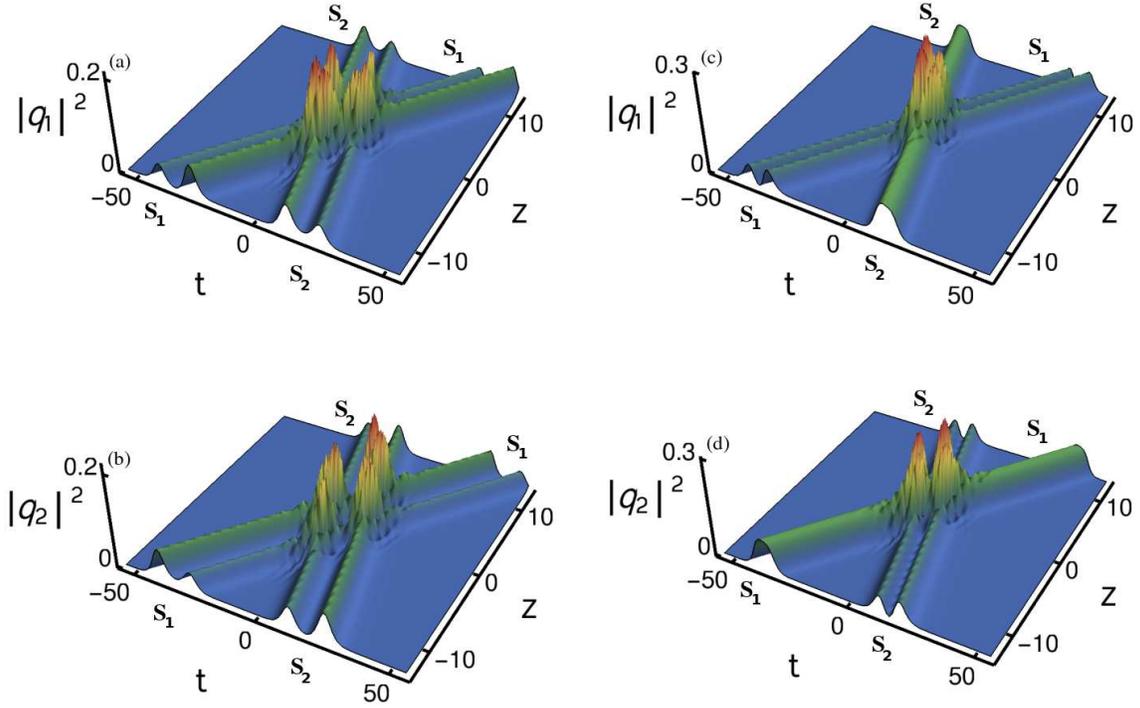}
	\caption{Shape preserving collision of asymmetric nondegenerate solitons: (a) and (b) represent two asymmetric soliton collision: $k_{1}=0.333-0.5i$, $l_{1}=0.315-0.5i$, $k_{2}=0.315+1.5i$, $l_{2}=0.333+1.5i$, $\alpha_{1}^{(1)}=0.65+0.45i$, $\alpha_{2}^{(1)}=0.49+0.5i$, $\alpha_{1}^{(2)}=0.49+0.5i$ and $\alpha_{2}^{(2)}=0.65+0.45i$ (c) and (d) denote asymmetric flattop-double-hump soliton: The corresponding parameter values are chosen as (a):  $k_{1}=0.425-0.5i$, $l_{1}=0.3-0.5i$, $k_{2}=0.3+1.5i$, $l_{2}=0.425+1.5i$, $\alpha_{1}^{(1)}=0.5+0.51i$, $\alpha_{2}^{(1)}=0.43+0.5i$, $\alpha_{1}^{(2)}=0.43+0.5i$ and $\alpha_{2}^{(2)}=0.5+0.51i$.}
	\label{f7}
\end{figure}

Next, we illustrate the shape preserving collision among the asymmetric solitons. As we pointed out earlier, the nondegenerate fundamental soliton also admits asymmetric profiles for $k_{1I}=l_{1I}$. To bring out one more asymmetric soliton  we set $k_{2I}=l_{2I}$ in the two-soliton solution (\ref{3.7a})-(\ref{3.7c}). In order to study the shape preserving collision of such two asymmetric solitons, first we locate asymmetric double-hump soliton $S_1$ along the line $\eta_{1R}=k_{1R}(t-2k_{1I}z)\simeq 0$, $\xi_{1R}=l_{1R}(t-2k_{1I}z)\simeq 0$ and another similar kind of soliton $S_2$  along the line $\eta_{2R}=k_{2R}(t-2k_{2I}z)\simeq 0$, $\xi_{2R}=l_{2R}(t-2k_{2I}z)\simeq 0$. These asymmetric structured double-hump solitons also preserve their structure after collision. This is clearly depicted in Figs. 7(a) and 7(b). To ensure the shape preserving nature of asymmetric solitons, we again explicitly calculate the relative separation distance values for both the asymmetric solitons $S_1$ and $S_2$ as $\Delta t_1=\Delta t_2=-0.0093$. These values again confirm the shape preserving property of the asymmetric double-hump solitons and they are indeed compatible with the shape preserving collision condition (\ref{19}).  As displayed in Figs. 7(c) and 7(d), the asymmetric flattop soliton also preserves its structure when it collides with an asymmetric double-hump soliton. In other cases also asymmetric solitons preserve their profiles. This can be confirmed from Fig. 8.  Very interestingly the shape preserving collision also occurs even when the asymmetric double-hump soliton interacts with the symmetric double-hump soliton. This is illustrated in Fig. \ref{f9}. During this collision also the standard position shift only occurs as a final outcome.

Then, we also come across another type of elastic collision, namely shape altering collision for certain set of parametric choices again with $k_{1I}=l_{1I}$ and $k_{2I}=l_{2I}$. We illustrate such collision scenario in Fig. \ref{f10}. We explain the profile alteration in the head-on collision between slowly moving symmetric double-hump soliton and fastly moving asymmetric double-hump soliton as displayed in Figs. 10(a)-(b). To draw this figure we fix the parametric choice as $k_{1}=0.41+0.5i$, $l_{1}=0.305+0.5i$, $k_{2}=0.305-2.2i$, $l_{2}=0.41-2.2i$, $\alpha_{1}^{(1)}=\alpha_{2}^{(2)}=0.44+0.499i$ and $\alpha_{2}^{(1)}=\alpha_{1}^{(2)}=0.44+0.5i$ in solution (\ref{3.7a})-(\ref{3.7c}). From this figure, we find that while symmetric double-hump soliton $S_1^{-}$ in the first mode slightly changes into an asymmetric structure, the asymmetric double-hump soliton $S_2^{-}$ becomes symmetric.  For this kind of shape altering collision the parameter values corresponding to Figs. 10(a)-(b) are inconsistent with the condition (\ref{19}), eventhough the unimodular condition of transition amplitudes is still preserved. Similar kind of profile alteration occurs in the second mode also. This is due to the incoherent interaction between the modes $q_1$ and $q_2$.  Again similar type of collision property has been observed when a symmetric (or asymmetric) flattop soliton collides with an asymmetric (or symmetric) double-hump soliton in the $q_1$ (or $q_2$) component, which is demonstrated in Figs. 10(c) and 10(d) for $k_{1}=0.425+0.5i$, $l_{1}=0.3+0.5i$, $k_{2}=0.3-2.2i$, $l_{2}=0.425-2.2i$, $\alpha_{1}^{(1)}=\alpha_{2}^{(2)}=0.5+0.5i$ and $\alpha_{2}^{(1)}=\alpha_{1}^{(2)}=0.45+0.5i$. In Figs. 10(e) and 10(f), we illustrate shape alteration collision between symmetric single-hump  and double-hump solitons in both the components by fixing the parameter values as $k_{1}=0.55-0.5i$, $l_{1}=0.333-0.5i$, $k_{2}=0.333+1.5i$, $l_{2}=0.55+1.5i$, $\alpha_{1}^{(1)}=\alpha_{2}^{(2)}=0.5+0.5i$ and $\alpha_{2}^{(1)}=\alpha_{1}^{(2)}=0.45+0.5i$. In each of the modes, the collision transforms the symmetric double-hump soliton into a slightly asymmetric double-hump soliton leaving the symmetric single-hump soliton unaltered. However, in all the above cases the energy does not get redistributed among the modes eventhough the shape of the solitons gets altered during the collision.  One can prove the unimodular nature of the transition amplitudes in these cases by following the procedure mentioned earlier in this section. As we pointed out earlier, the similar kind of shape preserving and shape altering collisions are also observed in the case of $k_{1I}\neq l_{1I}$ and $k_{2I}\neq l_{2I}$. Here, we have not displayed their plots and their corresponding asymptotic analysis for brevity.

Additionally, in Fig. \ref{f11}, we display another type of collision scenario for the velocity condition $k_{1I}=l_{1I}$, $k_{2I}\neq l_{2I}$. In this collision scenario the asymmetric double-hump solitons that are present in the two modes change dramatically. However, the single-hump solitons undergo collision without any change in their intensity profiles. Due to the incoherent coupling between the modes, the change occured only in the profile of the double-hump soliton. One can carry out an appropriate asymptotic analysis for this kind of collision process also. We also note here that this kind of shape changing collision is not observed in the degenerate case.
\begin{figure}
	\centering
	\includegraphics[width=0.95\linewidth]{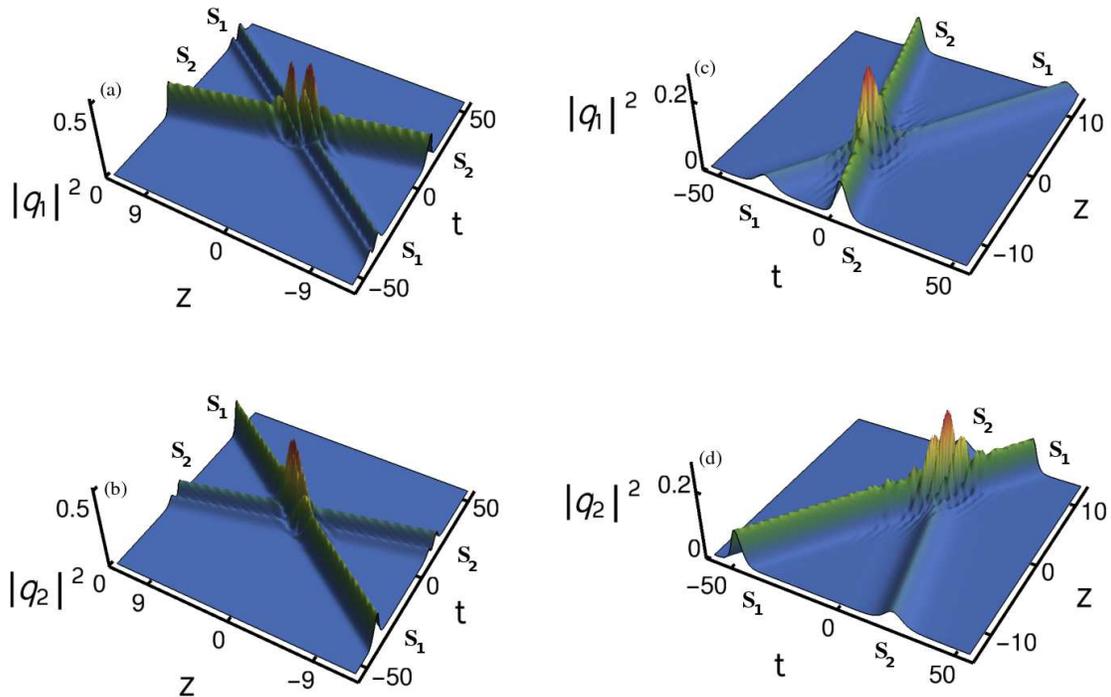}
	\caption{Shape preserving collision of asymmetric nondegenerate solitons: (a) and (b) represent asymmetric single-hump and double-hump soliton collision: $k_{1}=0.55-0.5i$, $l_{1}=0.333-0.5i$, $k_{2}=0.333+1.5i$, $l_{2}=0.55+1.5i$, $\alpha_{1}^{(1)}=1.2+0.5i$, $\alpha_{2}^{(1)}=0.5+0.45i$, $\alpha_{1}^{(2)}=0.5+0.45i$ and $\alpha_{2}^{(2)}=1.2+0.5i$.  (c) and (d) denote collision of two asymmetric single-hump solitons: The  parameter values of each figure are chosen as :  $k_{1}=0.333-0.5i$, $l_{1}=-0.2-0.5i$, $k_{2}=-0.2+1.5i$, $l_{2}=0.333+1.5i$, $\alpha_{1}^{(1)}=0.45+3.0i$, $\alpha_{2}^{(1)}=0.5+0.5i$, $\alpha_{1}^{(2)}=0.5+0.5i$ and $\alpha_{2}^{(2)}=0.45+3.0i$.}
	\label{f8}
\end{figure}
\begin{figure}
	\centering
	\includegraphics[width=0.95\linewidth]{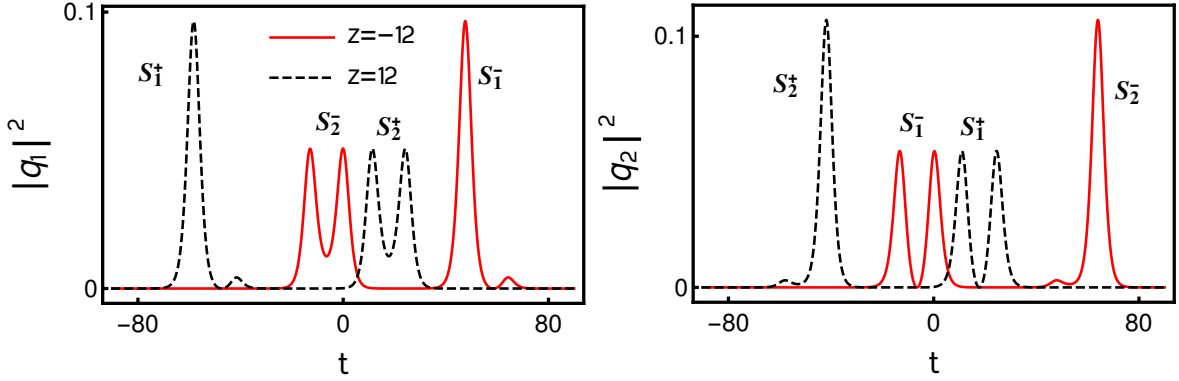}
	\caption{Shape preserving collision between symmetric double-hump soliton and asymmetric double-hump soliton: The parameter values are $k_{1}=0.333+0.5i$, $l_{1}=0.315+0.5i$, $k_{2}=0.315-2.2i$, $l_{2}=0.333-2.2i$, $\alpha_{1}^{(1)}=0.45+0.45i$, $\alpha_{2}^{(1)}=2.49+2.45i$, $\alpha_{1}^{(2)}=0.49+0.45i$ and $\alpha_{2}^{(2)}=0.45+0.45i$.}
	\label{f9}
\end{figure}
We remark that elastic collision is also noticed in the case of  dissipative solitons where a new soliton pair (doublet)  is formed when single soliton state (singlet) destroys initial doublet state. During this interaction, energy or momentum is not conserved in the fiber laser cavity \cite{dis1,dis3}. But the elastic collision observed in the present conservative system is entirely different from the above collision which has been observed in the dissipative system.  The vector solitons in dissipative systems exhibit several interesting dynamical features, especially in fiber lasers.  Fiber lasers are very useful nonlinear systems to study the formation and dynamics of temporal optical solitons experimentally. In fact several types of solitons were observed experimentally in fiber lasers. For instance, vector multi-soliton
operation and vector soliton interaction in an erbium doped fiber laser \cite{fbl01}, and  a novel type of vector dark domain wall soliton have been observed in a fiber ring
laser \cite{fbl02}. Also vector dissipative soliton operation of erbium-doped fiber lasers mode locked with atomic layer graphene was experimentally investigated \cite{fbl03} and the coexistence of polarization-locked and polarization rotating vector solitons in a fiber laser with a semiconductor saturable absorber mirror have been observed experimentally \cite{fbl04}.         


\subsection{Shape changing collision}
Further, here we demonstrate the shape changing collision scenario  of nondegenerate solitons for unequal velocities, that is   $k_{1I}\neq l_{1I}$ and $k_{2I}\neq l_{2I}$ (We also note here that for appropriate choices of parameters for this unequal velocity case as pointed out above both shape preserving and shape altering cases do occur). During this interaction, we observe that an intensity redistribution occurs among the modes of nondegenerate fundamental solitons along with profile change. We display such a collision dynamics in Figs. \ref{f12} and \ref{f13}. A typical intensity redistribution phenomenon is demonstrated in Fig. \ref{f12} when two asymmetric double-hump solitons collide with each other. To bring out this nonlinear phenomenon we choose the parameter values as $k_{1}=1.2-0.5i$, $l_{1}=0.8+0.5i$, $k_{2}=1.0+0.5i$, $l_{2}=1.5-0.5i$, $\alpha_{1}^{(1)}=\alpha_{2}^{(2)}=0.5+0.51i$ and $\alpha_{2}^{(1)}=\alpha_{1}^{(2)}=0.45+0.5i$. From Fig. \ref{f12}, one can easily observe that the profiles of asymmetric double-hump solitons $S_1$ and $S_2$ change dramatically after collision, where the initial asymmetric  solitons $S_1$ and $S_2$ lose their identities and reemerge with another set of asymmetric profiles. In addition to the profile changes, there is also a finite intensity redistribution which takes place between the two modes of the solitons. However, the total energy of the individual solitons as well as modes is conserved in order to hold the energy conservation of system (\ref{e1}). Similar kind of collision is also depicted in Fig. \ref{f13}, where
a drastic change only occurs in the profile of asymmetric double-hump soliton but without any change in the asymmetric single-hump soliton. This can be witnessed in   Fig. \ref{f13} by setting the values of the parameters as $k_{1}=0.36+0.5i$, $l_{1}=0.3-0.5i$, $k_{2}=0.5-2.1i$, $l_{2}=0.45-2.2i$, $\alpha_{1}^{(1)}=\alpha_{2}^{(2)}=0.5+0.5i$ and $\alpha_{2}^{(1)}=1.7+0.45i$, $\alpha_{1}^{(2)}=0.45+0.5i$ in the solution (\ref{3.7a})-(\ref{3.7c}).  From this figure one can confirm that the intensity redistribution only occurs among the modes of the asymmetric double-hump soliton. A detailed asymptotic analysis has been carried out in order to ensure this peculiar intensity redistribution, which we have given in Appendix B. We remark that the nondegenerate solitons also exhibit shape changing collision for the equal velocity case as well with $k_{1I}=l_{1I}$ and $k_{2I}=l_{2I}$ for appropriate choice of parameters, which are inconsistent with Eq. (\ref{19}).
\begin{figure*}
	\centering
	\includegraphics[width=0.65\linewidth]{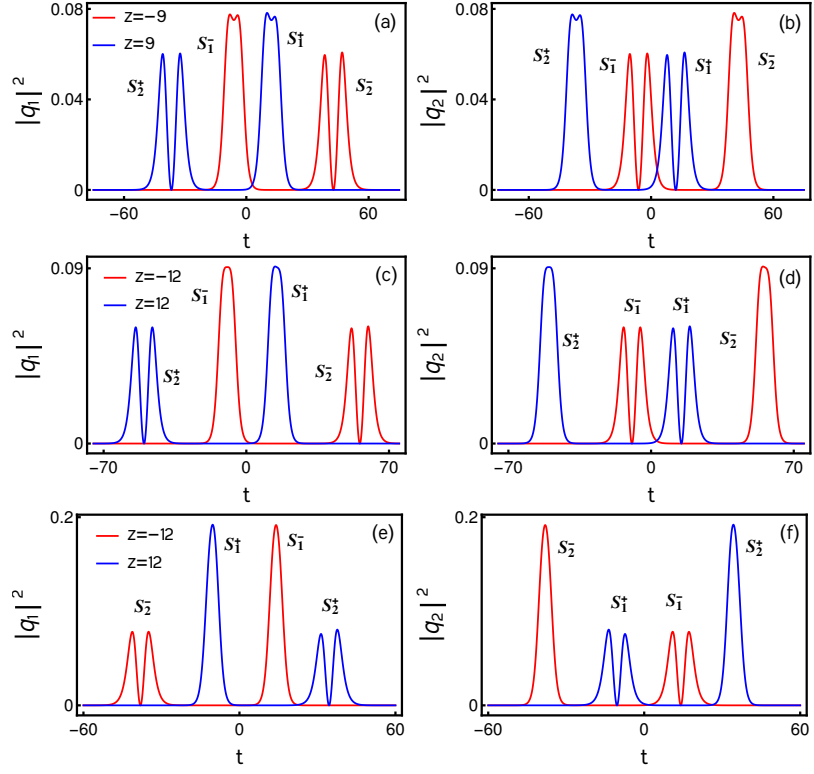}
	\caption{Shape altering collision: (a) and (b) denote shape altering collision between symmetric double-hump soliton and asymmetric double-hump soliton. (c) and (d) refer to collision between symmetric flattop and asymmetric double-hump soliton. (e) and (f) represent interaction between single-hump and asymmetric double-hump soliton.  }
	\label{f10}
\end{figure*} 
\begin{figure}
	\centering
	\includegraphics[width=0.95\linewidth]{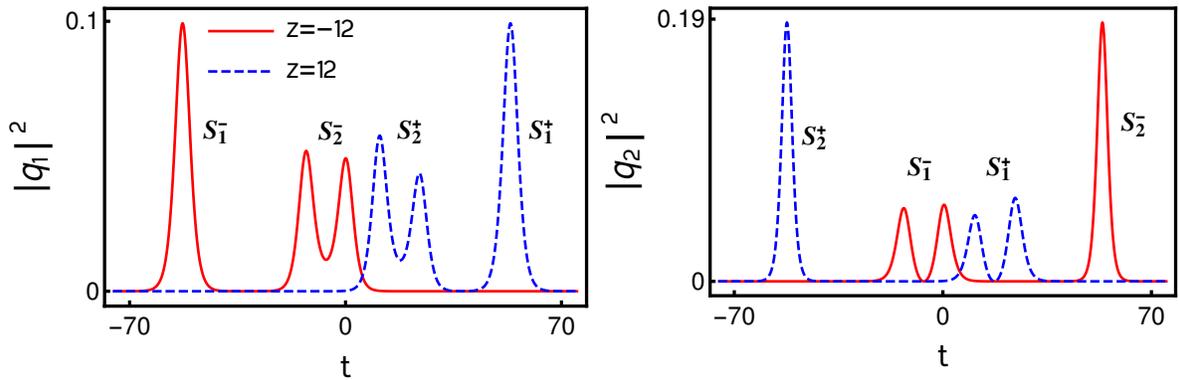}
	\caption{Shape changing collision between asymmetric double-hump soliton and single-hump soliton: $k_{1}=0.333+0.5i$, $l_{1}=0.315+0.5i$, $k_{2}=0.315+2.2i$, $l_{2}=0.433-2.2i$, $\alpha_{1}^{(1)}=\alpha_{2}^{(2)}=0.5+0.5i$, $\alpha_{2}^{(1)}=\alpha_{1}^{(2)}=0.45+0.5i$. }
	\label{f11}
\end{figure}
\begin{figure*}
	\centering
	\includegraphics[width=0.8\linewidth]{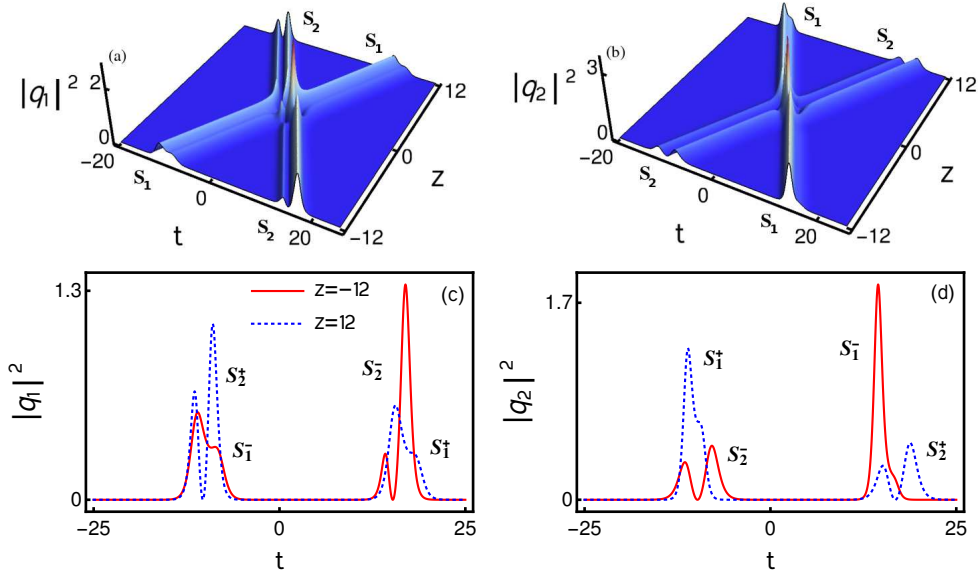}
	\caption{Shape changing collision between two asymmetric  double-hump solitons: $k_{1}=1.2-0.5i$, $l_{1}=0.8+0.5i$  $k_{2}=1.0+0.5i$, $l_{2}=1.5-0.5i$, $\alpha_{1}^{(1)}=\alpha_{2}^{(2)}=0.5+0.5i$,  $\alpha_{2}^{(1)}=\alpha_{1}^{(2)}=0.45+0.5i$. }
	\label{f12}
\end{figure*}
\begin{figure*}
	\centering
	\includegraphics[width=0.75\linewidth]{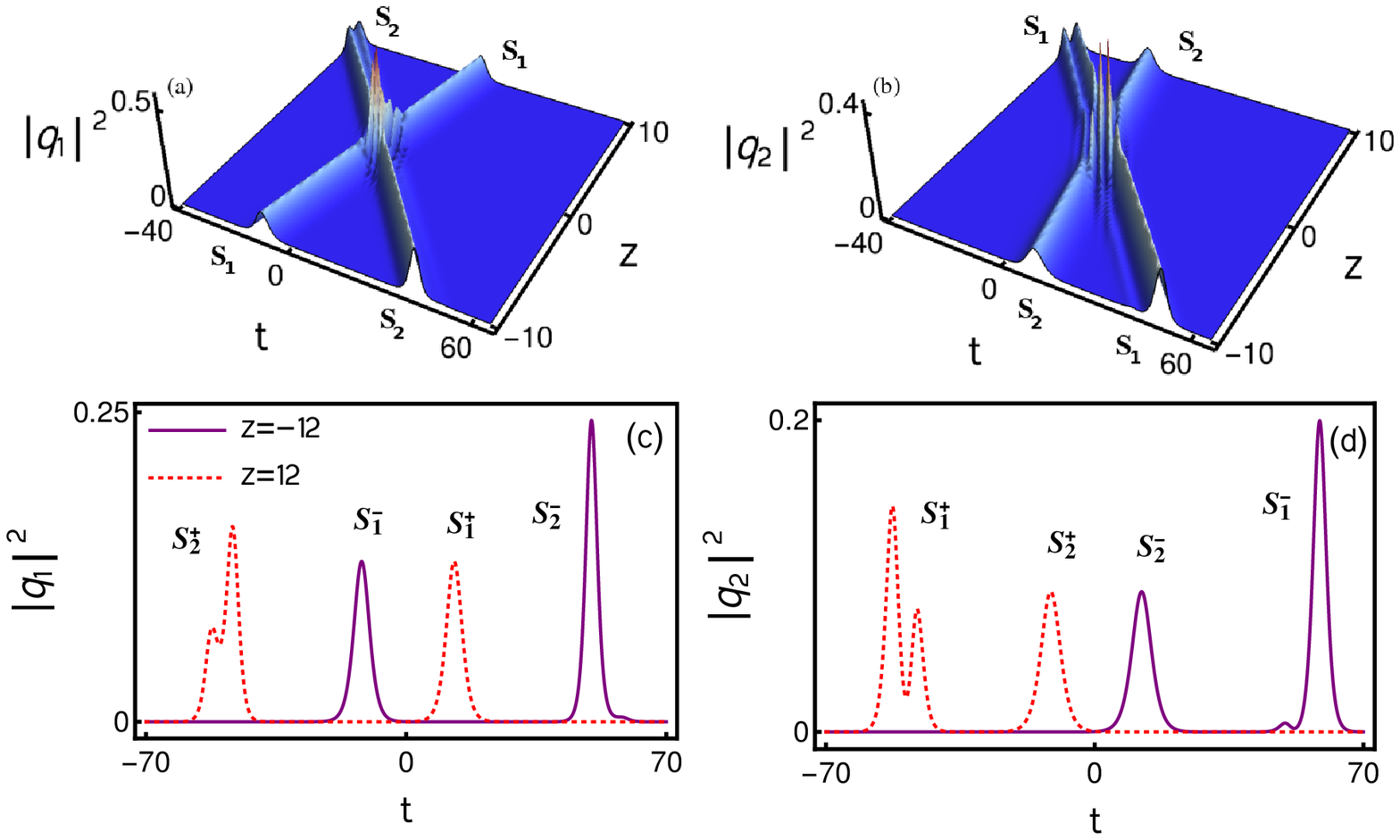}
	\caption{Shape changing collision between asymmetric single-hump and double-hump solitons: $k_{1}=0.36+0.5i$, $l_{1}=0.3-0.5i$  $k_{2}=0.5-2.1i$, $l_{2}=0.45-2.2i$, $\alpha_{1}^{(1)}=\alpha_{2}^{(2)}=0.5-0.5i$,  $\alpha_{2}^{(1)}=1.7+0.45i$, $\alpha_{1}^{(2)}=0.45+0.5i$. }
	\label{f13}
\end{figure*}
\section{Collision between nondegenerate and degenerate solitons}
In this section, we discuss the collision among degenerate and nondegenerate solitons admitted by the two-soliton solution (\ref{3.7a})-(\ref{3.7c}) of Manakov system (\ref{e1}) in the partial nondegenerate limit $k_1=l_1$ and $k_2\neq l_2$. The following asymptotic analysis assures that there is a definite energy redistribution occurs among the modes $q_1$ and $q_2$. 
\begin{figure*}
	\centering
	\includegraphics[width=0.75\linewidth]{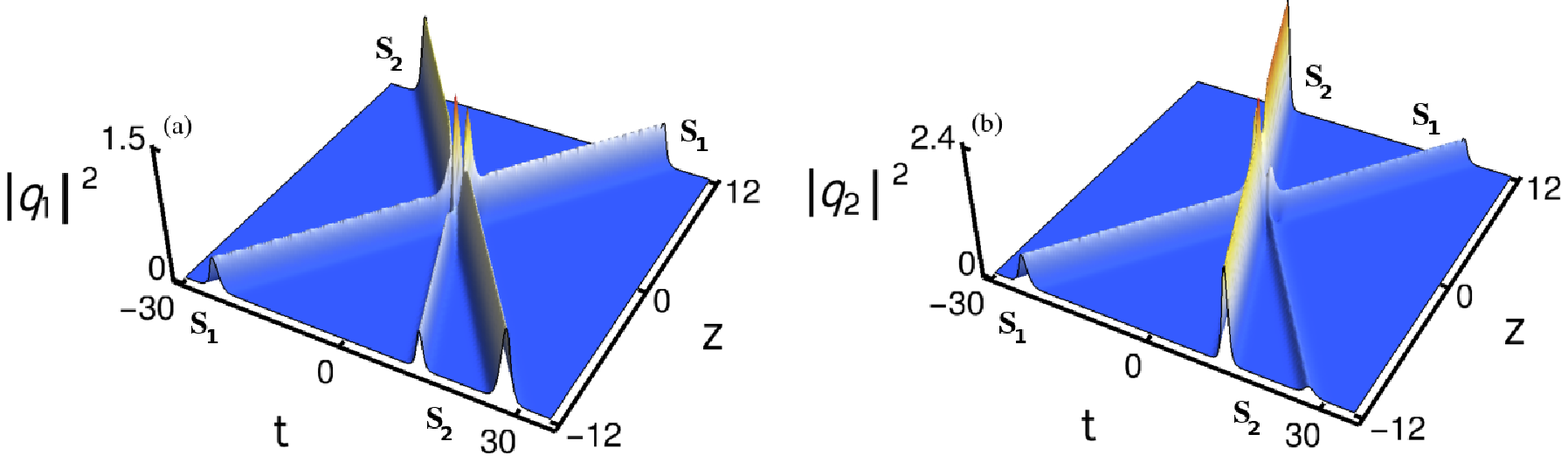}~~
	\caption{Energy sharing collision between degenerate and nondegenerate soliton: $k_{1}=l_1=1+i$,  $k_{2}=1-i$, $l_{2}=1.5-0.5i$, $\alpha_{1}^{(1)}=0.8+0.8i$, $\alpha_{2}^{(2)}=0.6+0.6i$, $\alpha_{2}^{(1)}=0.25+0.25i$, $\alpha_{1}^{(2)}=1+i$.}
	\label{f14}
\end{figure*}
\subsection{Asymptotic analysis}
To elucidate this new kind of collision behaviour, we analyze the partial nondegenerate two-soliton solution (\ref{3.8a})-(\ref{3.8c}) in the asymptotic limits $z\rightarrow \pm \infty$. The resultant action yields the asymptotic forms corresponding to degenerate and nondegenerate solitons. As we pointed out in the shape preserving collision case, to obtain the asymptotic forms for the present case we incorporate the asymptotic nature of the wave variables $\eta_{jR}=k_{jR}(t-2k_{Ij}z)$ and $\xi_{2R}=l_{2R}(t-2l_{2I}z)$, $j=1,2$, in the solution (\ref{3.8a})-(\ref{3.8c}). Here the wave variable $\eta_{1R}$ corresponds to the degenerate soliton and  $\eta_{2R}$, $\xi_{2R}$ correspond to the nondegenerate soliton.  In order to find the asymptotic behaviour of these wave variables we consider the parametric choice as $k_{1R},k_{2R},l_{2R}>0$, ~$k_{1I}>0$,~ $k_{2I},l_{2I}<0$, ~$k_{1I}>k_{2I}$,~$k_{1I}>l_{2I}$. For this choice, the wave variables behave asymptotically as follws: (i) degenerate soliton $S_1$: $\eta_{1R}\simeq0$, $\eta_{2R}$,$\xi_{2R}\rightarrow \mp\infty$ as $z\rightarrow \mp\infty$ (ii) nondegenerate soliton $S_2$: $\eta_{2R},\xi_{2R}\simeq 0$, $\eta_{1R}\rightarrow\pm \infty$ as $z\rightarrow\pm \infty$. By incorporating these asymptotic behaviours of wave variables in the solution (\ref{3.8a})-(\ref{3.8c}), we deduce the following asymptotic expressions for degenerate and nondegenerate solitons. \\
\underline{(a) Before collision}: $z\rightarrow -\infty$\\
\underline{Soliton 1}: In this limit, the asymptotic form for the degenerate soliton deduced from the partially nondegenerate two soliton solution (\ref{3.8a})-(\ref{3.8c}) is
\begin{eqnarray}
q_j\simeq\begin{pmatrix}
A_1^{1-}\\ \\
A_2^{1-}
\end{pmatrix} k_{1R}e^{i\eta_{1I}}\sech(\eta_{1R}+\frac{R}{2}), ~j=1,2,
\label{5.1}
\end{eqnarray}
where $A_j^{1-}=\al_1^{(j)}/(|\al_1^{(1)}|^2+|\al_1^{(2)}|^2)^{1/2}$, $j=1,2$, $R=\ln\frac{(|\al_1^{(1)}|^2+|\al_1^{(2)}|^2)}{(k_1+k_1^*)^2}$. Here, in $A_j^{1-}$ the superscript $1-$ denote soliton $S_1$ before collision and subscript $j$ refers to the mode number.  

\underline{Soliton 2}: The asymptotic expressions for the nondegenerate soliton $S_2$ which is present in the two modes before collision are obtained as
\begin{widetext}
	\begin{subequations}
		\begin{eqnarray}
		q_1&\simeq&\frac{2k_{2R}A_1^{2-}}{D}\bigg(	e^{i\xi_{2I}+\Lam_1}\cosh(\eta_{2R}+\frac{\Phi_{21}-\Del_{21}}{2})+e^{i\eta_{2I}+\Lam_{2}}\cosh(\xi_{2R}+\frac{\lam_{2}-\lam_{1}}{2})\bigg),\label{5.2}\\
		q_2&\simeq&\frac{2l_{2R}A_2^{2-}}{D}\bigg(e^{i\eta_{2I}+\Lam_7}\cosh(\xi_{2R}+\frac{\Gamma_{21}-\ga_{21}}{2})+e^{i\xi_{2I}+\Lam_{6}}\cosh(\eta_{2R}+\frac{\lam_{7}-\lam_{6}}{2})\bigg),\label{5.3}\\
		D&=&e^{\Lam_5}\cosh(\eta_{2R}-\xi_{2R}+\frac{\lam_3-\lam_4}{2})+e^{\Lam_3}\cosh(i(\eta_{2I}-\xi_{2I})+\frac{\vth_{12}-\varphi_{21}}{2})\nonumber\\
		&&+e^{\Lam_4}\cosh(\eta_{2R}+\eta_{3R}+\frac{\lam_5-R}{2}).\nonumber
		\end{eqnarray}
	\end{subequations}
	Here,  $A_{1}^{2-}=[\alpha_{2}^{(1)}/\alpha_{2}^{(1)^*}]^{1/2}$, $A_{2}^{2-}=[\alpha_{2}^{(2)}/\alpha_{2}^{(2)^*}]^{1/2}$. In the latter the superscript $2-$ denote nondegenerate soliton $S_2$ before collision.

\end{widetext}
\underline{(b) After collision}: $z\rightarrow +\infty$\\
\underline{Soliton 1}: The asymptotic forms for degenerate soliton $S_1$ after collision deduced from the solution (\ref{3.8a})-(\ref{3.8c}) as,
\begin{eqnarray}
q_j\simeq\begin{pmatrix}
A_1^{1+}\\ \\
A_2^{1+}
\end{pmatrix}e^{i(\eta_{1I}+\theta_j^+)}k_{1R}\sech(\eta_{1R}+\frac{R'-\vsa_{22}}{2}),~j=1,2,\label{5.4}
\end{eqnarray}
where  $A_1^{1+}=\al_1^{(1)}/(|\al_1^{(1)}|^2+\chi|\al_1^{(2)}|^2)^{1/2}$, $A_2^{1+}=\al_1^{(1)}/(|\al_1^{(1)}|^2\chi^{-1}+|\al_1^{(2)}|^2)^{1/2}$, $\chi=(|k_1-l_2|^2|k_1+k_2^*|^2)/(|k_1-k_2|^2|k_1+l_2^*|^2)$, $e^{i\theta_1^+}=\frac{(k_1-k_2)(k_1^*+k_2)(k_1-l_2)^{\frac{1}{2}}(k_1^*+l_2)^{\frac{1}{2}}}{(k_1^*-k_2^*)(k_1+k_2^*)(k_1^*-l_2^*)^{\frac{1}{2}}(k_1+l_2^*)^{\frac{1}{2}}}$, $e^{i\theta_2^+}=\frac{(k_1-k_2)^{\frac{1}{2}}(k_1^*+k_2)^{\frac{1}{2}}(k_1-l_2)(k_1^*+l_2)}{(k_1^*-k_2^*)^{\frac{1}{2}}(k_1+k_2^*)^{\frac{1}{2}}(k_1^*-l_2^*)(k_1+l_2^*)}$. Here $1+$ in $A_1^{1+}$ refers to degenerate soliton $S_1$ after collision. \\
\underline{Soliton 2}: Similarly the expression for the nondegenerate soliton, $S_2$, after collision deduced from the two soliton solution (\ref{3.8a})-(\ref{3.8c}) is
\begin{eqnarray}
&&q_{1}\simeq \frac{2k_{2R}A_1^{2+}e^{i\eta_{2I}}\cosh(\xi_{2R}+\frac{\Lam_{22}-\rho_1}{2})}{\big[\frac{(k_{2}^{*}-l_{2}^{*})^{\frac{1}{2}}}{(k_{2}^{*}+l_{2})^{\frac{1}{2}}}\cosh(\eta_{2R}+\xi_{2R}+\frac{\vsa_{22}}{2})+\frac{(k_{2}+l_{2}^{*})^{\frac{1}{2}}}{(k_{2}-l_{2})^{\frac{1}{2}}}\cosh(\eta_{2R}-\xi_{2R}+\frac{R_3-R_6}{2})\big]},\\
&&q_2\simeq \frac{2l_{2R}A_2^{2+}e^{i\xi_{2I}}\cosh(\eta_{2R}+\frac{\mu_{22-\rho_2}}{2})}{\big[\frac{(k_{2}^{*}-l_{2}^{*})^{\frac{1}{2}}}{(k_{2}+l_{2}^*)^{\frac{1}{2}}}\cosh(\eta_{2R}+\xi_{2R}+\frac{\vsa_{22}}{2})+\frac{(k_{2}^*+l_{2})^{\frac{1}{2}}}{(k_{2}-l_{2})^{\frac{1}{2}}}\cosh(\eta_{2R}-\xi_{2R}+\frac{R_3-R_6}{2})\big]}.
\end{eqnarray}
where $\rho_j=\log\al_2^{(j)}$, $j=1,2$, $A_{1}^{2+}=[\alpha_{2}^{(1)}/\alpha_{2}^{(1)^*}]^{1/2}$, $A_{2}^{2+}=i[\alpha_{2}^{(2)}/\alpha_{2}^{(2)^*}]^{1/2}$. The explicit expressions of all the constants are given in Appendix C. 
\subsection{Degenerate soliton collision induced shape changing scenario of nondegenerate soliton} 
The coexistence of nondegenerate and degenerate solitons can be brought out from the partially nondegenerate soliton solution (\ref{3.8a})-(\ref{3.8c}). Such coexisting solitons undergo a novel collision property, which has been illustrated in Fig. \ref{f14}. From this figure, one can observe that the intensity of the degenerate soliton $S_1$ is enhanced after collision in the first mode and it gets suppressed in the second mode. As we expected the degenerate soliton undergoes energy redistribution among the modes $q_1$ and $q_2$.  In the degenerate soliton case, the polarization vectors, $A_j^{l}=\alpha_l^{(j)}/(|\alpha_1^{(1)}|^2+|\alpha_1^{(2)}|^2)^{1/2}$, $l,j=1,2$, play crucial role in changing the shape of the degenerate solitons under collision,  where the intensity/energy redistribution happens between the modes $q_1$ and $q_2$. As we have pointed out in the next section, the shape preserving collision arises in the pure degenerate case when the polarization parameters obey the condition, $\frac{\alpha_1^{(1)}}{\alpha_2^{(1)}}=\frac{\alpha_1^{(2)}}{\alpha_2^{(2)}}$ where $\alpha _i^{(j)}$'s, $i, j = 1,2$, are complex numbers related to the polarization vectors as given above. 
The above collision is similar to the one which occurs in the completely degenerate case \cite{i,j}. 
However, this is not true in the case of nondegenerate solitons. The nondegenerate asymmetric double-hump soliton $S_2$ exhibits a novel collision property as depicted in Fig. \ref{f14}. In both the modes, the nondegenerate soliton $S_2$  experiences strong effect when it interacts with a degenerate soliton. As a result the nondegenerate soliton swtiches its asymmetric double-hump profile into single-hump profile with an enhancement of intensity along with a phase shift. In addition to the latter case, we also noticed that the nondegenerate soliton loses its asymmetric double-hump profile into another form of asymmetric double-hump profile when it interacts with a degenerate soliton.  In the nondegenerate case, the relative separation distances (or phases) are in general not preserved during the collision. Therefore the mechanism behind the occurence of shape preserving and changing collisions in the nondegenerate solitons is quite new. These novel collision properties can be understood from the corresponding asymptotic analysis given in the previous subsection. The asymptotic analysis reveals that energy redistribution occurs between modes $q_1$ and $q_2$. In order to confirm the shape changing nature of this interesting collision process we obtain the following expression for the transition amplitudes, 
\begin{eqnarray}
T_1^1=\frac{(|\al_1^{(1)}|^2+|\al_1^{(2)}|^2)^{1/2}}{(|\al_1^{(1)}|^2+\chi|\al_1^{(2)}|^2)^{1/2}},~
T_2^1=\frac{(|\al_1^{(1)}|^2+|\al_1^{(2)}|^2)^{1/2}}{(|\al_1^{(1)}|^2\chi^{-1}+|\al_1^{(2)}|^2)^{1/2}}.
\end{eqnarray} 
In general, the transition amplitudes are not equal to unity. If the quantity $T_j^l$ is not unimodular (for this case the constant $\chi\neq 1$)  then the degenerate and nondegenerate solitons always exhibit shape changing collision.
The standard elastic collision can be recovered when $\chi=1$. One can calculate the shift in the positions of both degenerate and nondegenerate solitons after collision from the asymptotic analysis.  This new kind of collision property has not been observed in the degenerate vector bright solitons of Manakov system \cite{i,j}. The property of enhancement of intensity in both the components of nondegenerate soliton is similar to the one observed earlier in the mixed coupled nonlinear Schr\"{o}dinger system \cite{mix}.  The amplification process of a single-humped nondegenerate soliton 
in both the modes can be viewed as an application for signal amplification where the degenerate soliton acts as a pumping wave.  
\section{ Degenerate vector bright soliton solutions and their collision dynamics}
\begin{figure}
	\centering
	\includegraphics[width=0.5\linewidth]{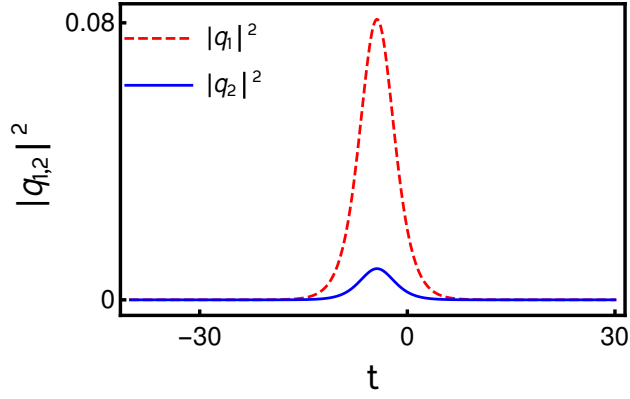}
	\caption{Degenerate one-soliton: The values are $k_1=0.3+0.5i$, $\alpha_{1}^{(1)}=1.5+1.5i$, $\alpha_{1}^{(2)}=0.5+0.5i$.}
	\label{f15}
\end{figure}

The already reported degenerate vector one-bright soliton solution of Manakov system (\ref{e1}) can be deduced from the one-soliton solution (\ref{3.3}) by imposing $k_{1}=l_{1}$ in it. The forms of $q_j$ given in Eq. (\ref{3.3}) degenerates into the standard bright soliton form \cite{d,j}   
\begin{equation}
q_{j}=\frac{\alpha_{1}^{(j)}e^{\eta_1}}{1+e^{\eta_1+\eta_1^*+R}}, ~j=1,2,\label{6.1}\\
\end{equation}
which can be rewritten as
\begin{equation}
q_{j}=k_{1R} \hat{A_{j}}e^{i\eta_{1I}}\sech(\eta_{1R}+\frac{R}{2}),\label{6.2}
\end{equation}
where $\eta_{1R}=k_{1R}(t-2k_{1I}z)$, $\eta_{1I}=k_{1I}t+(k_{1R}^{2}-k_{1I}^{2})z$, $\hat{A_{j}}=\frac{\alpha_{1}^{(j)}}{\sqrt{(|\alpha_{1}^{(1)}|^2+|\alpha_{1}^{(2)}|^2)}}$, $e^{R}=\frac{(|\alpha_{1}^{(1)}|^2+|\alpha_{1}^{(2)}|^2)}{(k_{1}+k_{1}^*)^2}$, $j=1,2$. 
Note that the above fundamental bright soliton always propagates in both the modes $q_{1}$ and $q_{2}$ with the same velocity $2k_{1I}$. The polarization vectors $(\hat{A}_1,\hat{A}_2)^{\dagger}$ have different amplitudes and phases, unlike the case of nondegenerate solitons where they have only different phases. The presence of single wave number $k_1$ in the solution (\ref{6.2}) restricts the degenerate soliton to have a single-hump form only.  A typical profile of the degenerate soliton is shown in Fig. \ref{f15}. As already pointed out in \cite{i,j} the amplitude and central position of the degenerate vector bright soliton are obtained as $2k_{1R}\hat{A}_j$, $j=1,2$ and $\frac{R}{2k_{1R}}$, respectively.
   
Further, the degenerate two-soliton solution can be deduced from the nondegenerate two-soliton solution (\ref{3.7a})-(\ref{3.7c}) by applying the degenerate limits $k_{1}=l_{1}$ and $k_{2}=l_{2}$. This results in the following standard degenerate two-soliton solution \cite{j}, that is 
\begin{widetext}\vspace{-0.21cm}
	\begin{equation}
	q_{j}(t,z)= \frac{\alpha_{1}^{(j)}e^{\eta_{1}}+\alpha_{2}^{(j)}e^{\eta_{2}}+e^{\eta_{1}+\eta_{1}^*+\eta_{2}+\delta_{1j}}+e^{\eta_{1}+\eta_{2}+\eta_{2}^*+\delta_{2j}}}{1+e^{\eta_{1}+\eta_{1}^*+R_{1}}+e^{\eta_{1}+\eta_{2}^*+\delta_{0}}+e^{\eta_{1}^*+\eta_{2}+\delta_{0}^*}+e^{\eta_{2}+\eta_{2}^*+R_{2}}+e^{\eta_{1}+\eta_{1}^*+\eta_{2}+\eta_{2}^*+R_{3}}},\label{6.3}
	\end{equation} \end{widetext}
where $j=1,2$, $\eta_{j}=k_{j}(t+ik_{j}z)$, $e^{\delta_{0}}=\frac{k_{12}}{k_{1}+k_{2}^*}$, $e^{R_{1}}=\frac{k_{11}}{k_{1}+k_{1}^*}$, $e^{R_{2}}=\frac{k_{22}}{k_{2}+k_{2}^*}$,
$e^{\delta_{1j}}=\frac{(k_{1}-k_{2})(\alpha_{1}^{(j)}k_{21}-\alpha_{2}^{(j)}k_{11})}{(k_{1}+k_{1}^*)(k_{1}^*+k_{2})}$, $e^{\delta_{2j}}=\frac{(k_{2}-k_{1})(\alpha_{2}^{(j)}k_{12}-\alpha_{1}^{(j)}k_{22})}{(k_{2}+k_{2}^*)(k_{1}+k_{2}^*)}$,
$e^{R_{3}}=\frac{|k_{1}-k_{2}|^2}{(k_{1}+k_{1}^*)(k_{2}+k_{2}^*)|k_{1}+k_{2}^*|^2}(k_{11}k_{22}-k_{12}k_{21})$ and
$k_{il}=\frac{\mu~\sum_{n=1}^{2}~\alpha_{i}^{(n)}\alpha_{i}^{(n)^*}}{(k_{i}+k_{l}^*)}$, $i,l=1,2$, $\mu=+1$.  The $N$ degenerate vector bright soliton solution can be recovered from the nondegenerate $N$-soliton solutions by fixing the wave numbers as $k_{i}=l_{i}, i=1,2,...,N$. In passing we  also note that the nondegenerate fundamental soliton solution (7) can arise when we fix the parameters $\alpha_{2}^{(1)}=\alpha_{1}^{(2)}=0$ in Eq. (34)  and rename the constants $k_2$ as $l_1$ and $\alpha_{2}^{(2)}$ as $\alpha_{1}^{(2)}$ in the resultant solution. We also note that the above degenerate two-soliton solution (\ref{6.3}) can also be rewritten using Gram determinants from the Gram determinant forms of nondegenerate two-soliton solution (\ref{3.7a})-(\ref{3.7c}). Such Gram determinant forms of degenerate two-soliton solution are new to the literature.  

As reported in \cite{i,j}, the degenerate fundamental solitons ($k_i=l_i$, $i=1,2$) in the Manakov system undergo  shape changing collision due to intensity redistribution among the modes. The energy redistribution occurs in the degenerate case because of the polarization vectors of the two modes combine with each other. This shape changing collision illustrated in Fig. \ref{f3} where the intensity redistribution occurs because of the enhancement of soliton $S_1$ in the first mode and the corresponding intensity of the same soliton is suppressed in the second mode.   
To hold the conservation of energy between the modes the intensity of the solitons $S_{2}$ gets suppressed in the first mode and it is enhanced in the second mode. The standard elastic collision has already been brought out in the degenerate case for the very special case $\frac{\alpha_1^{(1)}}{\alpha_2^{(1)}}=\frac{\alpha_1^{(2)}}{\alpha_2^{(2)}}$ \cite{i,m}. 
\begin{figure}
	\centering
	\includegraphics[width=0.45\linewidth]{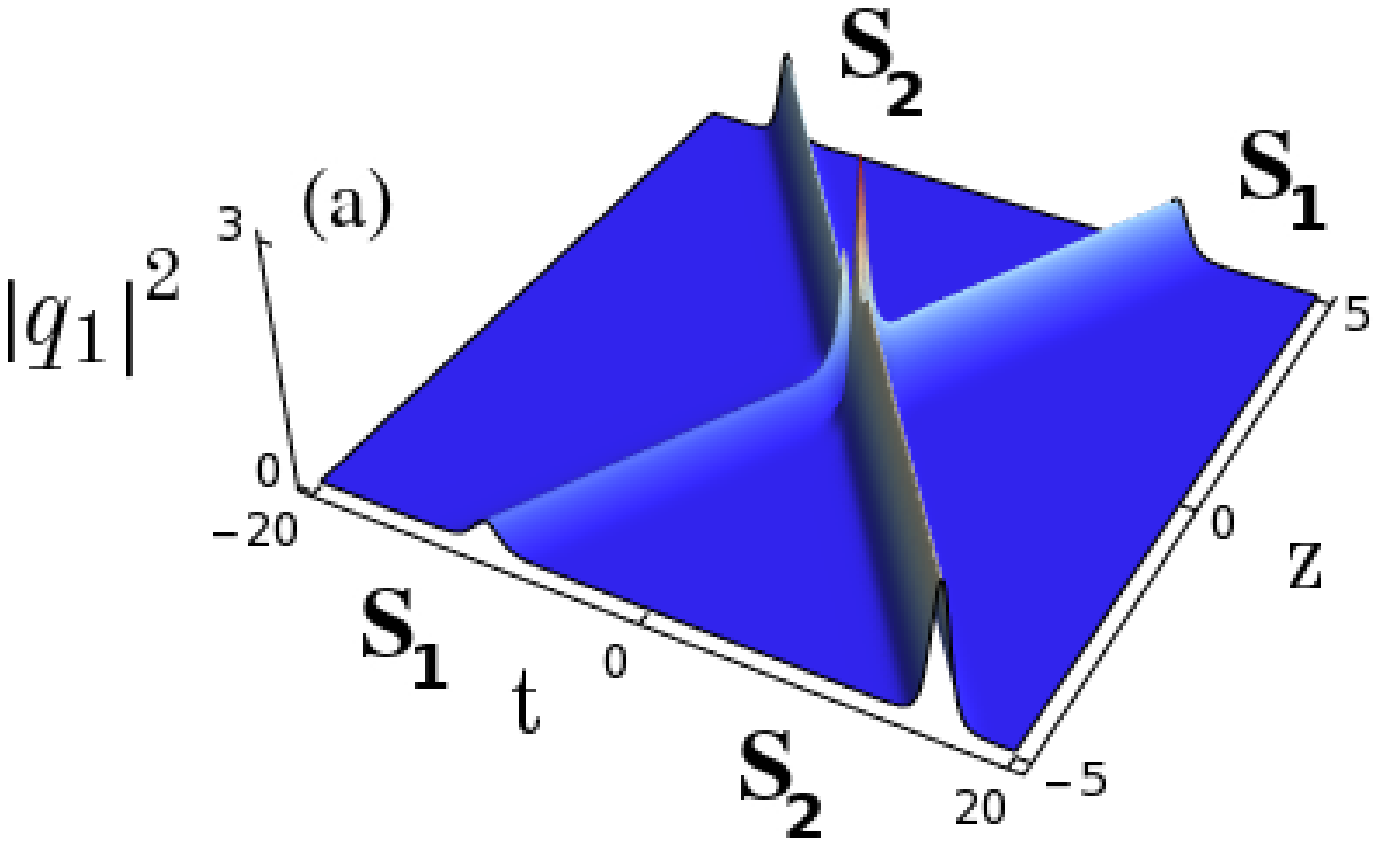}~~
	\includegraphics[width=0.45\linewidth]{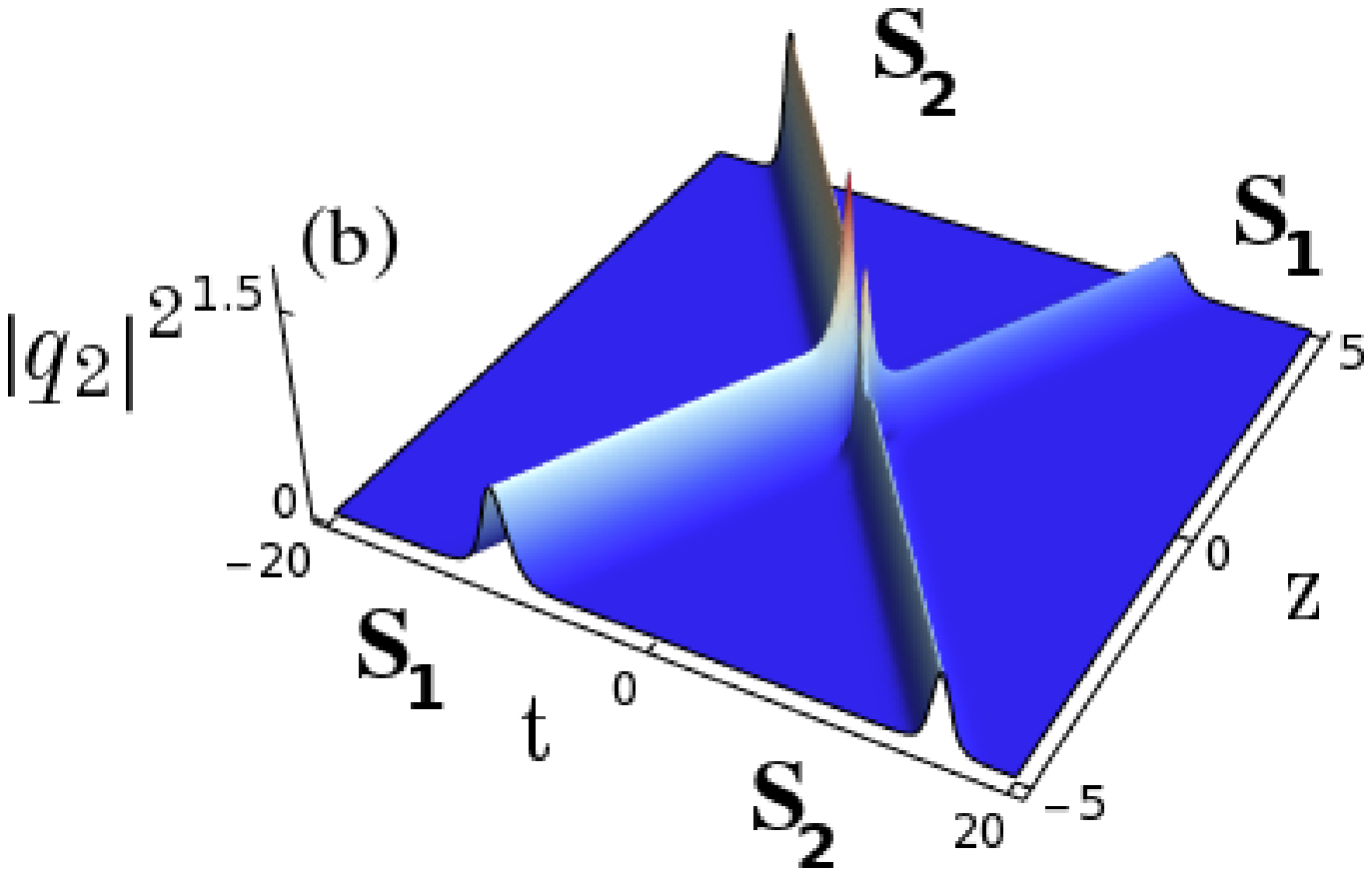}
	\caption{Shape changing collision of degenerate two-solitons: $k_{1}=l_1=1+i$, $k_{2}=l_2=1.51-1.51i$,  $\alpha_{1}^{(1)}=0.5+0.5i$, $\alpha_{2}^{(1)}=\alpha_{1}^{(2)}=\alpha_{2}^{(2)}=1$.}
	\label{f16}
\end{figure}



\section{Possible Experimental Observations of Nondegenerate Solitons}
To experimentally observe the nondegenerate vector solitons (single hump/double hump solitons)  one may adopt the mutual-incoherence method which has been used to observe the multi-hump multi-mode solitons experimentally (please see Ref. [36]). The Manakov solitons (degenerate solitons) can also  be  observed by the same experimental procedure with appropriate modifications (please see Ref. [24]). In the following, we briefly envisage how the procedure given in Ref. [36] can be modified to generate the single hump/double hump soliton (nondegenerate soliton) discussed in our work.  
	
	To generate the nondegenerate vector solitons it is essential to consider two laser sources of different characters, so that the wavelength of the first laser beam is different from the second one. Using polarizing beam splitters, each one of the laser beams can be split into ordinary and extraordinary beams. The extraordinary beam coming out from the first source can be further split into two individual fields $F_{11}$ and $F_{12}$ by allowing it to fall on a beam splitter. These two fields are nothing but the reflected and transmitted extraordinary beams coming out from the beam splitter. The intensities of these two fields are different. Similarly the second beam which is coming out from the second source  can also be split into two fields $F_{21}$ and $F_{22}$ by passing through another beam splitter. The intensities of these two fields are also different. As a result one can generate four fields that are incoherent to each other. To set the incoherence in phase among these four fields one should allow them to travel sufficient distance before coupling is performed. The fields  $F_{11}$ and $F_{12}$ now become nondegenerate two individual solitons in the first mode whereas $F_{21}$ and $F_{22}$ form another set of two nondegenerate solitons in the second mode. The coupling between the fields $F_{11}$ and $F_{21}$ can be performed by combining them using another beam splitter. Similarly, by suitably locating another beam splitter, one can combine the fields $F_{12}$ and $F_{22}$, respectively.  After appropriate coupling is performed the resultant optical field beams can now be focused through two individual cylindrical lenses and the output may be recorded in an imaging system, which consists of a crystal and CCD camera. The collision between the nondegenerate two-solitons in both the modes can now be seen from the recorded images. 
	
	To observe the elastic collision between nondegenerate solitons (single hump/double hump solitons), one must make arrangements to vanish the mutual coherence property between the solitons $F_{11}$ and $F_{12}$ in the first mode $q_1$ and $F_{21}$ and $F_{22}$ in the second mode $q_2$ (please see Ref. [24]). The four optical beams are now completely independent and incoherent with one another. The collision angle at which the nondegenerate solitons interact should be sufficiently large enough. Under this situation, no energy exchange is expected to occur between the nondegenerate solitons of the two modes.

\section{ Conclusion}
From the present study, we point out a few applications of our above reported soliton solutions. The shape preserving collision property of the nondegenerate solitons can be used for  optical communication applications. The nondegenerate solitons of Manakov system can be seen as a soliton molecule when $k_{1I}\approx k_{2I}$ and $l_{1I}\approx l_{2I}$. Therefore as explained in the context of soliton molecule, the double hump (or multi-hump) structure of the nondegenerate solitons can be useful for sending information of densely packed data \cite{new3}. Degenerate soliton collision induced enhancement of intensity property of nondegenerate soliton is considered as signal amplification application. Recently the various properties associated with soliton molecule have been explored in the literature \cite{new3,new3a,new4,new4a,new4b}. Also breather wave molecule has been identified in \cite{bre}. The interesting collision property of degenerate soliton has already been shown that it is useful for optical computing \cite{l1,m}. Our results provide a new possibility to investigate nondegenerate type solitons in both integrable and non-integrable systems. The present study can also be extended to fiber arrays and multi-mode fibers where Manakov type equations describe the pulse propagation. Recently we have investigated the novel dynamics of nondegenerate solitons in $N$-coupled system and the results will be published elsewhere.   

We have derived a general form of nondegenerate one-, two-  and three-soliton solutions for the Manakov model through Hirota bilinear method. Such new class of solitons admit various interesting profile structures. The double-hump formation  is elucidated by analysing the relative velocities of the modes of the solitons. Then we have pointed out the coexistence of degenerate and nondegenerate solitons in the Manakov system by imposing a wave number restriction on the obtained two-soliton solution. We have found that nondegenerate solitons undergo shape preserving, shape altering and shape changing collision scenarios for both equal velocities and unequal velocities cases. However, for partially equal velocity case, we have demonstrated shape changing collision. By performing appropriate asymptotic analysis, the novel shape changing collision has been explained while the degenerate soliton interacts with the nondegenerate soliton. Finally we recovered the well known energy exchanging collision exhibiting  degenerate soliton solutions from the newly identified nondegenerate one and two-soliton solutions. We have also verified the stability nature of double hump solitons even during collision using Crank-Nicolson method as explained in Appendix D.  It is also very interesting to investigate many possibilities of collision dynamics using three-soliton solution as deduced in Appendix A. Now we are investigating what will happen when (i) two degenerate solitons interact with a nondegenerate soliton and (ii) two nondegenerate solitons collide with a degenerate soliton and so on. The results will be published elsewhere.       

\section*{Acknowledgement}
The authors are  thankful to Prof. P. Muruganandam, Department of Physics, Bharathidasan University, Tiruchirapalli - 620 024 for his personal help in verifying the shape preserving collision nature of symmetric double-hump solitons numerically with white noise and Gaussian noise as perturbations. The work of R.R., S.S. and M.L. are supported by DST-SERB Distinguished Fellowship program (SB/DF/04/2017) of M.L.
\appendix
\section{Three-soliton solution}
The explicit form of nondegenerate three-soliton solution of Eq. (\ref{e1}) can be deduced by proceeding with the Eqs. (\ref{2.4}) using the series representation upto orders $\epsilon^{11}$ for $g^{(N)}$ and $\epsilon^{12}$ for $f$. Then the solution can be expressed using Gram determinant in the following way:
\begin{subequations}
\begin{eqnarray}
g^{(N)}=\begin{vmatrix}
A & I & \phi \\ 
-I &B & {\bf 0}^T \\ 
{\bf 0}  & C_N & 0
\end{vmatrix},~~f=\begin{vmatrix}
A & I \\ 
-I &B \\ 
\end{vmatrix},~~N=1,2.\label{A.1a}
\end{eqnarray}
Here the matrices $A $ and $B$ are of the order $(6\times 6)$ defined as
\begin{eqnarray}
A=\begin{pmatrix}
A_{mm'} & A_{mn} \\ 
A_{nm}& A_{nn'}\\ 
\end{pmatrix},~~B=\begin{pmatrix}
\kappa_{mm'} & \kappa_{mn} \\ 
\kappa_{nm} &\kappa_{nn'} \\ 
\end{pmatrix}, m, m' ,n ,n'=1,2,3.
\end{eqnarray}
The various elements of matrix $A$ are obtained from the following, 
\begin{eqnarray}
A_{mm'}=\frac{e^{\eta_m+\eta_{m'}^*}}{(k_m+k_{m'}^*)},~ A_{mn}=\frac{e^{\eta_m+\xi_{n}^*}}{(k_m+l_{n}^*)},
\end{eqnarray}
\begin{eqnarray}
A_{nn'}=\frac{e^{\xi_n+\xi_{n'}^*}}{(l_n+l_{n'}^*)}, ~A_{nm}=\frac{e^{\eta_n^*+\xi_{m}}}{(k_n^*+l_{m})},~m,m',n,n'=1,2,3.
\end{eqnarray}
The elements of matrix $B$ is defined as
\begin{eqnarray}
\kappa_{mm'}=\frac{\psi_m^{\dagger}\sigma\psi_{m'}}{(k_m^*+k_{m'})},~\kappa_{mn}=\frac{\psi_m^{\dagger}\sigma\psi'_{n}}{(k_m^*+l_{n})},~\kappa_{nm}=\frac{\psi_n^{'\dagger}\sigma\psi_{m}}{(l_n^*+k_{m})},~\kappa_{nn'}=\frac{\psi_n^{'\dagger}\sigma\psi'_{n'}}{(l_n^*+l_{n'})}.\label{A.1e}
\end{eqnarray} \end{subequations}
In (\ref{A.1e}) the column matrices are 
$\psi_j=\begin{pmatrix}
\alpha_j^{(1)}\\
0
\end{pmatrix}$, ~$\psi'_j=\begin{pmatrix}
0\\
\alpha_j^{(2)}
\end{pmatrix}$, $j=m,m',n,n'=1,2,3$, $\eta_j=k_jt+ik_j^2z$ and $\xi_j=l_jt+il_j^2z$, $j=1,2,3$.
The other matrices in Eq. (\ref{A.1a}) are defined below: \\
$\phi=\begin{pmatrix}
e^{\eta_1} & e^{\eta_2}  & e^{\eta_3}  & e^{\xi_1}  & e^{\xi_2} &e^{\xi_3}
\end{pmatrix}^T$, $C_1=-\begin{pmatrix}
\alpha_1^{(1)} & \alpha_2^{(1)} & \alpha_3^{(1)} & 0 & 0 &0
\end{pmatrix}$, $C_2=-\begin{pmatrix}
0 & 0 & 0 &\alpha_1^{(2)} & \alpha_2^{(2)} & \alpha_3^{(2)} \end{pmatrix}$, ${\bf 0} =\begin{pmatrix}
0 & 0 & 0 &0 & 0& 0 \end{pmatrix}$ and $\sigma=I$ is a $(6\times 6)$ identity matrix.
\section{Asymptotic analysis of shape changing collision of nondegenerate solitons in the unequal velocity case: $k_{1I}\neq l_{1I}$, $k_{2I}\neq l_{2I}$}
To carry out the asymptotic analysis for the shape changing collision we fix the parameters as $k_{1I}<k_{2I}$, $l_{1I}>l_{2I}$, $k_{jR}, l_{jR}>0$, $j=1,2$ and $k_{1I}\neq l_{1I}$, $k_{2I}\neq l_{2I}$. For this choice the nondegenerate two-soliton solution (\ref{3.7a})-(\ref{3.7c}) reduces to the following asymptotic forms:  \\
\underline{(a) Before collision}: $z\rightarrow -\infty$\\
\underline{Soliton 1}: $(\eta_{1R}, \xi_{1R}\simeq 0,~\eta_{2R} \rightarrow +\infty, \xi_{2R}\rightarrow -\infty)$
\begin{subequations}
	\begin{eqnarray}
	&&q_{1}\simeq \frac{2A_1^{1-}k_{1R}e^{i(\eta_{1I}+\theta_1^{1-})}\cosh(\xi_{1R}+\psi_1^-)}{\big[{\frac{(k_{1}^{*}-l_{1}^{*})^{\frac{1}{2}}}{(k_{1}^{*}+l_{1})^{\frac{1}{2}}}}\cosh(\eta_{1R}+\xi_{1R}+\psi_3^-)+\frac{(k_{1}+l_{1}^{*})^{\frac{1}{2}}}{(k_{1}-l_{1})^{\frac{1}{2}}}\cosh(\eta_{1R}-\xi_{1R}+\psi_4^-)\big]},\label{B.1a}\\
	&&q_{2}\simeq\frac{2A_2^{1-}l_{1R}e^{i(\xi_{1I}+\theta_2^{1-})}\cosh(\eta_{1R}+\psi_2^-)}{\big[\frac{(k_{1}^{*}-l_{1}^{*})^{\frac{1}{2}}}{(k_{1}+l_{1}^{*})^{\frac{1}{2}}}\cosh(\eta_{1R}+\xi_{1R}+\psi_3^-)+\frac{(k_{1}^{*}+l_{1})^{1/2}}{(k_{1}-l_{1})^{1/2}}\cosh(\eta_{1R}-\xi_{1R}+\psi_4^-)\big]}.\label{B.1b}
	\end{eqnarray}\end{subequations}\\
Here,  $\psi_1^-=\frac{1}{2}\log\frac{(k_1-l_1)|k_2-l_1|^2|\alpha_1^{(2)}|^2}{(k_1+l_1^*)|k_2+l_1^*|^2(l_1+l_1^*)^2}$,  $\psi_2^-=\frac{1}{2}\log\frac{(l_1-k_1)|k_1-k_2|^4|\alpha_1^{(1)}|^2}{(k_1^*+l_1)|k_1+k_2^*|^4(k_1+k_1^*)^2}$, $e^{i\theta_1^{1-}}=\frac{(k_{1}-k_{2})(k_{1}^*+k_{2})}{(k_{1}^{*}-k_{2}^{*})(k_{1}+k_{2}^*)}$, $\psi_4^-=\frac{1}{2}\log\frac{|k_1-k_2|^4|k_2+l_1^*|^2|\alpha_1^{(1)}|^2(l_1+l_1^*)^2}{|\alpha_1^{(2)}|^2|k_1+k_2^*|^4|k_2-l_1|^2(k_1+k_1^*)^2}$, $\psi_3^-=\frac{1}{2}\log\frac{|k_1-k_2|^4|k_1-l_1|^2|k_2-l_1|^2|\alpha_1^{(2)}|^2|\alpha_1^{(1)}|^2}{|k_1+k_2^*|^4|k_1+l_1^*|^2|k_2+l_1^*|^2(k_1+k_1^*)^2(l_1+l_1^*)^2}$, 
 $e^{i\theta_2^{1-}}=\frac{(k_{2}-l_{1})^{\frac{1}{2}}(k_{2}^{*}+l_{1})^{\frac{1}{2}}}{(k_{2}^*-l_{1}^*)^{\frac{1}{2}}(k_{2}+l_{1}^*)^{\frac{1}{2}}}$, $A_{1}^{1-}=[\alpha_{1}^{(1)}/\alpha_{1}^{(1)^*}]^{1/2}$ and $A_{2}^{1-}=i[\alpha_{1}^{(2)}/\alpha_{1}^{(2)^*}]^{1/2}$. \\
\underline{Soliton 2}: $(\eta_{2R}, \xi_{2R}\simeq 0,~\eta_{1R} \rightarrow -\infty, \xi_{1R}\rightarrow +\infty)$
\begin{subequations}
	\begin{eqnarray}
	&&q_{1}\simeq \frac{2k_{2R}A_1^{2-}e^{i(\eta_{2I}+\theta_1^{2-})}\cosh(\xi_{2R}+\chi_1^-)}{\big[\frac{(k_{2}^{*}-l_{2}^{*})^{\frac{1}{2}}}{(k_{2}^{*}+l_{2})^{\frac{1}{2}}}\cosh(\eta_{2R}+\xi_{2R}+\chi_3^-)+\frac{(k_{2}+l_{2}^{*})^{\frac{1}{2}}}{(k_{2}-l_{2})^{\frac{1}{2}}}\cosh(\eta_{2R}-\xi_{2R}+\chi_4^-)\big]},\label{B.2a}\\
	&&q_2\simeq \frac{2l_{2R}A_2^{2-}e^{i(\xi_{2I}+\theta_2^{2-})}\cosh(\eta_{2R}+\chi_2^-)}{\big[\frac{(k_{2}^{*}-l_{2}^{*})^{\frac{1}{2}}}{(k_{2}+l_{2}^*)^{\frac{1}{2}}}\cosh(\eta_{2R}+\xi_{2R}+\chi_3^-)+\frac{(k_{2}^*+l_{2})^{\frac{1}{2}}}{(k_{2}-l_{2})^{\frac{1}{2}}}\cosh(\eta_{2R}-\xi_{2R}+\chi_4^-)\big]}.\label{B.2b}
	\end{eqnarray} \end{subequations}
In the above,
\begin{eqnarray}
&&\chi_1^-=\frac{1}{2}\log\frac{|l_1-l_2|^4(k_2-l_2)|\alpha_2^{(2)}|^2}{|l_1+l_2^*|^4(k_2+l_2^*)(l_2+l_2^*)^2},~ \chi_2^-=\frac{1}{2}\log\frac{|k_2-l_1|^2(l_2-k_2)(l_2+l_1^*)^2|\alpha_2^{(1)}|^2}{|k_2+l_1^*|^2(k_2^*+l_2)(k_2+k_1^*)^2(k_2+k_2^*)^2},\nonumber\\
&&e^{i\theta_1^{2-}}=\frac{(k_{2}-l_{1})^{\frac{1}{2}}(k_{2}^*+l_{1})^{\frac{1}{2}}}{(k_{2}^*-l_{1}^*)^{\frac{1}{2}}(k_{2}+l_{1}^*)^{\frac{1}{2}}},~e^{i\theta_2^{2-}}=\frac{(l_{1}-l_{2})(l_{1}+l_{2}^{*})}{(l_{1}^{*}-l_{2}^{*})(l_{1}^{*}+l_{2})},~A_{1}^{2-}=[\alpha_{2}^{(1)}/\alpha_{2}^{(1)^*}]^{1/2},\nonumber \\
&&\chi_3^-=\frac{1}{2}\log\frac{|l_1-l_2|^4|k_2-l_1|^2|k_2-l_2|^2|\alpha_2^{(1)}|^2|\alpha_2^{(2)}|^2}{|l_1+l_2^*|^4|k_2+l_1^*|^2|k_2+l_2^*|^2(k_2+k_2^*)^2(l_2+l_2^*)^2},  ~A_{2}^{2-}=[\alpha_{2}^{(2)}/\alpha_{2}^{(2)^*}]^{1/2},\nonumber\\
&&\chi_4^-=\frac{1}{2}\log\frac{|k_2-l_1|^2|l_1+l_2^*|^4|\alpha_2^{(1)}|^2(l_2+l_2^*)^2}{|\alpha_2^{(2)}|^2|k_2+l_1^*|^2|l_1-l_2|^4(k_2+k_2^*)^2}.\nonumber
\end{eqnarray}
\underline{(b) After collision}: $z\rightarrow +\infty$\\
\underline{Soliton 1}: $(\eta_{1R}, \xi_{1R}\simeq 0,~\eta_{2R} \rightarrow -\infty, \xi_{2R}\rightarrow +\infty)$
\begin{subequations}
	\begin{eqnarray}
	&&q_{1}\simeq \frac{2k_{1R}A_1^{1+}e^{i(\eta_{1I}+\theta_1^{1+})}\cosh(\xi_{1R}+\psi_1^+)}{\big[\frac{(k_{1}^{*}-l_{1}^{*})^{\frac{1}{2}}}{(k_{1}^{*}+l_{1})^{\frac{1}{2}}}\cosh(\eta_{1R}+\xi_{1R}+\psi_3^+)+\frac{(k_{1}+l_{1}^{*})^{\frac{1}{2}}}{(k_{1}-l_{1})^{\frac{1}{2}}}\cosh(\eta_{1R}-\xi_{1R}+\psi_4^+)\big]},\label{B.3a}\\
	&&q_2\simeq \frac{2l_{1R}A_1^{2+}e^{i(\xi_{1I}+\theta_2^{1+})}\cosh(\eta_{1R}+\psi_2^+)}{\big[\frac{(k_{1}^{*}-l_{1}^{*})^{\frac{1}{2}}}{(k_{1}+l_{1}^*)^{\frac{1}{2}}}\cosh(\eta_{1R}+\xi_{1R}+\psi_3^+)+\frac{(k_{1}^*+l_{1})^{\frac{1}{2}}}{(k_{1}-l_{1})^{\frac{1}{2}}}\cosh(\eta_{1R}-\xi_{1R}+\psi_4^+)\big]}.\label{B.3b}
	\end{eqnarray} \end{subequations}
Here,
\begin{eqnarray}
&&\psi_1^+=\frac{1}{2}\log\frac{|l_1-l_2|^4(k_1-l_1)|\alpha_1^{(2)}|^2}{|l_1+l_2^*|^4(k_1+l_1^*)(l_1+l_1^*)^2},~\psi_2^+=\frac{1}{2}\log\frac{|k_1-l_2|^2(l_1-k_1)|\alpha_1^{(1)}|^2}{|k_1+l_2^*|^2(k_1^*+l_1)(k_1+k_1^*)^2},\nonumber\\
&&e^{i\theta_1^{1+}}=\frac{(k_{1}-l_{2})^{\frac{1}{2}}(k_{1}^{*}+l_{2})^{\frac{1}{2}}}{(k_{1}^{*}-l_{2}^{*})^{\frac{1}{2}}(k_{1}+l_{2}^{*})^{\frac{1}{2}}},~e^{i\theta_2^{1+}}=\frac{(l_{1}-l_{2})(l_{1}^*+l_{2})}{(l_{1}^{*}-l_{2}^{*})(l_{1}+l_{2}^*)},~A_{1}^{1+}=[\alpha_{1}^{(1)}/\alpha_{1}^{(1)^*}]^{1/2}
\nonumber\\
&&\psi_3^+=\frac{1}{2}\log\frac{|k_1-l_1|^2|k_1-l_2|^2|l_1-l_2|^4|\alpha_1^{(1)}|^2|\alpha_1^{(2)}|^2}{|k_1+l_1^*|^2|k_1+l_2^*|^2|l_1+l_2^*|^4(k_1+k_1^*)^2(l_1+l_1^*)^2},~A_{2}^{1+}=[\alpha_{1}^{(2)}/\alpha_{1}^{(2)^*}]^{1/2}\nonumber\\
&&~\psi_4^+=\frac{1}{2}\log\frac{|k_1-l_2|^2|l_1+l_2^*|^4|\alpha_1^{(1)}|^2(l_1+l_1^*)^2}{|\alpha_1^{(2)}|^2|k_1+l_2^*|^2|l_1-l_2|^4(k_1+k_1^*)^2}. \nonumber
\end{eqnarray} 
\underline{Soliton 2}: $(\eta_{2R}, \xi_{2R}\simeq 0,~\eta_{1R} \rightarrow +\infty, \xi_{1R}\rightarrow -\infty)$
\begin{subequations}
	\begin{eqnarray}
	&&q_{1}\simeq \frac{2A_2^{1+}k_{2R}e^{i(\eta_{2I}+\theta_1^{2+})}\cosh(\xi_{2R}+\chi_1^+)}{\big[{\frac{(k_{2}^{*}-l_{2}^{*})^{\frac{1}{2}}}{(k_{2}^{*}+l_{2})^{\frac{1}{2}}}}\cosh(\eta_{2R}+\xi_{2R}+\chi_3^+)+\frac{(k_{2}+l_{2}^{*})^{\frac{1}{2}}}{(k_{2}-l_{2})^{\frac{1}{2}}}\cosh(\eta_{2R}-\xi_{2R}+\chi_4^+)\big]},\label{B.4a}\\
	&&q_{2}\simeq \frac{2A_2^{2+}l_{2R}e^{i(\xi_{2I}+\theta_2^{2+})}\cosh(\eta_{2R}+\chi_2^+)}{\big[\frac{i(k_{2}^{*}-l_{2}^{*})^{\frac{1}{2}}}{(k_{2}+l_{2}^{*})^{\frac{1}{2}}}\cosh(\eta_{2R}+\xi_{2R}+\chi_3^+)+\frac{(k_{2}^{*}+l_{2})^{\frac{1}{2}}}{(l_{2}-k_{2})^{\frac{1}{2}}}\cosh(\eta_{2R}-\xi_{2R}+\chi_4^+)\big]},\label{B.4b}
	\end{eqnarray} \end{subequations} 
where $\chi_1^+=\frac{1}{2}\log\frac{(k_2-l_2)|k_1-l_2|^2|\alpha_2^{(2)}|^2}{(k_2+l_2^*)|k_1+l_2^*|^2(l_2+l_2^*)^2}$, $\chi_2^+=\frac{1}{2}\log\frac{\al_1^{(2)}|k_1-k_2|^4(k_1-l_1)(k_2-l_1)(k_1^*+l_2)|\alpha_2^{(1)}|^2}{\al_2^{(2)}|k_1+k_2^*|^4(k_1^*+l_1)(k_2^*+l_1)(l_2-k_1)(k_2+k_2^*)^2}$, $e^{i\theta_1^{2+}}=\frac{(k_{1}-k_{2})(k_{1}+k_{2}^*)}{(k_{1}^{*}-k_{2}^{*})(k_{1}^*+k_{2})}$, $e^{i\theta_2^{2+}}=\frac{(k_{1}-l_{2})^{\frac{1}{2}}(k_{1}+l_{2}^*)^{\frac{1}{2}}}{(k_{1}^*-l_{2}^*)^{\frac{1}{2}}(k_{1}^*+l_{2})^{\frac{1}{2}}}$, $\chi_3^+=\frac{1}{2}\log\frac{|k_1-k_2|^4|k_1-l_2|^2|k_2-l_2|^2|\alpha_2^{(1)}|^2|\alpha_2^{(2)}|^2}{|k_1+k_2^*|^4|k_1+l_2^*|^2|k_2+l_2^*|^2(k_2+k_2^*)^2(l_2+l_2^*)^2}$, $A_{1}^{2+}=[\alpha_{2}^{(1)}/\alpha_{2}^{(1)^*}]^{1/2}$, $\chi_4^+=\frac{1}{2}\log\frac{|k_1-k_2|^4|k_1+l_2^*|^2|\alpha_2^{(1)}|^2(l_2+l_2^*)^2}{|\alpha_2^{(2)}|^2|k_1+k_2^*|^4|k_1-l_2|^2(k_2+k_2^*)^2}$ and $A_{2}^{2+}=i[\alpha_{2}^{(2)}/\alpha_{2}^{(2)^*}]^{1/2}$. 

From the above analysis, we find that the structures of individual solitons are invariant before and after collisions except for the terms corresponding to the various phases $\psi_j^-$, $\chi_j^-$, $\psi_j^+$, $\chi_j^+$, $j=1,2,3,4$. For instance, from Eqs. (\ref{B.1a}) and (\ref{B.3a}), the phase terms $\psi_j^-$, $j=1,2,3,4$ corresponding to the first soliton in the $q_1$ mode change into $\psi_j^+$, $j=1,2,3,4$, respectively. Similar phase changes take place in the second component of the first soliton and in the structure of the second soliton as well. Consequently the phase changes leads to the occurrence of shape changing collision in the unequal velocity case. Therefore in general, the shape preserving collision does not occur in the unequal velocity case. However, it  can arise when the phase terms obey the following conditions, \begin{equation}
\psi_j^-=\psi_j^+,~\chi_j^-=\chi_j^+,~j=1,2,3,4.\label{A6}
	\end{equation}
	  
	  Using the complicated shape changing collision property of nondegenerate solitons we could not identify a linear fractional transformation (as in the case of the degenerate case) in order to construct optical logic gates.   
\section{Constants which appear in the asymptotic expressions in Section V }
The various constants which arise in the asymptotic analysis of collision between degenerate and nondegenerate solitons in Sec. V are given below.
	\begin{eqnarray}
&&e^{\Lam_1}=\frac{i\alpha_{1}^{(1)}(k_1-k_2)^{\frac{1}{2}}(k_1-l_2)^{\frac{1}{2}}(k_1^*+k_2)^{\frac{1}{2}}(k_1+k_1^*)(k_2+l_2^*)^{\frac{1}{2}}|k_1+l_2^*|^2}{\alpha_{2}^{(1)}(k_1^*-l_2^*)^{\frac{1}{2}}(k_2^*-l_2^*)^{\frac{1}{2}}}e^{R_5^*+\frac{R_3-R_6}{2}},\nonumber\\
&&e^{\Lam_2}=\frac{(k_1-k_2)^{\frac{1}{2}}(k_2^*+l_2)^{\frac{1}{2}}(k_1+k_2^*)\hat{\Lam}_1\hat{\Lam}_2}{(k_1^*-k_2^*)^{\frac{1}{2}}(k_2^*-l_2^*)^{\frac{1}{2}}(k_1^*+k_2)},~ e^{\Lam_3}=\frac{|\alpha_{1}^{(1)}||\alpha_{1}^{(2)}|(k_1+k_1^*)(k_2+k_2^*)(l_2+l_2^*)}{|k_2-l_2|},\nonumber\\
&&e^{\Lam_4}=(|\alpha_{1}^{(1)}|^2+|\alpha_{1}^{(2)}|^2)^{1/2}(|\alpha_{1}^{(1)}|^2|k_1-k_2|^2|k_1+l_2^*|^2+|\alpha_{1}^{(2)}|^2|k_1-l_2|^2|k_1+k_2^*|^2)^{1/2},\nonumber\\
&&e^{\Lam_5}=\frac{|k_2+l_2^*|}{|k_2-l_2|}(|\alpha_{1}^{(1)}|^2|k_1+l_2^*|^2+|\alpha_{1}^{(2)}|^2|k_1-l_2|^2)^{1/2}(|\alpha_{1}^{(1)}|^2|k_1-k_2|^2+|\alpha_{1}^{(2)}|^2|k_1+k_2^*|^2)^{1/2},\nonumber\\
&&e^{\Lam_6}=\frac{(k_1-l_2)^{\frac{1}{2}}(k_2+l_2^*)^{\frac{1}{2}}(k_1+l_2^*)\hat{\Lam}_3\hat{\Lam}_4}{(k_1^*-l_2^*)^{\frac{1}{2}}(k_2^*-l_2^*)^{\frac{1}{2}}(k_1^*+l_2)},~\hat{\Lam}_1=(|\alpha_{1}^{(1)}|^2(k_1-k_2)-|\alpha_{1}^{(2)}|^2(k_1^*+k_2))^{1/2},\nonumber\\
&&e^{\Lam_7}=\frac{\alpha_{1}^{(2)}(k_1-k_2)^{\frac{1}{2}}(k_1-l_2)^{\frac{1}{2}}(k_1^*+l_2)^{\frac{1}{2}}(k_1+k_1^*)(k_2^*+l_2)^{\frac{1}{2}}|k_1+k_2^*|^2}{\alpha_{2}^{(2)}(k_1^*-k_2^*)^{\frac{1}{2}}(k_2^*-l_2^*)^{\frac{1}{2}}}e^{R_2^*+\frac{R_6-R_3}{2}},\nonumber\\
&&\hat{\Lam}_2=(|\alpha_{1}^{(1)}|^2(k_1-k_2)|k_1+l_2^*|^2-|\alpha_{1}^{(2)}|^2|k_1-l_2|^2(k_1^*+k_2))^{1/2},\nonumber\end{eqnarray}\begin{eqnarray}
&&\hat{\Lam}_4=(|\alpha_{1}^{(1)}|^2|k_1-k_2|^2(k_1^*+l_2)-|\alpha_{1}^{(2)}|^2(k_1-l_2)|k_1+k_2^*|^2)^{1/2},\nonumber\\
&&\hat{\Lam}_3=(|\alpha_{1}^{(2)}|^2(k_1-l_2)-|\alpha_{1}^{(1)}|^2(k_1^*+l_2))^{1/2}, \nonumber\\
&&e^{\frac{\Phi_{21}-\Del_{21}}{2}}=\frac{|\alpha_{2}^{(1)}|(k_1-k_2)(k_2^*-k_1^*)^{\frac{1}{2}}(k_2-l_2)^{\frac{1}{2}}}{(k_1+k_2^*)(k_2+k_2^*)(k_2+k_1^*)^{\frac{1}{2}}(k_2^*+l_2)^{\frac{1}{2}}},~e^{\frac{\lam_2-\lam_1}{2}}=\frac{|\alpha_{2}^{(2)}||k_1-l_2|(k_2-l_2)^{\frac{1}{2}}\hat{\Lam}_2}{(k_2+l_2^*)^{\frac{1}{2}}|k_1+l_2^*|^2(l_2+l_2^*)\hat{\Lam}_1}, \nonumber\\
&&e^{\frac{\lam_5-R}{2}}=\frac{|k_1-k_2||k_1-l_2||k_2-l_2|\hat{\Lam}_5}{|k_1+k_2^*|^2|k_1+l_2^*|^2|k_2+l_2^*|(|\alpha_{1}^{(1)}|^2+|\alpha_{1}^{(2)}|^2)^{1/2}}e^{\frac{R_3+R_6}{2}},\nonumber\\
&&e^{\frac{\vth_{12}-\varphi_{21}}{2}}=\frac{(k_2-k_1)^{\frac{1}{2}}(k_1^*-l_2^*)^{\frac{1}{2}}(k_2^*+l_2)^{\frac{1}{2}}}{(k_2+l_2^*)^{\frac{1}{2}}(k_2^*-k_1^*)^{\frac{1}{2}}(k_1-l_2)^{\frac{1}{2}}}e^{\frac{R_2^*+R_5-(R_2+R_5^*)}{2}},~e^{\frac{\lam_3-\lam_4}{2}}=\frac{|k_1-k_2|\hat{\Lam}_6|k_1+l_2^*|^2e^{\frac{R_3-R_6}{2}}}{|k_1+k_2^*|^2|k_1-l_2|\hat{\Lam}_7},\nonumber\\
&&e^{\frac{\Gamma_{21}-\ga_{21}}{2}}=\frac{(k_2-l_2)^{\frac{1}{2}}(k_1-l_2)(k_1^*-l_2^*)^{\frac{1}{2}}}{(k_2+l_2^*)^{\frac{1}{2}}(k_1+l_2^*)(k_1^*+l_2)^{\frac{1}{2}}}e^{\frac{R_6}{2}},~e^{\frac{\lam_7-\lam_6}{2}}=\frac{(k_1-k_2)(k_2-l_2)^{\frac{1}{2}}\hat{\Lam}_4}{|k_1+k_2^*|^2(k_2^*+l_2)^{\frac{1}{2}}\hat{\Lam}_3}e^{\frac{R_3}{2}},\nonumber\\
&&\hat{\Lam}_5= (|\alpha_{1}^{(1)}|^2|k_1-k_2|^2|k_1+l_2^*|^2+|\alpha_{1}^{(2)}|^2|k_1-l_2|^2|k_1+k_2^*|^2)^{1/2},\nonumber\\
&&e^{\frac{R'-\vsa_{22}}{2}}=\frac{|k_1-k_2||k_1-l_2|\hat{\Lam}_5}{|k_1+k_2^*|^2|k_1+l_2^*|^2(k_1+k_1^*)},~e^{\frac{\vsa_{22}}{2}}=\frac{|k_2-l_2|}{|k_2+l_2^*|}e^{\frac{R_3+R_6}{2}},~e^{\frac{R_3-R_6}{2}}=\frac{|\alpha_{2}^{(1)}|(l_2+l_2^*)}{|\alpha_{2}^{(2)}|(k_2+k_2^*)},\nonumber\\
&&\hat{\Lam}_6=(|\alpha_{1}^{(1)}|^2|k_1-k_2|^2+|\alpha_{1}^{(2)}|^2|k_1+k_2^*|^2)^{1/2},~\hat{\Lam}_7=(|\alpha_{1}^{(1)}|^2|k_1+l_2^*|^2+|\alpha_{1}^{(2)}|^2|k_1-l_2|^2)^{1/2},\nonumber\\
&&e^{\frac{\Lam_{22}-\rho_1}{2}}=\frac{(k_2-l_2)^{\frac{1}{2}}}{(k_2+l_2^*)^{\frac{1}{2}}}e^{\frac{R_6}{2}},~e^{\frac{\mu_{22}-\rho_2}{2}}=\frac{(l_2-k_2)^{\frac{1}{2}}}{(k_2^*+l_2)^{\frac{1}{2}}}e^{\frac{R_3}{2}},~e^{R_1}=\frac{|\alpha_{1}^{(1)}|^2}{(k_1+k_1^*)^2},~e^{R_2}=\frac{\alpha_{1}^{(1)}\alpha_{2}^{(1)*}}{(k_1+k_2^*)^2},\nonumber\\
&&e^{R_3}=\frac{|\alpha_{2}^{(1)}|^2}{(k_2+k_2^*)^2},~e^{R_4}=\frac{|\alpha_{1}^{(2)}|^2}{(k_1+k_1^*)^2},~e^{R_5}=\frac{\alpha_{1}^{(2)}\alpha_{2}^{(2)*}}{(k_1+l_2^*)^2},~e^{R_6}=\frac{|\alpha_{2}^{(2)}|^2}{(l_2+l_2^*)^2}.\nonumber
\end{eqnarray}
\section{Numerical stability analysis corresponding to Figs. 5(a) and 5(b) under perturbation}
In this appendix, we wish to point out the stability nature of the obtained nondegenerate soliton solutions numrerically using Crank-Nicolson procedure even under the addition of suitable white noise or Gaussian noise to the initial conditions. Specifically we consider the shape preserving collision of symmetric double hump solitons discussed in Figs. 5. For this purpose, we have considered the Manakov system (\ref{e1}) with the initial conditions, 
	\begin{eqnarray}
	q_j(0,t)=[1+ A \zeta(t) ]q_{j,0}(t), ~j=1,2.
	\end{eqnarray}
	In the above, $q_{j,0}$'s, $j=1,2$, are the initial conditions obtained from the nondegenerate two-soliton solution Eqs. (\ref{3.7a})-(\ref{3.7c}) at  $z=-10$. Here $A$ is the amplitude of the white noise and $\zeta(t)$  represents the noise or fluctuation function. The white noise was created by generating  random numbers in the interval $[-1,1]$.  To fix the initial conditions in the numerical algorithm, we consider the same complex parameter values which are given for the figures 5(a)-5(b) in Sec. IV. We also consider the space and time step sizes, respectively, as $dz=0.1$ and $dt=0.001$  in the numerical algorithm. To  study the collision scenario of double-hump solitons (Figs. 17(a) and 17(b)) under perturbation we fix the domain ranges for $t$ and $z$ as $[-45,45]$ and $[-10,10]$, respectively.  
	
	First, we consider 10\% ($A=0.1$) of random perturbation on the intial solution of Manakov system.  For this strength of perturbation, we do not observe any significant change in the profile as well as in the dynamics of the nondegenerate solitons apart from a slight change, which is insignificant, in the amplitudes of double-hump solitons after the collision. This is illustrated in Figs. 17(c) and 17(d). Then we study the stability with 20\% white noise ($A = 0.2$), which is a stronger perturbation, for the double-hump solitons. Such a study is demonstrated in Figs. 17(e) and 17(f). The numerical analysis shows that the double-hump soliton profiles still survive after the collision under as strong as 20\% perturbation apart from a slight distortion in the amplitudes. This ensures the stability of nondegenerate solitons against perturbations of the above type of noise.
	
	Similarly we have also verified the stability of nondegenerate solitons with Gaussian noise perturbation as well. 
	\begin{figure}
		\centering
		\includegraphics[width=0.95\linewidth]{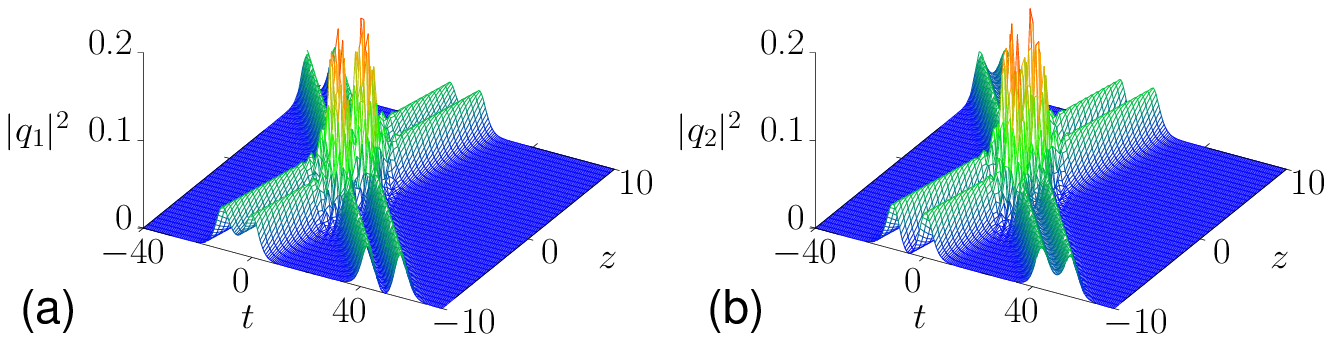}\\
		\includegraphics[width=0.95
		\linewidth]{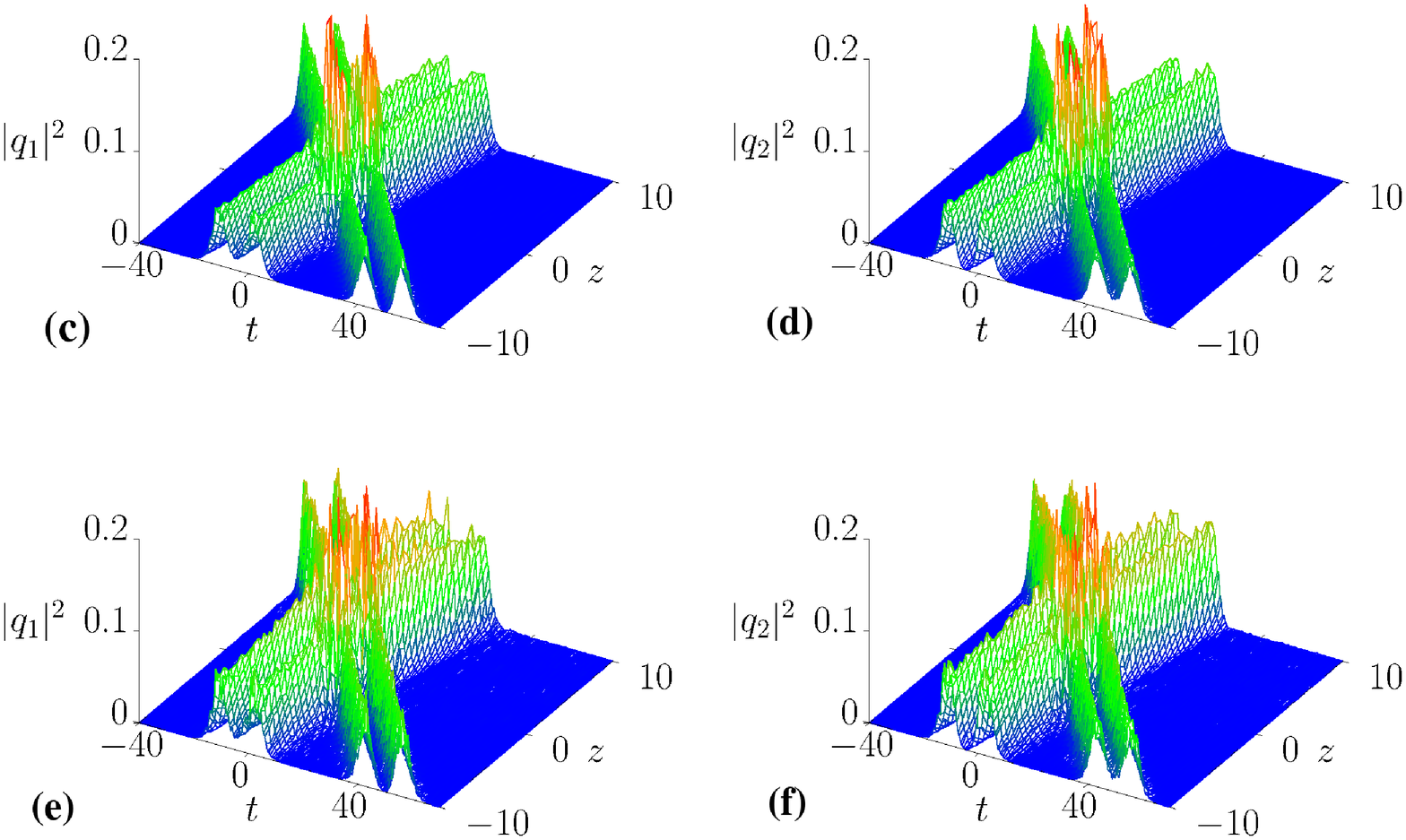}
		\caption{Numerical plots of shape preserving collision of nondegenerate symmetric double hump solitons  with 10\% and 20\% white noise as perturbations: (a) and (b) denote the elastic collision of two symmetric double hump solitons without perturbation.  (c) and (d) denote the collision with 10\% white noise. (e) and (f) represent the collision with 20\% strong white noise as perturbation.   }
		\label{f1}
	\end{figure}

\end{document}